\documentclass[letterpaper,11pt,leqno]{article}
\usepackage{paper}
\usepackage{lscape}
\hypersetup{pdftitle={Market Beliefs about Open vs. Closed AI}}

\begin{document}

\title{Market Beliefs about Open vs. Closed AI}

\author{Daniel Bj\"orkegren
%
\thanks{Bj\"orkegren: Columbia University, dan@bjorkegren.com. Thanks to Isaiah Andrews, Fernando Cirelli, Maryam Farboodi, Basil Halperin, Stephen Terry, and Tom Zimmermann for helpful conversations. Disclosure: D.B. serves as an Advising Fellow for the Microsoft AI Economy Institute and Nonresident Fellow at the Center for Global Development. }}

\date{\today}

\begin{titlepage}
\maketitle

Market expectations about AI's economic impact may influence interest rates. Previous work has shown that US bond yields decline around the release of a sample of mostly proprietary AI models \citep{andrews_markets_2025}. I extend this analysis to include also open weight AI models that can be freely used and modified. I find long-term bond yields shift in opposite directions following the introduction of open versus closed models. Patterns are similar for treasuries, corporate bonds, and TIPS. The different movements suggest that  that markets may anticipate open and closed AI advances to have different economic implications, and that the cumulative impact of AI releases on bond yields may be more muted.

\end{titlepage}

\section{Introduction}\label{s:introduction}

There is wide disagreement about the economic implications of Artificial Intelligence. Some suggest that AI will have muted effects on economic growth \citep{acemoglu_simple_2025}; others have considered the possibility that AI will lead to substantial growth or material abundance \citep{brynjolfsson_artificial_2017,aghion_artificial_2019,trammell_economic_2023}, or even existential risk to humankind \citep{yudkowsky_if_2025}. \citet{chow_transformative_2025} suggest that interest rates are a measurable object that could partly reveal what market participants believe will happen. In either extreme scenario, theory suggests that long-term real interest rates would rise: it is less important to save resources for futures that either have complete abundance or in which we do not exist. 

A pioneering study by \citet{andrews_markets_2025} develops an empirical test based on news about AI model releases. It documents that long-run US bond yields do indeed shift around the release of a set of AI Large Language Models (LLMs) by frontier labs. The average estimated effects are robust, and downward. A puzzle is that they are cumulatively large. They imply that a set of AI model releases in 2023-4 cumulatively shifted long-term treasury yields 1.5 percentage points lower than they otherwise would have been (relative to an average of around 4.25\%). When interpreted as being driven by changes in expected consumption using an asset pricing model, these estimates suggest that these releases have led to a total 3 percentage point drop in annual arrival probability of bliss or doom, or a 0.6-3 percentage point drop in expected consumption growth. That estimated effects are in the same direction is consistent with AI advances continuing to surprise in the same direction on average. The events coincide with forecasts of Artificial General Intelligence (AGI) moving sooner in time, so that direction is the opposite of a common prediction that extreme outcomes or increased growth becomes more likely as capabilities of Artificial Intelligence grow.

\citet{andrews_markets_2025} considers major releases by the labs that produced the highest rated models. The release of those models and the understanding of their capabilities is informative of the frontier of AI capabilities. Because the labs producing the most advanced models tend to keep their internals secured, almost all the models included in that analysis are proprietary models. However, there is another broad category of AI models that is mostly omitted from this
exercise: open weight models that have similar architectures to the proprietary models but more permissive licensing which allows them to be freely downloaded, modified, and run on any computer.%
\footnote{Some refer to these models as `open source'; however, labs typically release only the estimated model weights, not a clear description of what content the model was trained on, or the code to produce it, and so are more accurately referred to as open weight \citep{open_source_initiative_open_nodate}.}

\begin{figure}
     \centering
     \includegraphics[width=\textwidth]{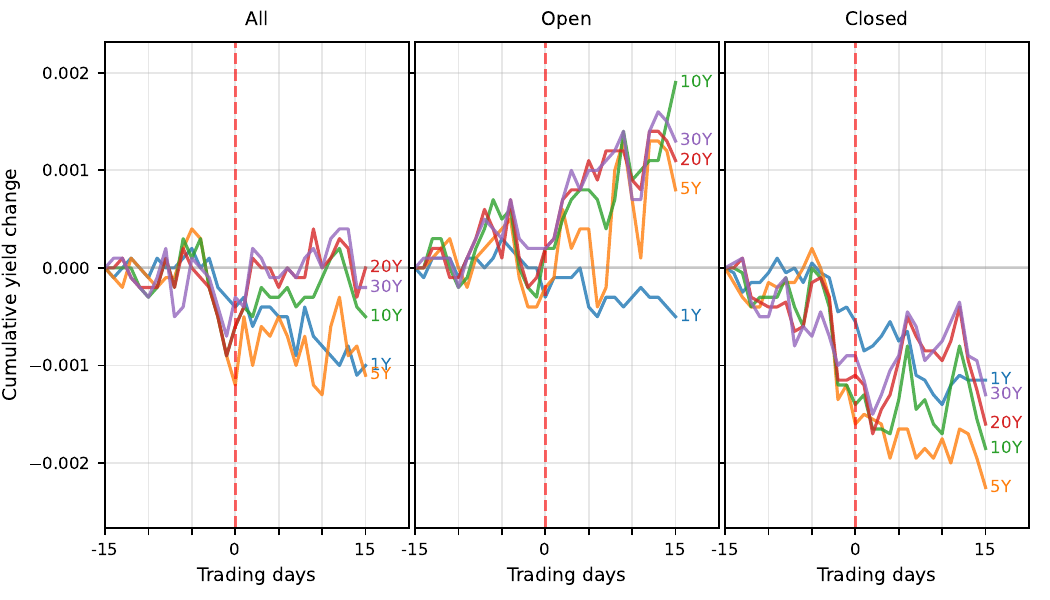}     
    \caption{US treasury bond movements around AI releases}
        \label{fig:yield_event_study}
\note{Median change across release events relative to 15 business days before release. 0=event day. Panels consider (a) all AI releases, (b) all open weight releases, and (c) all closed/proprietary releases.  Figure~\ref{fig:yield_event_study2023_4} shows the same plot restricted to 2023-4 releases. Constant maturity duration noted in years.}
\end{figure}

I revisit the exercise of \citet{andrews_markets_2025}, also considering the release of open weight models, over the time span late 2022-5. I replicate their finding that the yields of long dated U.S. treasury bonds decline around the median release of proprietary AI models, by roughly 10 basis points over a $\pm$15 trading day window. However, the pattern is opposite around the median release of open weight AI models: bond yields \emph{increase}, by roughly 10 basis points in this simplest specification. These median shifts that differ by a magnitude of roughly 20 basis points are shown in the latter two panels of Figure~\ref{fig:yield_event_study}. When both types of releases are pooled, the net effect on is roughly zero, as shown in the first panel.

These results suggest additional nuance to the phenomenon, in two respects. First, when open releases are included, the cumulative effect of AI releases on bond yields is much smaller, and hard to distinguish from zero. If interpreted as arising from changes in expected consumption using the asset pricing model, that suggests that beliefs about probabilities of doom or bliss, or consumption growth, may not have moved cumulatively around this broader set of AI releases. It is still possible that these beliefs have been altered by news occurring between releases. Second, that open and closed releases have opposite effects suggests that markets may believe that the two types of releases have different implications. Opposing impacts would be consistent with bond yields moving up or down depending on the respective rates of progress in different classes of AI models. That could be consistent with markets responding to a `race' between open and closed systems. If one applies the asset pricing interpretation, that would suggest that open advances have opposite implications for beliefs about consumption growth, or the probability of doom or bliss. But given that the two kinds of releases both represent advances in similar AI technology, the opposing pattern admits the possibility that the phenomenon may have alternate explanations.

There is reason to believe that open AI systems may be economically important. Following Meta's release of its Llama LLM in early 2023, the market became flooded with variants of Llama, and other labs released open weight models, including Mistral, Alibaba, DeepSeek, and eventually also Google and OpenAI. Open weight models tend to be behind the frontier, but can be replicated and more flexibly integrated into processes. There has been demand for these types of models: Meta announced that its Llama open-weight models have been downloaded nearly 350 million times (\citeyear{meta_10x_2024}), and a pair of papers find that open weight models can be used at 70-90\% lower prices than equivalently intelligent closed models \citep{demirer_emerging_2025,nagle_latent_2025}.

The proliferation of open weight models has been the object of debate within the AI community. Some suggest that open models will diffuse the benefits of AI, and could result in gains accruing widely through the economy, not just to select tech firms. Others claimed that open proliferation of AI technology would cause extreme or even existential risks and that these systems should be regulated \citep{harris_ai_2023,seger_open-sourcing_2023}. Practically, because open models can be adapted and operated by all types of firms, integrating open models into the economy may require different investments; for example, individual firms may train their own models or build private compute clusters. How open models advance relative to closed models thus could have important economic implications.

I replicate and extend the analysis in \citet{andrews_markets_2025}, including both proprietary and open releases. In addition to reproducing estimates using median differences across time, I use regression specifications that account for overlaps between event windows and estimate deviations relative to a prevailing trend. The resulting mean changes follow the same pattern and are slightly larger. I find similar patterns in corporate bonds. Patterns are similar in inflation protected bonds (TIPS) but more muted: the different model releases also coincide with opposing changes in inflation expectations and a US dollar currency index. Interpreting these estimates as causal effects of AI model announcements relies on the original assumption of \citet{andrews_markets_2025} that the release dates of models are random with respect to other variables that would influence yields.

Results are similar when the sample is restricted to model releases between 2023-4, though statistical power decreases. The pattern does not appear to be driven by a particular model release, and is also present in shorter windows around events. Across robustness tests, the most precisely estimated findings tend to be that the yield response is differential between open and closed releases, and that yields decline around the release of closed models relative to trend. In some tests, the yield increases around open releases are harder to differentiate statistically from the prevailing trend. However, the estimates admit the possibility that both the differential of the effects between open and closed releases, and the individual effects of each type of release, are large in magnitude.

Closed and open models differ in more ways than just their licensing: closed models tend to be closer to the frontier, and Chinese labs tend to release their model weights while many US labs keep their models closed. However, the strongest factor along which I find heterogeneity in bond yield responses is licensing, both in terms of magnitude and in terms of statistical significance. I do not find meaningful heterogeneity in bond responses by performance of the model or by distance to the frontier at time of release (both measured using scores on LMArena), or by whether forecasters moved the anticipated date of AGI arrival sooner or later around the date of model release (using estimates from Metaculus). Within open releases, I do not find meaningful heterogeneity by whether models are developed by Chinese or non-Chinese labs.

There is some suggestive indication that other market indicators may also shift around these release dates. It is possible that open releases coincide with an increase in uncertainty (VIX) relative to closed releases; and lower equity returns for US tech firms closely associated with AI, particularly in 2023-4. However, results are too noisy to say with much confidence.

These findings suggest that the pattern of market movements around AI releases may not be due to the \emph{technology} per se: the technology is similar between open- and closed-weight LLMs. However, there remain many possible explanations for the patterns documented here. It could suggest that movements arise due to beliefs about who the gains of the technology will accrue to, or how the technology will diffuse. It could suggest that open licensing induces firms to differentially seek access to credit to finance investments. \citet{andrews_markets_2025} interpret their results as possibly suggesting lower growth in consumption, or a lower probability of extremely good or bad outcomes. In that interpretation of these results, that would suggest that proprietary model releases suggest a lower probability of `bliss' or `doom', and open models a higher probability.

It could also be that the dates of AI model releases are systematically correlated with market movements for other reasons, so that the estimated effects are not attributable to AI advances. That would require that open- and closed-weight model release dates be systematically different. An important caveat to all of these results is that they are based on the small amounts of data available as of this writing. We are likely to learn much more about these underlying phenomena as the technology continues to advance and diffuse.

\paragraph{Related literature}
This paper relates to work considering how economies may balance the risks and economic growth potential of AI. \citet{jones_ai_2024} models an optimal strategy to experiment with AI to learn about its risks, and \citet{jones_how_2025} considers how much a society should be willing to pay to avert existential risk from AI. In a cross-country exercise, \citet{chow_transformative_2025} finds that growth expectations correlate with higher long-term real interest rates, suggesting that the latter can be used as an indicator of the former. \citet{andrews_markets_2025} implement such a test and find statistically robust impacts of AI releases on financial markets. This paper builds on that important study.

This paper also relates to a rich literature analyzing the effects of macroeconomic announcements on bonds and equities (for example, \citet{nakamura_high-frequency_2018} and \citet{bauer_alternative_2023}).

This paper also joins micro evidence on the market for AI models \citep{demirer_emerging_2025,nagle_latent_2025}, and papers describing the tradeoffs and considerations in making AI models open \citep{anderljung_frontier_2023,bommasani_considerations_2024,kapoor_societal_2024}. We describe more of this debate in the next section.

\section{Background}\label{s:background}

Openness has been a point of contention in the development of general AI models. On OpenAI's founding in 2015, the organization said it would encourage its researchers to publish code, and it would share any patents with the world. Its first LLM, GPT-1, was released openly, with  code and model weights freely available. However, as it developed larger and more capable models, the organization described concerns that the models would be powerful enough to do harm, such as conduct destabilizing influence campaigns. OpenAI released GPT-2 in stages: first an initial smaller, less capable version of the model, and only later the full version. However, with GPT-3 and beyond, the organization and much of the industry kept the weights of models closed and training procedures secret.

Some observers and closed model creators warned that open weight models can be customized to remove safeguards, or may have vulnerabilities that cannot be repaired centrally. Eliezer Yudkowsky, a commentator concerned about the risks of AI, said `open sourcing, you know, that's just sheer catastrophe' \citep{lexteam_368_2023}. (Many commentators use the term open source rather than open weight, but most open models provide only the final estimated weights, not the data or source code used to create them.) Ilya Sutskever, a co-founder and former chief scientist at OpenAI said, `I fully expect that in a few years it’s going to be completely obvious to everyone that open-sourcing AI is just not wise.' But other observers of the industry noted that the shift to secrecy coincided with the development of models that were economically useful, and that if models were kept secret, more of the value they generated might accrue to their creators.

Amidst this backdrop, on February 24, 2023, Meta announced an LLM (`Llama') for which it shared code and training details, and for which weights could be obtained through an application process for approved researchers and officials. However, the weights of the model were leaked to the public on March 3, 2023, allowing access outside the application process. While Meta initially fought to take down the model weights that had begun to spread, a few months later on July 18, it released Llama 2 with openly downloadable weights. These releases spurred the creation of many variants of models built on Llama. These were followed by additional open weight models released by Mistral, a French lab, and further releases by Chinese labs Alibaba and DeepSeek.

As more open weight models became available, commentators continued to debate the risks and benefits of open models, and policymakers debated restrictions. Harris wrote, `open-source AI is uniquely dangerous', and Gary Marcus, an influential voice in AI policy at the time wrote, `how we regulate open-source AI is THE most important unresolved issue in the immediate term' \citep{harris_open-source_2024}. The Biden administration asked AI firms to secure model weights \citep{house_fact_2023}, and proposals to prohibit release of model weights for sufficiently powerful systems have circulated widely, including for the South Korea AI summit \citep{cfg_policy_2024}. However, others argued that open weight AI could act as a check on large tech firms, or disperse the economic benefits of AI further \citep{brooks_open-source_2024}. Notably, a poll of AI experts is split on the question, `Should we ban future open-source AI models that can be used to create weapons of mass destruction?' \citep{youcongress_should_nodate}.

Despite these concerns, powerful open models have become a standard part of the market. Since 2023, both the number of open weight models, and the number of open weight model creators, has grown faster than the equivalents for closed models \citep{demirer_emerging_2025}. Some labs that primarily released proprietary models also released open models, including Google's Gemma family and OpenAI's gpt-oss. The performance of open weight models has also increased, shrinking the gap with the proprietary model frontier. \citet{demirer_emerging_2025} finds that the market share of open vs. closed models in one marketplace fluctuates with their relative price and capabilities, suggesting that open models are generating competition for closed models.

\begin{figure}[t]
\includegraphics[width=\columnwidth]{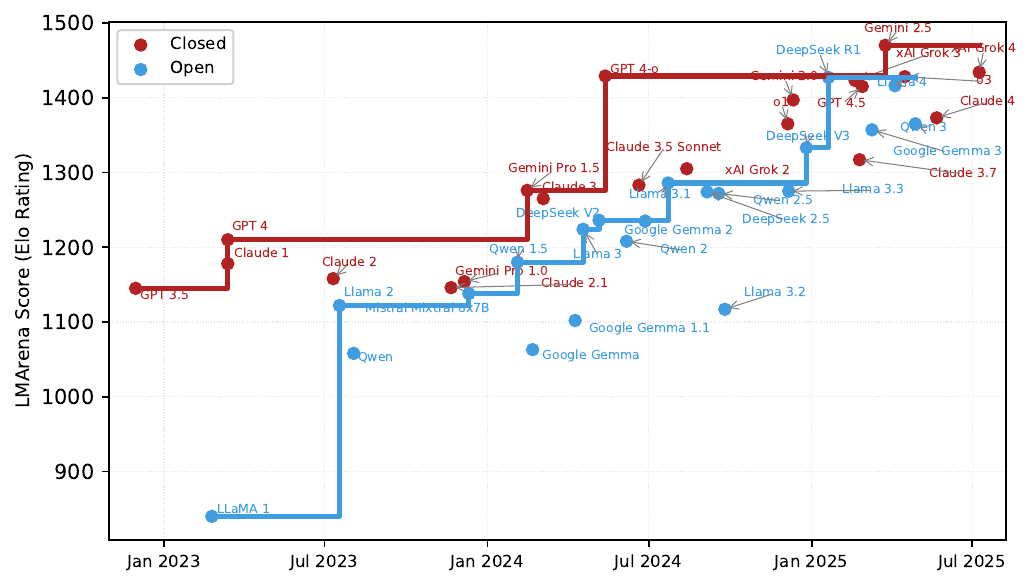}
\caption{AI model releases and frontiers}
\note{Release dates of AI models in the sample, with associated LMArena score on text (Elo rating) as of August 4, 2025 \citep{llmleaderboard2025}, for models that could be matched. Solid lines represents the highest performance to date among open and closed models (frontiers).}
\label{f:lmarena_scores}\end{figure}

Figure~\ref{f:lmarena_scores} shows a performance score for each model in the data (as rated by LMArena), and the frontier of the best score for open and proprietary models over this span of time.

\subsection{Data}\label{s:data}

I analyze the effects of model releases using data from multiple sources.

\paragraph{Model releases}

\citet{andrews_markets_2025} considers major LLM releases from OpenAI, Google, Anthropic, xAI, and DeepSeek. While DeepSeek's models are open weights, the only other models included are proprietary. To these I add open weight releases from Google (the Gemma family) and OpenAI (gpt-oss), and two new closed releases that came out after the posting of that paper (xAI Grok 4 and OpenAI GPT 5). I also add the major releases of major labs that released primarily open weight models during this time, including Meta, Alibaba, and Mistral.%
\footnote{These represent the builders of the top 50 models on LMArena as of August 4, 2025 \citep{llmleaderboard2025}; omitting Chinese labs Moonshot AI, Z.ai, MiniMax, and Tencent that have made more minor releases for English audiences, and omitting labs producing mostly derivatives of other labs' open weight models.} Following \citet{andrews_markets_2025}, I use the date of announcement on each lab's web page where available. The set of models includes 47 models; 23 open and 24 closed; dates are presented in Table~\ref{tab:model_release_dates}.

\paragraph{Model performance}
The LMArena creates a leaderboard of models. Users of the website may submit a query, and then are given two alternate responses generated by different models. They are asked to rate the two responses, blinded to which models were used. The pairwise ratings are turned into an Elo rating score. I use a snapshot of the scores as of August 4, 2025 \citep{llmleaderboard2025}. When a release includes multiple versions of a model (such as different sizes), I match to the highest rating of the versions.

\paragraph{Forecasts of AI capabilities}
The Metaculus website allows users to submit predictions of future events. We use the median answer to the question `When will the first general AI system be devised, tested, and publicly announced?' \citep{metaculus_when_2025}, as do \citet{andrews_markets_2025}.

\paragraph{Financial data}

I use bond yields from FRED as well as equity prices from Yahoo Finance.

\subsection{Descriptives}
An overview of model releases, AGI forecasts, and best LMArena score to date is shown in Appendix~Figure~\ref{f:event_timeline}. 

Open weight models differ from closed models in several ways. They tend to be further from the frontier, both absolutely given models available in August 2025, and relative to the best model at the time of release, as shown in Figure~\ref{f:perf_comparison_open_closed}(A) and (B). However, forecasts of AGI arrival shift similarly around the release of either type of model, as shown in Figure~\ref{f:perf_comparison_open_closed}(C). The sequence of these shifts is shown in Figure~\ref{f:agi_shift_histogram}.

\section{Analysis}\label{s:analysis}

Building on the workhorse specifications of \citet{andrews_markets_2025}, I consider event studies in price series $y$ (which is measured in levels for bond yields, and logs for equity prices). Each observation is a business date $t$. Define daily returns as,
\[
\Delta y_t = y_{t} - y_{t-1}.
\]

I seek to understand how cumulative returns are affected by the release of AI models. Let $t_i$ denote
the nearest business day on or before event $i$. There are $N$ events.

In the expanded set of events that includes open releases, a nontrivial number of the event windows overlap, as shown in Figure~\ref{f:event_timeline}. When event windows overlap, the pre-periods of some events overlap with the post-periods of other events. To address this, I estimate event-day fixed effects using a regression framework that accounts for possibility that some business days may be affected by multiple AI releases. As a robustness check, I also replicate the simpler median difference estimates from \citet{andrews_markets_2025}. 

\paragraph{Basic regression framework}
To build intuition, consider a simple version of the regression where all events are pooled. For each asset $j$, this includes window-day fixed effects,
\begin{equation*}
\Delta y_t = \sum_{i=1}^{N} \sum_{s=-W}^{W} \alpha_{s} \cdot \mathbb{1}_{t = t_{i} + s} + \mu + \varepsilon_t
\end{equation*}
where $\alpha_{s}$ is the fixed effect (daily return) for relative day $s$, and $W$ is the size of the window, which I set to 15 days in the main specifications to account for possible leakage in the preperiod, following \citet{andrews_markets_2025}.%
\footnote{Prediction markets offer one way to assess leakage; for example, an active market predicting the week in which OpenAI would release GPT-5 began to tick up on August 3, 4 days in advance of its actual release \citep{polymarket_gpt-5_2025}. However, information about the relative magnitude of improvement had trickling out for months.} 
The double sum accounts for all events and all relative days within each event. The constant term $\mu$ accounts for the trend in daily returns over the period. I include observations of $t$ from the $W$ days before the first event to $W$ days after the last event. When date $t$ falls within multiple event windows at the same relative day, that observation contributes to the estimation of each $\alpha_{s}$ for all relevant windows, accounting for overlap under the assumption that effects are additive. 

Cumulative returns from day $-W$ to day $r$ are obtained by summing daily estimates:
\begin{align}
    \Delta\hat{y}_{W,r} &= \sum_{s=-W}^{r} \hat{\alpha}_s
    \label{eq:cum_return_single}
\end{align}

I minimize sum of squared residuals, so that estimate $\Delta\hat{y}_{W,r}$ can be interpreted as the mean cumulative return as of day $r$, relative to trend. I report results for all values of $r$ within the windows; the later values may incorporate more of the time needed for the market to understand the implications of a model release.%
\footnote{For example, a substantial selloff of Nvidia on January 27, 2025, was attributed to the January 20 release of DeepSeek R1 \citep{bratton_nvidia_2025}.}

\paragraph{Comparison regression framework}

For comparing two groups of events (e.g., open vs. closed releases), I extend
the framework to estimate the contributions of both types of events jointly. Let $A$ and $B$ denote the two
groups, with event dates $t_i$ for $i \in A$ and $t_j$ for $j \in B$:
\begin{equation*}
\Delta y_{t} = \sum_{i \in A} \sum_{s=-W}^{W} \alpha_{s} \cdot \mathbb{1}_{t = t_{i} + s} +
         \sum_{j \in B} \sum_{s=-W}^{W} \beta_{s} \cdot \mathbb{1}_{t = t_{j} + s} + \mu +
         \varepsilon_{t}
\end{equation*}
where $\alpha_{s}$ and $\beta_{s}$ are separate day fixed effects for groups A and B.

Cumulative returns for each group from day $-W$ to day $r$ are obtained by summing daily estimates:
\begin{align}
    \Delta\hat{y}_{W,r}^A &= \sum_{s=-W}^{r} \hat{\alpha}_s \label{eq:cum_return_paired}\\
    \Delta\hat{y}_{W,r}^B &= \sum_{s=-W}^{r} \hat{\beta}_s \nonumber
\end{align}
and the difference between cumulative returns of the two groups up to $r$ can be computed as $\sum_{s=-W}^{r} \left[\hat{\alpha}_s - \hat{\beta}_s \right]$.

This joint estimation handles dates that overlap between the two groups and provides coefficients that can be directly compared. To account for autocorrelation, I compute heteroskedasticity- and autocorrelation-consistent (Newey–West) standard errors with 30-business-day lag length.

\paragraph{Results}

\begin{figure}
     \centering
        \begin{subfigure}[b]{\textwidth}
         \centering
         \includegraphics[width=0.4\textwidth]{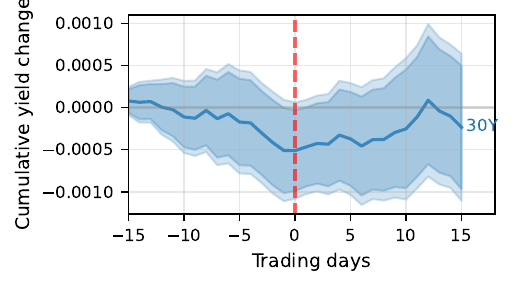}
     \end{subfigure}
     \\
        \begin{subfigure}[b]{\textwidth}
         \centering
         \includegraphics[width=0.4\textwidth]{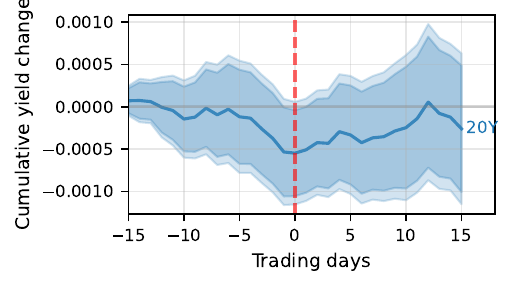}
     \end{subfigure}
     \\
        \begin{subfigure}[b]{\textwidth}
         \centering
         \includegraphics[width=0.4\textwidth]{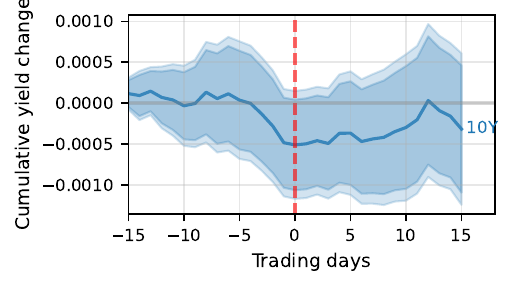}
     \end{subfigure}
     \\
        \begin{subfigure}[b]{\textwidth}
         \centering
         \includegraphics[width=0.4\textwidth]{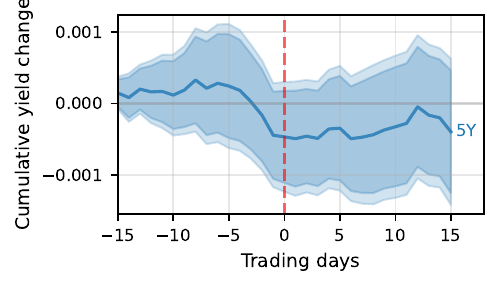}
     \end{subfigure}
     \\
        \begin{subfigure}[b]{\textwidth}
         \centering
         \includegraphics[width=0.4\textwidth]{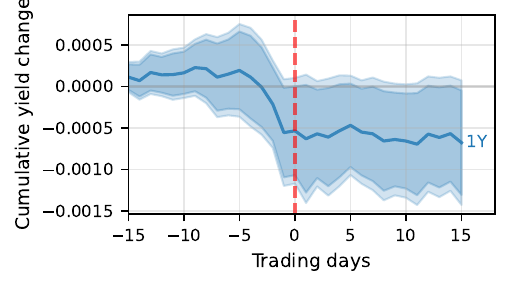}
     \end{subfigure}
     
     \caption{US treasury bond yield event study regression estimates: all events pooled}
     \label{fig:regression_pooled}
    \note{Estimates of cumulative returns. Cumulative returns from 15 days prior to release, using equation~\eqref{eq:cum_return_single}. Constant maturity duration noted in years. 90\% and 95\% confidence intervals are shaded.}
\end{figure}

Results for all events pooled for US treasury bonds are shown in Figure~\ref{fig:regression_pooled}, given the estimates of equation~\eqref{eq:cum_return_single}. As suggested by Figure~\ref{fig:yield_event_study}, the yield impact is smaller and not statistically significant from zero when open and closed releases are pooled. The remainder of the paper focuses on the split between open and closed releases.

\begin{figure}
     \centering
        \begin{subfigure}[b]{\textwidth}
         \centering
         \includegraphics[width=0.9\textwidth]{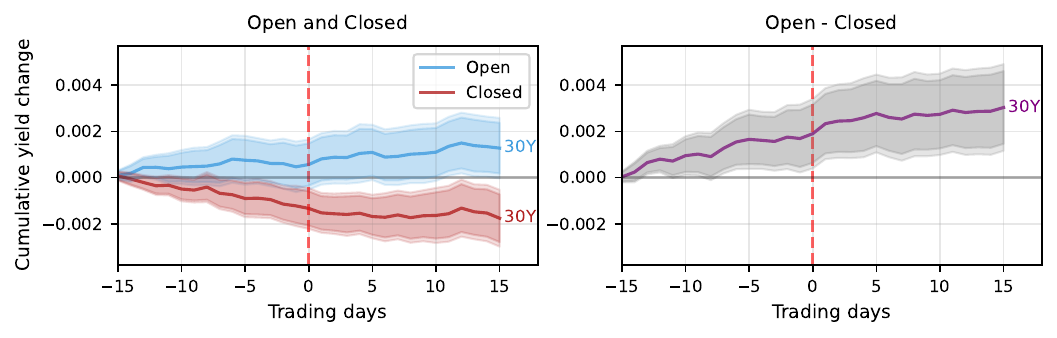}
     \end{subfigure}
     \\
        \begin{subfigure}[b]{\textwidth}
         \centering
         \includegraphics[width=0.9\textwidth]{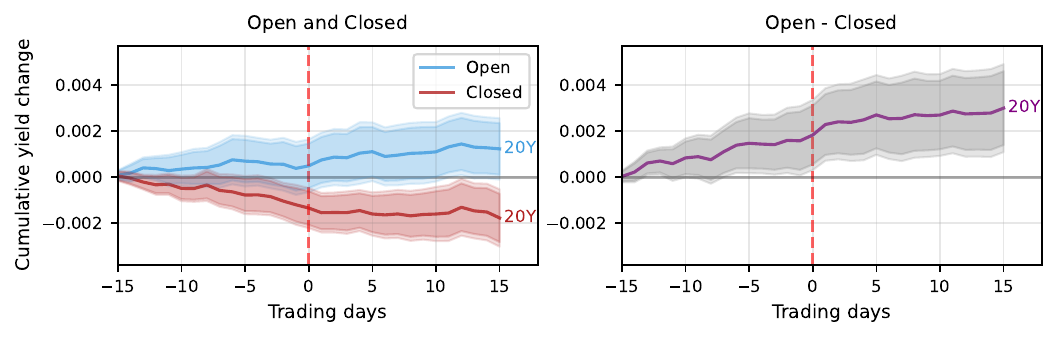}
     \end{subfigure}
     \\
        \begin{subfigure}[b]{\textwidth}
         \centering
         \includegraphics[width=0.9\textwidth]{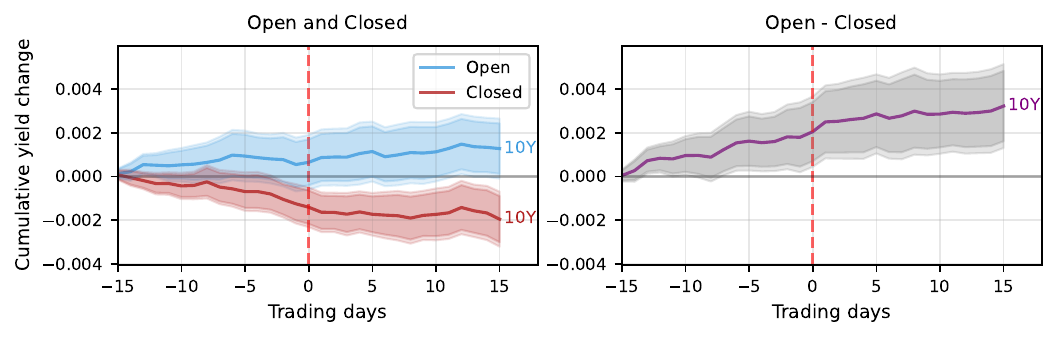}
     \end{subfigure}
     \\
        \begin{subfigure}[b]{\textwidth}
         \centering
         \includegraphics[width=0.9\textwidth]{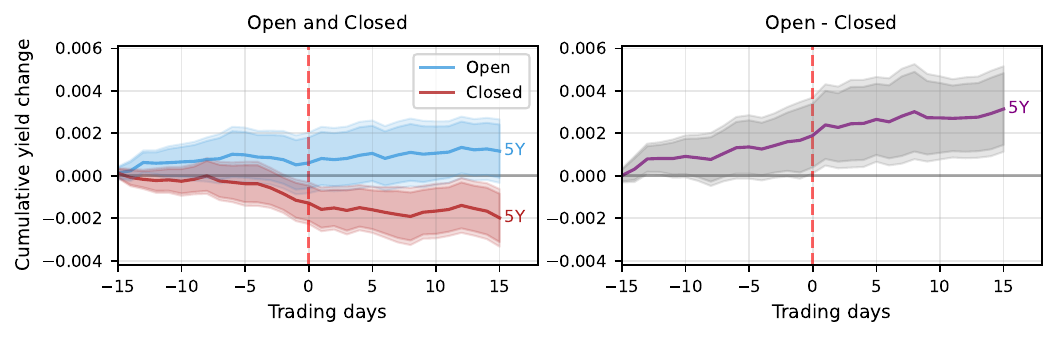}
     \end{subfigure}
     \\
        \begin{subfigure}[b]{\textwidth}
         \centering
         \includegraphics[width=0.9\textwidth]{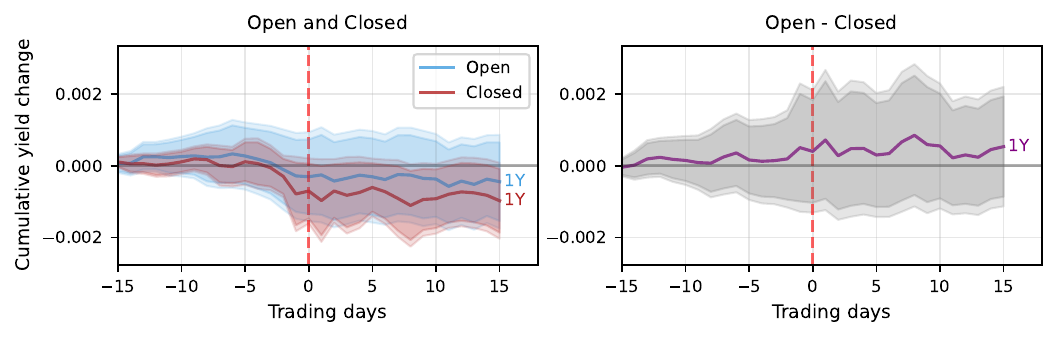}
     \end{subfigure}
     
     \caption{US treasury bond yield event study regression estimates: split}
     \label{fig:regression_difference}
    \note{Estimates of cumulative returns. The left panels show cumulative returns from 15 days prior to release, using equation~\eqref{eq:cum_return_paired}. The right panels show the estimated difference. Figure~\ref{fig:regression_difference2023_24} shows the same plot restricted to 2023-4 releases. Constant maturity duration noted in years. 90\% and 95\% confidence intervals are shaded.}
\end{figure}

Results for US treasury bonds are shown in Figure~\ref{fig:regression_difference}, and in table format in Table~\ref{tab:regression_estimates_bonds}. In the figure, each row corresponds to a different maturity term. For both types of releases there is some suggestive movement in the 7 business days leading up to a release, which widens until the release and then becomes stable. Intermediate and long maturity bonds are most affected; changes on 1 year maturity bonds are very slightly negative for both types of releases and not significantly different. On 30 year bonds, yields decrease by 13-18 basis points (bps) 15 days after a closed release (p-values 0.01 or below); and increase by 11-15 bps after open releases (p-values 0.04-0.08). The difference between these responses is 27-30 bps (p-values $<$ 0.01).

For corporate bonds above 3 year duration, results are similar, as shown in Figure~\ref{fig:regression_difference_corporate} and Table~\ref{tab:regression_estimates_corporate}. For inflation protected bonds (TIPS), the movement on both open and closed releases are less extreme, shown in Figure~\ref{fig:regression_difference_tips} and Table~\ref{tab:regression_estimates_tips}: for 30 year, a estimated decrease of 7-10 bps following closed releases (p-values 0.03-0.16) and increase of 7-9 bps following open releases (p-values 0.08-0.19). There remains a difference but it is narrower, 15-18 bps (p-values 0.01-0.04). Corroborating this, Figure~\ref{fig:regression_difference_other} panel (a) finds that inflation expectations move up on the release of open models, and down on the release of closed models. (Figure~\ref{fig:otherindicators} shows that increases in inflation expectations coincide with increased gaps between treasuries and TIPS.)

\subsection{Robustness}

The same pattern is evident if one restricts the sample to 2023-4 releases as \citet{andrews_markets_2025} do in the main portion of their analysis, shown in Figure~\ref{fig:regression_difference2023_24}.%
\footnote{For simplicity I include all 2023-4 releases, not just those that include $\pm W$ day windows fully within those years.} 
The signs and magnitudes remain similar. Confidence intervals around yield changes for open releases include zero, but yield changes corresponding to closed releases and the difference between open and closed are significantly different from zero.

I plot the path of yields for each individual event in Figures~\ref{fig:yields_by_event}, \ref{fig:yields_by_event2}, and \ref{fig:yields_by_event3}. The entire distribution of paths is shifted between open and closed releases for longer maturity bonds, and the pattern is not driven by particular releases.

In small samples, permutation tests can provide more robust measures of uncertainty. I follow the procedure outlined by \citet{andrews_markets_2025}, developing extensions for the regression procedure, and to compute differences, in Appendix~\ref{a:robustness}. Results for the regression specification are presented in Figure~\ref{f:permutation_open_and_closed_ols}, and are similar to the main results.

Point estimates are similar when changes are estimated using median changes, $Median_i\left(y_{t_i+W} - y_{t_i-W}\right)$, following \citet{andrews_markets_2025}. Median changes for the entire period are shown in Figure~\ref{fig:yield_event_study} and for 2023-4 releases in Figure~\ref{fig:yield_event_study2023_4}. This approach may be less sensitive to extreme outcomes, but does not account for overlap. Permutation test results for median changes are presented in Figure~\ref{f:permutation_open_and_closed_medianchange}, which are similar to the main results except that while yield increases around open model releases tend to lie above the placebo average for long duration bonds, but most of the paths lie within the 90\% range of the placebo distribution. The difference in significance of open releases between the two estimators may result from a combination of interference from overlaps and measuring median versus mean changes.%
\footnote{If I estimate the regression specification by minimizing the absolute deviation, and then accumulate the estimated median daily deviations, results are  similar but attenuated, as shown in Figure~\ref{f:permutation_open_and_closed_lad}. Note that the sum of median daily differences can differ from the median cumulative difference. For example, if we believed that all releases had a step function impact but leakage varied so the timing of the step was different between releases, all of the median daily differences could be zero even if the median long difference were large.} 

The divergence is also clear when events are left out using a jackknife test, though for some subsamples the difference loses statistical significance. Regression results are shown in Appendix Figure~\ref{f:permutation_open_and_closed_ols_jackknife_drop1} (dropping one pair of events; one from open and one from closed) and Appendix Figure~\ref{f:permutation_open_and_closed_ols_jackknife_drop2} (dropping two pairs of events). Median changes are shown in Appendix Figure~\ref{f:permutation_open_and_closed_medianchange_jackknife_drop1} (one pair) and Figure~\ref{f:permutation_open_and_closed_medianchange_jackknife_drop2} (two pairs).

The divergence is also clear under different window lengths. It is similar for $W=30$ day windows around the events, using OLS (Figure~\ref{f:permutation_open_and_closed_ols_30day}) or the median change estimator (Figure~\ref{f:permutation_open_and_closed_medianchange_30day}). Under these longer windows, the yield paths are approximately zero until approximately day -15, which supports the window length of the main specification. A smaller divergence is present in a $W=5$ day windows around the events, whether using OLS (Figure~\ref{f:permutation_open_and_closed_ols_5day}) or the median change estimator (Figure~\ref{f:permutation_open_and_closed_medianchange_5day}). In those specifications it is statistically harder to distinguish the yield path of open releases from trend, but the difference between the paths of open and closed releases and the path of closed releases are relatively less likely in the placebo distributions.

Another robustness exercise controls for news that may have happened simultaneously, using the residualization procedure of \citet{andrews_markets_2025}, in Appendix~\ref{a:residualization}. AI releases may themselves contribute to these indices, but nonetheless I find that it can be harder to distinguish the yield path of open releases from trend but the magnitude remains large, and the difference between open and closed releases remains large and statistically significant.

\subsubsection{Alternate dimensions of heterogeneity}

\begin{figure}
    \centering
     \begin{subfigure}[b]{\textwidth}
         \centering
         \includegraphics[width=0.9\textwidth]{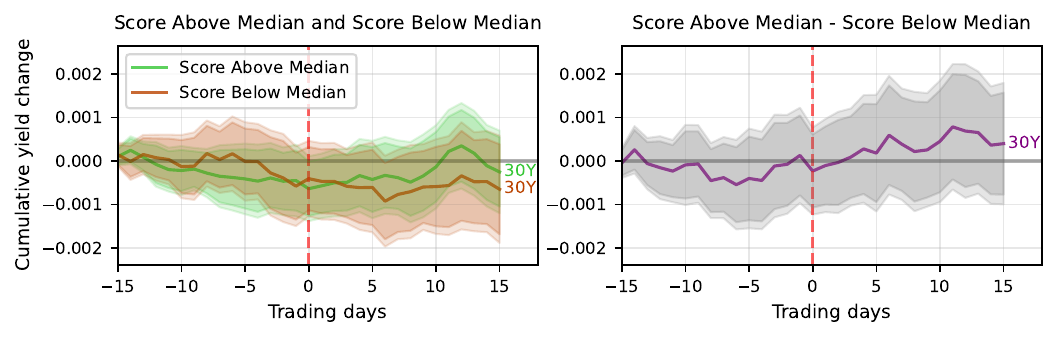}
         \caption{LMArena score}
     \end{subfigure}
     \\
     \begin{subfigure}[b]{\textwidth}
         \centering
         \includegraphics[width=0.9\textwidth]{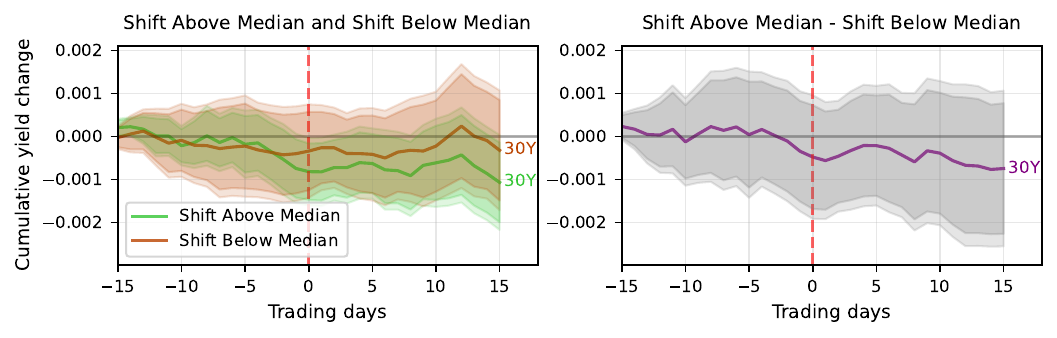}
         \caption{LMArena shift relative to frontier}
     \end{subfigure}
     \\
     \begin{subfigure}[b]{\textwidth}
         \centering
         \includegraphics[width=0.9\textwidth]{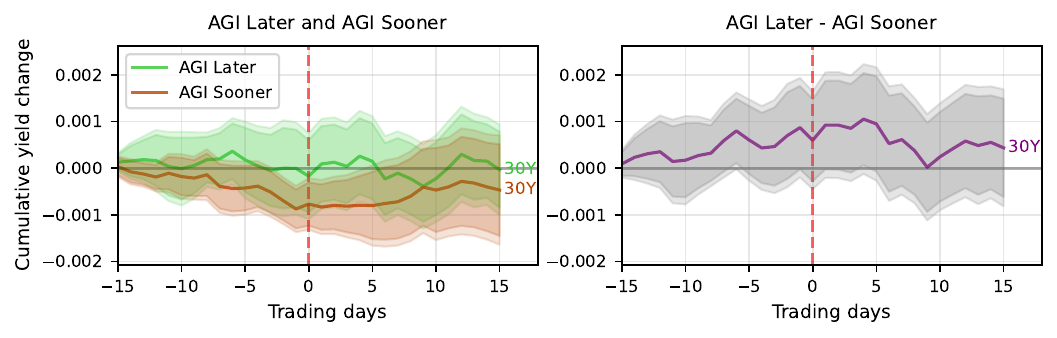}
         \caption{AGI forecast impact}
     \end{subfigure}
     \\
     \begin{subfigure}[b]{\textwidth}
         \centering
         \includegraphics[width=0.9\textwidth]{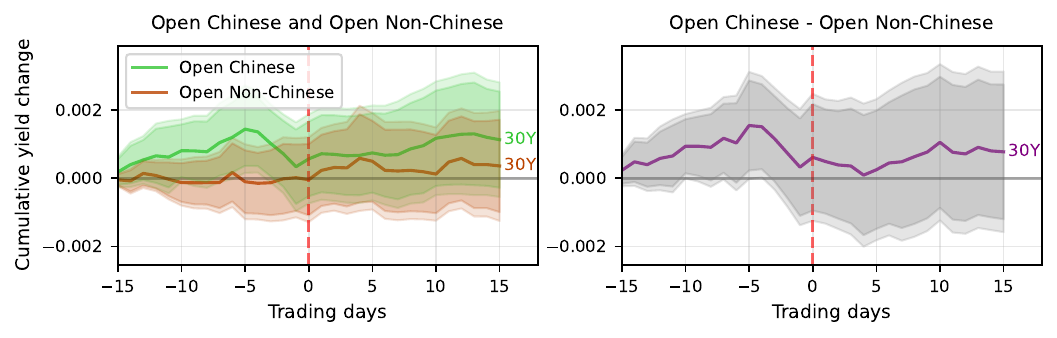}
         \caption{Country origin (open models)}
     \end{subfigure}
     \caption{US treasury bond (30Y) yield event study regression: other splits}
     \label{fig:regression_other_splits} 
    \note{Estimates of cumulative returns. The left panels show cumulative returns from 15 days prior to release, using equation~\eqref{eq:cum_return_paired} and splitting by the designated characteristic. The right panels show the estimated difference. 30 year maturity duration. 90\% and 95\% confidence intervals are shaded.}
\end{figure}

Open and closed model releases differ in more than just licensing: open models also tend to be further from the frontier, and Chinese labs are more likely than US labs to release models under open licenses. Thus it could be that the documented differences arise from correlated factors. Figure~\ref{fig:regression_other_splits} shows the equivalent tests when the sample of events is divided based on other factors, including whether the LMArena score is above or below median (panel A), whether the distance to the LMArena frontier at time of release is above or below median (panel B), whether forecasters anticipated date of AGI arrival shifts sooner or later around the date of model release (panel C), and the national origin of open models (panel D). None of these attributes explain differences consistently. 

It would be reasonable to expect that larger shifts in expectations correlate with larger changes in yields. However, if open and closed releases induce opposing effects, large positive expectation shifts could have counterveiling effects for the two types of releases. Figure~\ref{fig:regression_interacted_splits} thus considers the interaction. The point estimates suggest that releases that induce forecasters to anticipate sooner arrival of AGI have a very slightly more positive effect on 30-year treasury yields for open releases, and a stronger negative effect for closed releases. However, these estimates are extremely imprecise.%
\footnote{They also do not account for overlap with the categories omitted from the plot.}

These results suggest that these associated factors are less likely to explain the opposing shifts. However, it is possible that these performance measures imperfectly capture the economic importance of a model, and that the heterogeneity I observe is driven by some other associated factor.

\subsection{Other impacts}\label{s:equities}

\begin{figure}
     \centering
        \begin{subfigure}[b]{\textwidth}
         \centering
         \includegraphics[width=\textwidth]{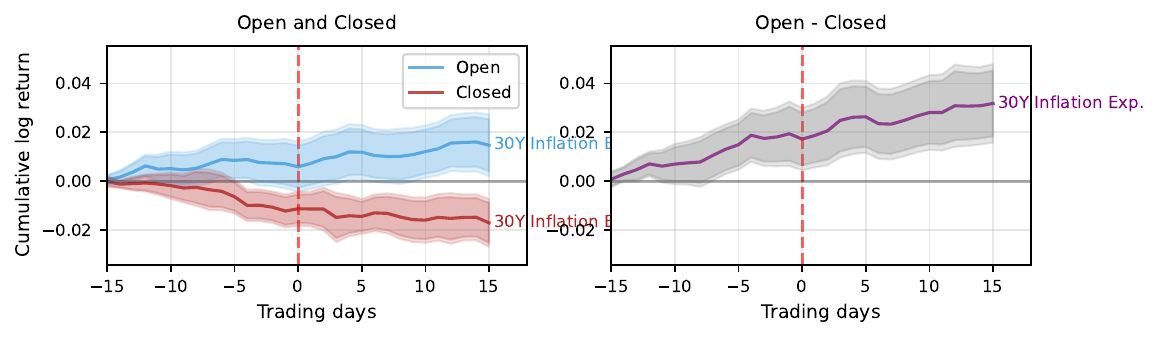}
         \caption{Inflation expectations (RINF)}
     \end{subfigure}
     \\
        \begin{subfigure}[b]{\textwidth}
         \centering
         \includegraphics[width=\textwidth]{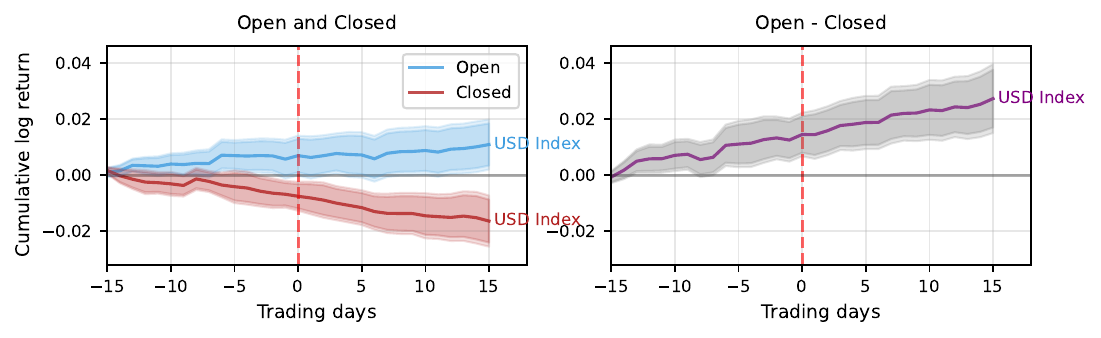}
         \caption{US dollar currency index (UUP)}
     \end{subfigure}
     \\
        \begin{subfigure}[b]{\textwidth}
         \centering
         \includegraphics[width=\textwidth]{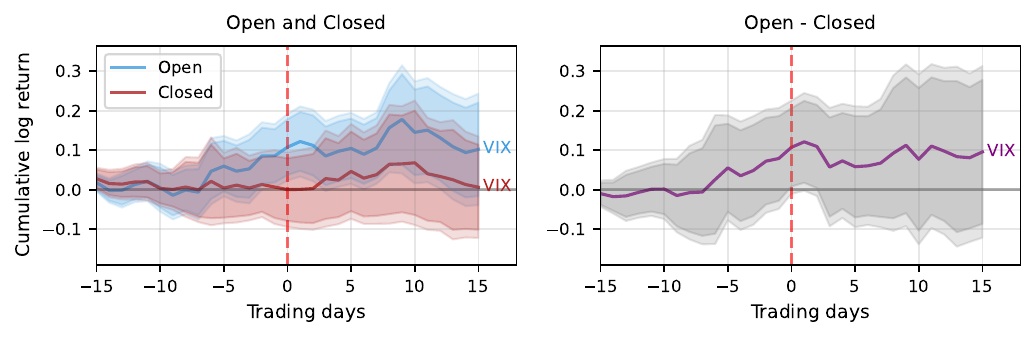}
         \caption{Uncertainty index (VIX)}
     \end{subfigure}
     \caption{Other outcomes event study regression estimates}
     \label{fig:regression_difference_other}
    \note{Estimates of cumulative returns. The left panels show cumulative returns from 15 days prior to release, using equation~\eqref{eq:cum_return_paired}. The right panels show the estimated difference. 90\% and 95\% confidence intervals are shaded.}
\end{figure}

Impacts on indicators of inflation, the US dollar, and volatility are shown in Figure~\ref{fig:regression_difference_other}, and I repeat these for the set of 2023-4 releases in Figure~\ref{fig:regression_difference_other_2023_4}. The different types of releases coincide with differential impact on the US dollar, with open strengthening and closed weakening, as shown in Figure~\ref{fig:regression_difference_other} panel (B). There is some indication that open releases coincide with increases in indicators of volatility. This is weakly evident in the full sample in Figure~\ref{fig:regression_difference_other} panel (C) but is stronger among 2023-4 releases (Figure~\ref{fig:regression_difference_other_2023_4} panel (C)).

I also assess the impact on equities. There is suggestive evidence that open releases correspond with decreases in returns to AI-focused US tech firms, but it is imprecise. Figure~\ref{fig:regression_equity} plots impacts on equities for the full sample of releases and Figure~\ref{fig:regression_equity_2023_4} for 2023-4 releases. In the period 2023-4, the returns of the S\&P 500, Nvidia, Meta, Microsoft, Google, and Amazon all decline around open releases relative to closed releases. When all events are included, the differential effect on the S\&P 500 shrinks to roughly zero, and on Google and Amazon becomes slightly positive. However, many of these patterns are not statistically significant. It is thus unclear from this data whether open vs. closed releases lead to consistently differential returns for US tech firms, but they admit the possibility of lower returns.

\section{Discussion}\label{s:ccl}

There is wide disagreement about the economic implications of advances in machine intelligence. This paper builds on a literature that attempts to better understand market beliefs about these implications through financial decisions \citep{jones_ai_2024,chow_transformative_2025}. This paper investigates the puzzle documented by \citet{andrews_markets_2025}. It finds evidence that long term bond markets shift in opposite directions following open versus closed AI model releases. The patterns documented here could be explained by many different stories. One possibility is that firms make different investments depending on the two types of advancements. While established tech firms have substantial internal resources to fund AI investments, open models allow for wider categories of firms to invest in developing and integrating AI technologies. It is possible that this could increase demand for resources in the present. It could also result from other economic decisions impacted by whether advances in AI are open or closed. It could also be that markets believe that open advances lead to higher economic growth or higher existential risk than closed advances. Regardless of the underlying mechanism, these results suggest the possibility that market participants believe that whether advances in AI are open or closed could have important economic implications.

\bibliography{paper}

\clearpage

\appendix

\section{Appendix}\label{a:appendix1}

\subsection{Permutation tests}\label{a:robustness}
I assess the robustness of results using permutation tests. Given a test statistic $x$ computed on actual event dates $\boldsymbol{\tau}$, I consider where the result lies relative to the results that would be attained by draws from a set of placebo event dates $\tilde{\boldsymbol\tau}$.

I consider four test statistics:
\begin{itemize}
    \item \emph{Regression coefficients.} Based on the joint regression specification in Section~\ref{s:analysis}, equation~\eqref{eq:cum_return_paired}, 
    \begin{itemize}
        \item Group level: $x_{b,r,g} = \Delta\hat{y}^g_{W,r}$ for group $g\in\{A,B\}$
        \item Group comparison: $\Delta x_{b,r} = \Delta\hat{y}^A_{W,r} - \Delta\hat{y}^B_{W,r}$
    \end{itemize}
    \item \emph{Median changes.}
    \begin{itemize}
        \item Group level: $x_{b,g,r} = \Delta y_{t,W,b}^g$ for group $g\in\{A,B\}$, following \citet{andrews_markets_2025}
        \item Group comparison: $\Delta x_{b,r} = \Delta y^A_{t,W,b} - \Delta y^B_{t,W,b}$
    \end{itemize}
\end{itemize}

For group levels, I set $\tilde{\boldsymbol{\tau}}$ to all business days in the data. For each replication $b$, I draw a placebo sample of $K$ events from $\tilde{\boldsymbol{\tau}}$ randomly without replacement. Then $x_b$ is computed based on that placebo sample. 

For comparisons between groups, I set $\tilde{\boldsymbol{\tau}}$ to be the union of the dates of the events in the two groups. For each replication $b$, two placebo samples are drawn from $\tilde{\boldsymbol{\tau}}$, one of $K_A$ events (called sample A), and one of $K_B$ events (called sample B), randomly without replacement. I then compute comparison statistic $\Delta x_{b,r}$ based on samples A and B. This considers the concern that given a small sample of events, splitting the events into two groups could result in spurious differences. We would like to rule out that the differences are arbitrary. For any set of events A and B, we compute the test statistic on A vs. B, and compare where each lie relative to the distribution of statistics that could result from arbitrary splits of events sampled from the pooled events of A and B. If A and B are an arbitrary split, then the statistic computed on either sample would be expected to be near the center of the distribution. But if A and B result in meaningfully different impacts, then the realized differences may differ substantially from the distribution of placebo draws.

I repeat for $B=5,000$ replications to form placebo distributions, and plot the mean, and 90\% and 95\% percentile.

\begin{figure}[ht]
     \begin{subfigure}[b]{\textwidth}
         \centering
         \includegraphics[width=0.4\textwidth]{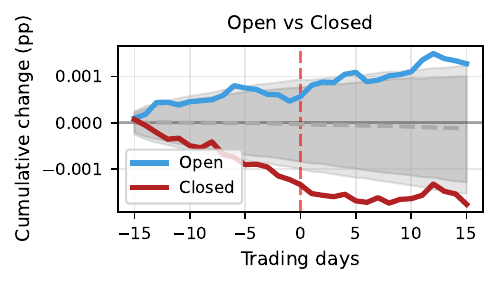}
         \includegraphics[width=0.4\textwidth]{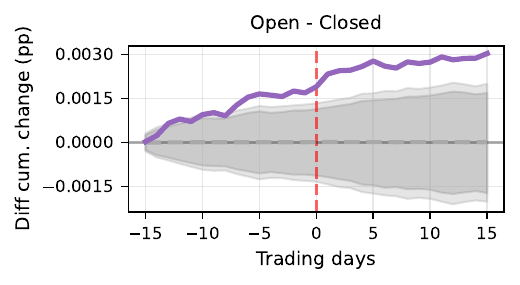}
         \caption{30 year}
     \end{subfigure}
     \\
     \begin{subfigure}[b]{\textwidth}
         \centering
         \includegraphics[width=0.4\textwidth]{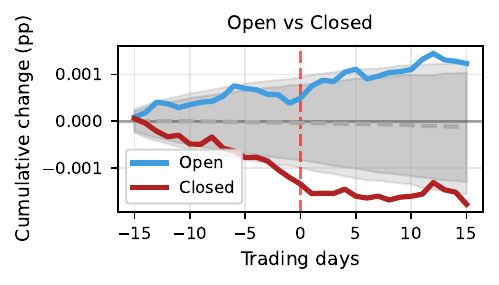}
         \includegraphics[width=0.4\textwidth]{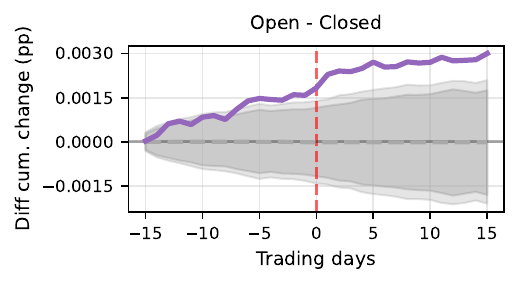}
         \caption{20 year}
     \end{subfigure}
     \\
     \begin{subfigure}[b]{\textwidth}
         \centering
         \includegraphics[width=0.4\textwidth]{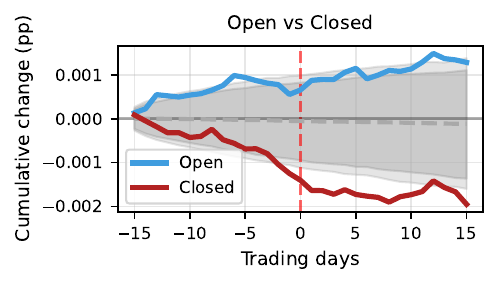}
         \includegraphics[width=0.4\textwidth]{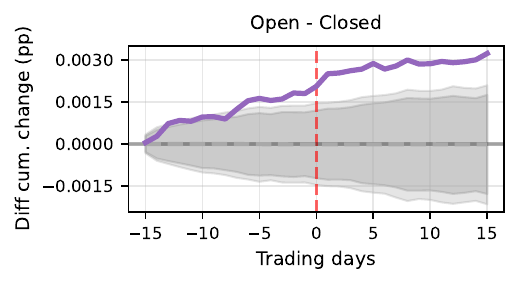}
         \caption{10 year}
     \end{subfigure}
     \\
     \begin{subfigure}[b]{\textwidth}
         \centering
         \includegraphics[width=0.4\textwidth]{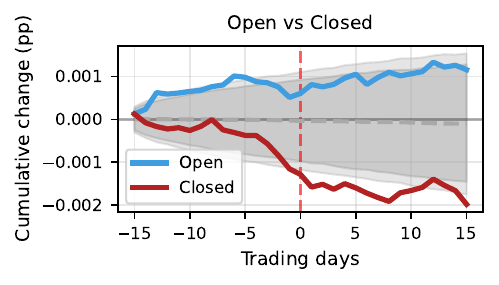}
         \includegraphics[width=0.4\textwidth]{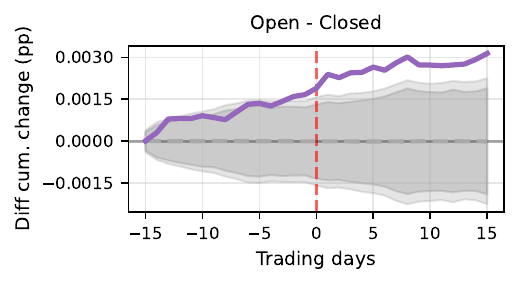}
         \caption{5 year}
     \end{subfigure}
     \\
     \begin{subfigure}[b]{\textwidth}
         \centering
         \includegraphics[width=0.4\textwidth]{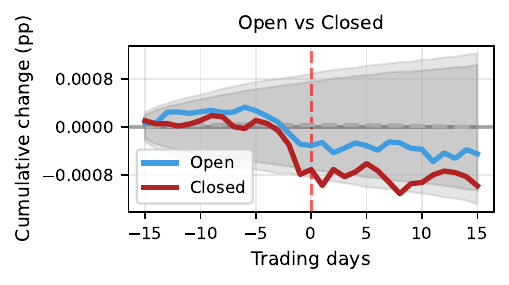}
         \includegraphics[width=0.4\textwidth]{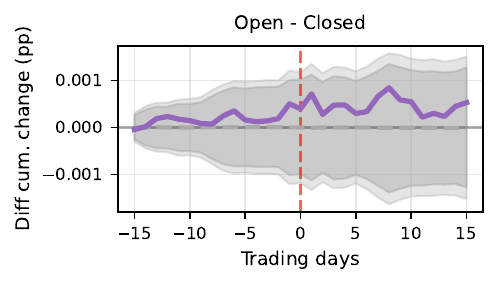}
         \caption{1 year}
     \end{subfigure}
     \\
\caption{Estimated yield change in US treasury bonds around major AI releases (OLS) and permutation test}
\note{OLS coefficient estimates. The shaded regions represent 90\% and 95\% of a placebo distribution, over 5,000 permutation draws. Dashed line indicates mean of the placebo distribution. For the first column, the placebo includes all dates in data. For last column, the placebo distribution includes the pooled event dates of closed and open AI model releases.}
\label{f:permutation_open_and_closed_ols}\end{figure}

\paragraph{Regression coefficients (OLS)}
Figure~\ref{f:permutation_open_and_closed_ols} plots the results of permutation tests on the regression specifications. The left column shows the jointly estimated cumulative yield change for open and closed model releases. The placebo distribution represents all business days in the dataset, from which a sample of the smaller of the two groups of events is drawn.

The right column shows the resulting estimated group comparison between open and closed releases. Here placebo days $\tilde{\boldsymbol{\tau}}$ are drawn from the union of dates with open and closed model releases.

\begin{figure}[ht]
     \begin{subfigure}[b]{\textwidth}
         \centering
         \includegraphics[width=0.4\textwidth]{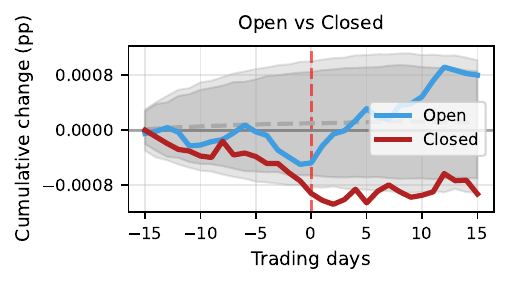}
         \includegraphics[width=0.4\textwidth]{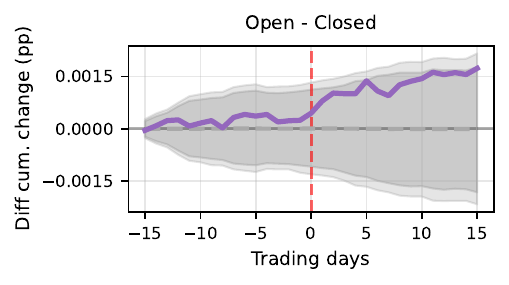}
         \caption{30 year}
     \end{subfigure}
     \\
     \begin{subfigure}[b]{\textwidth}
         \centering
         \includegraphics[width=0.4\textwidth]{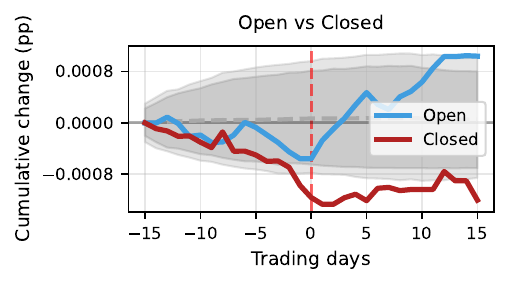}
         \includegraphics[width=0.4\textwidth]{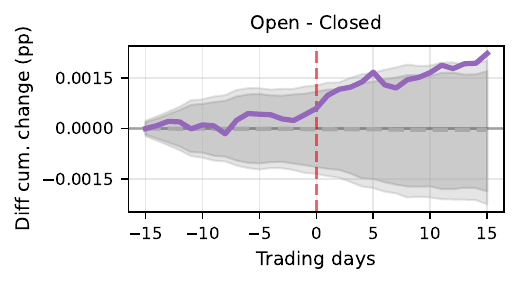}
         \caption{20 year}
     \end{subfigure}
     \\
     \begin{subfigure}[b]{\textwidth}
         \centering
         \includegraphics[width=0.4\textwidth]{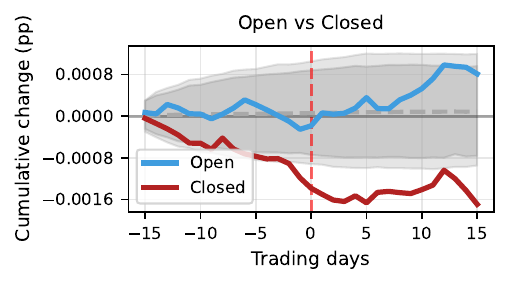}
         \includegraphics[width=0.4\textwidth]{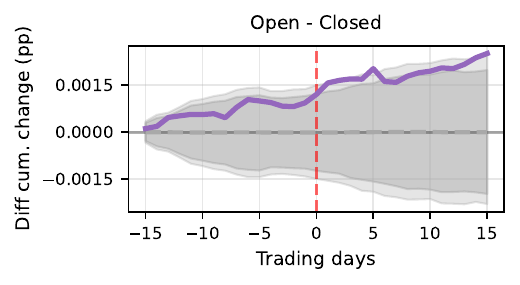}
         \caption{10 year}
     \end{subfigure}
     \\
     \begin{subfigure}[b]{\textwidth}
         \centering
         \includegraphics[width=0.4\textwidth]{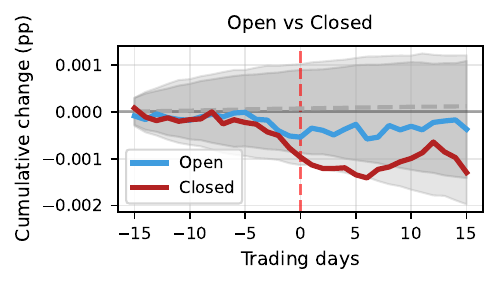}
         \includegraphics[width=0.4\textwidth]{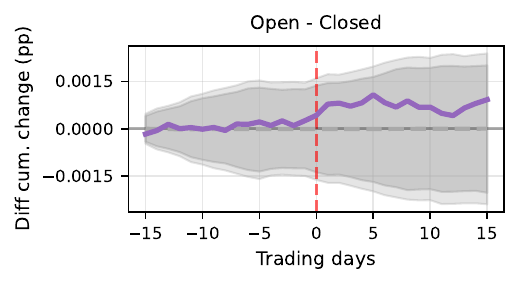}
         \caption{5 year}
     \end{subfigure}
     \\
     \begin{subfigure}[b]{\textwidth}
         \centering
         \includegraphics[width=0.4\textwidth]{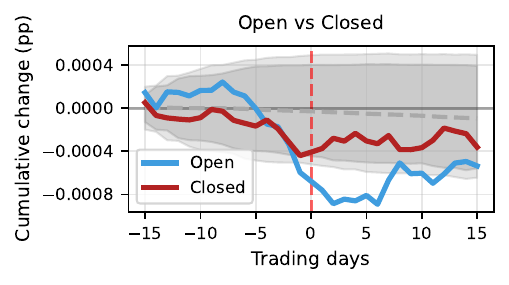}
         \includegraphics[width=0.4\textwidth]{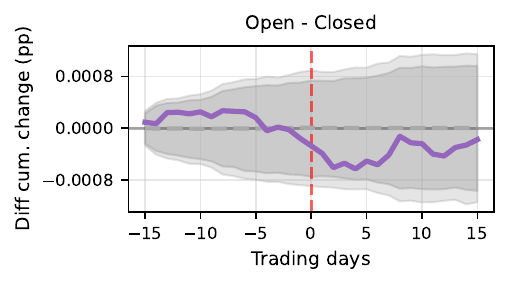}
         \caption{1 year}
     \end{subfigure}
     \\
\caption{Estimated yield change in US treasury bonds around major AI releases (LAD, accumulated) and permutation test}
\note{Cumulative sum of least absolute deviation coefficient estimates. The shaded regions represent 90\% and 95\% of a placebo distribution, over 5,000 permutation draws. Dashed line indicates mean of the placebo distribution. For the first column, the placebo includes all dates in data. For last column, the placebo distribution includes the pooled event dates of closed and open AI model releases.}
\label{f:permutation_open_and_closed_lad}\end{figure}

\paragraph{Regression coefficients (LAD)}
Figure~\ref{f:permutation_open_and_closed_lad} repeats this exercise but estimating the regression minimizing absolute deviation (LAD) rather than  squares of residuals (OLS).  A similar gap between open and closed is observed. Note that the sum of median daily changes may not equal the median change over a longer span.

\begin{figure}[ht]
     \begin{subfigure}[b]{\textwidth}
         \centering
         \includegraphics[width=0.4\textwidth]{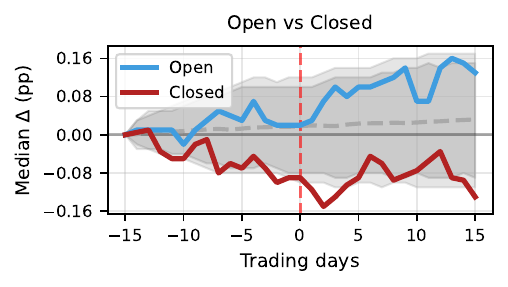}
         \includegraphics[width=0.4\textwidth]{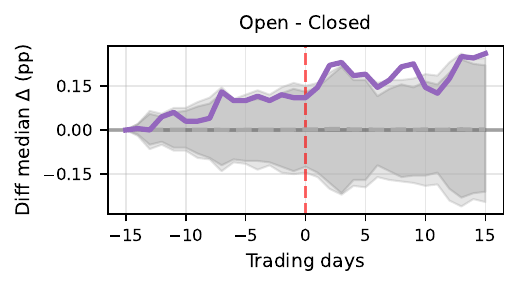}
         \caption{30 year}
     \end{subfigure}
     \\
     \begin{subfigure}[b]{\textwidth}
         \centering
         \includegraphics[width=0.4\textwidth]{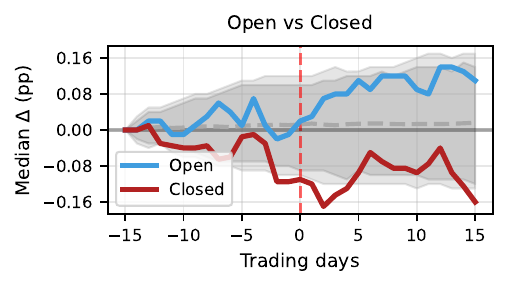}
         \includegraphics[width=0.4\textwidth]{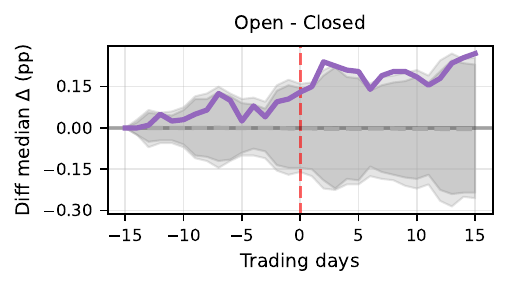}
         \caption{20 year}
     \end{subfigure}
     \\
     \begin{subfigure}[b]{\textwidth}
         \centering
         \includegraphics[width=0.4\textwidth]{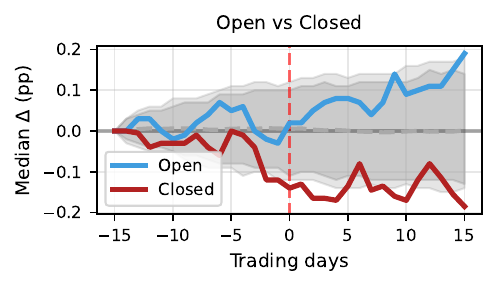}
         \includegraphics[width=0.4\textwidth]{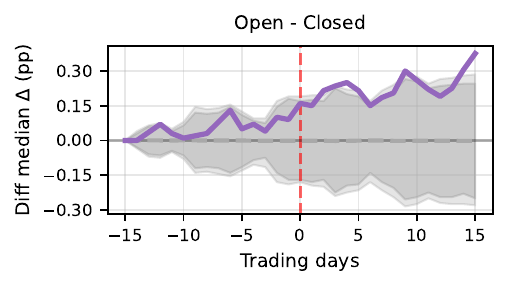}
         \caption{10 year}
     \end{subfigure}
     \\
     \begin{subfigure}[b]{\textwidth}
         \centering
         \includegraphics[width=0.4\textwidth]{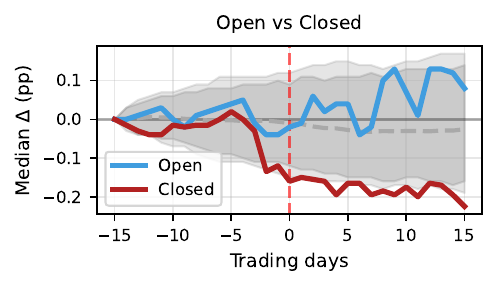}
         \includegraphics[width=0.4\textwidth]{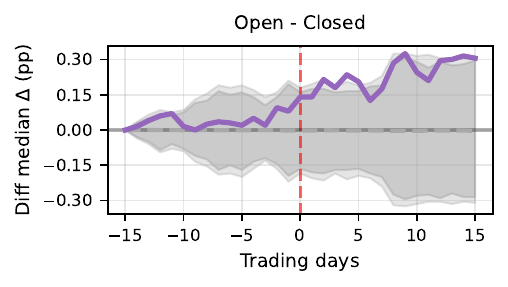}
         \caption{5 year}
     \end{subfigure}
     \\
     \begin{subfigure}[b]{\textwidth}
         \centering
         \includegraphics[width=0.4\textwidth]{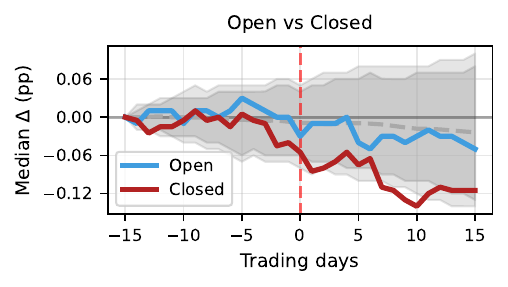}
         \includegraphics[width=0.4\textwidth]{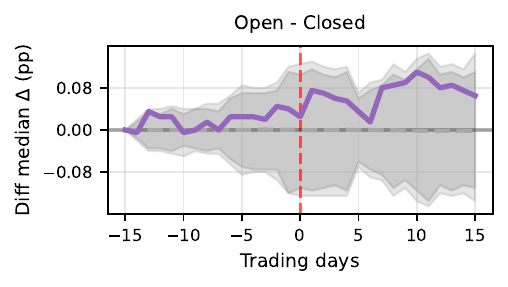}
         \caption{1 year}
     \end{subfigure}
     \\
\caption{Median yield change in US treasury bonds around major AI releases and permutation test}
\note{Solid line shows the estimated change for the median event, relative to 15 days before the event. The shaded regions represent 90\% and 95\% of a placebo distribution, over 5,000 permutation draws. Dashed line indicates mean of the placebo distribution. For left column, the placebo includes all dates in data. For right column, the placebo distribution includes the pooled event dates of closed and open AI model releases.}
\label{f:permutation_open_and_closed_medianchange}\end{figure}

\paragraph{Median changes}
Figure~\ref{f:permutation_open_and_closed_medianchange} plots the median change in treasury yield from 15 days before release for open models (left column) and closed models (middle column), using as placebo dates all business days in the dataset.

When we compute the difference between median yield changes for open vs. closed models, we see reasonably strong evidence of a difference, in the right column of Figure~\ref{f:permutation_open_and_closed_medianchange}. Here placebo days $\tilde{\boldsymbol{\tau}}$ are drawn from the union of dates with open and closed model releases.

\paragraph{Regression coefficients: assessment of coverage}
As an additional check, I assess the coverage of the HAC standard errors using permutation tests. I draw a placebo set of events of the same size as one of the group of events (open or closed), and compute the resulting regression, that yields estimates $x_{b,r,g} = \Delta\hat{y}^g_{W,r}$ with associated HAC standard errors. These permutation tests for 10 and 30 year treasury yields suggest that the 95\% confidence intervals derived from those standard errors have approximately 91-92\% coverage and the 90\% confidence intervals have approximately 85-86\% coverage.

\subsection{Controlling for other news}\label{a:residualization}

As an additional robustness check, I follow \citet{andrews_markets_2025} in residualizing the main series based on other news, with the Citigroup US Economic Surprise Index \citep{citigroup_global_markets_citigroup_2025}, the 
VIX volatility index, and the Federal Reserve Bank of San Francisco Daily
News Sentiment Index \citep{shapiro_measuring_2022}.

\begin{figure}
     \centering
        \begin{subfigure}[b]{\textwidth}
         \centering
         \includegraphics[width=0.9\textwidth]{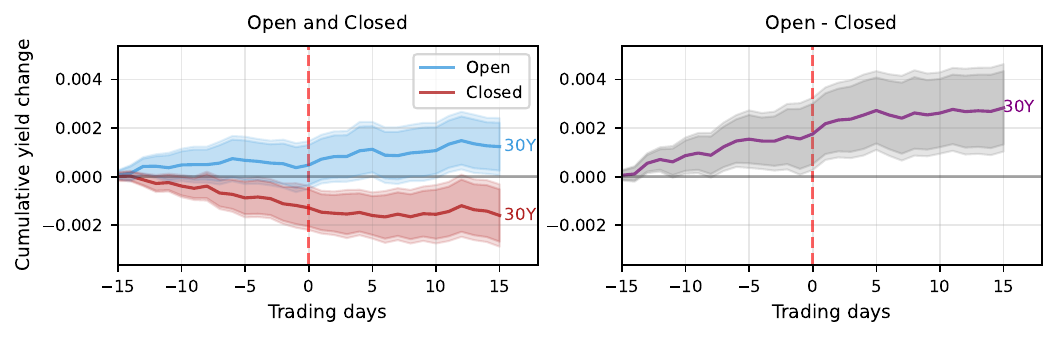}
     \end{subfigure}
     \\
        \begin{subfigure}[b]{\textwidth}
         \centering
         \includegraphics[width=0.9\textwidth]{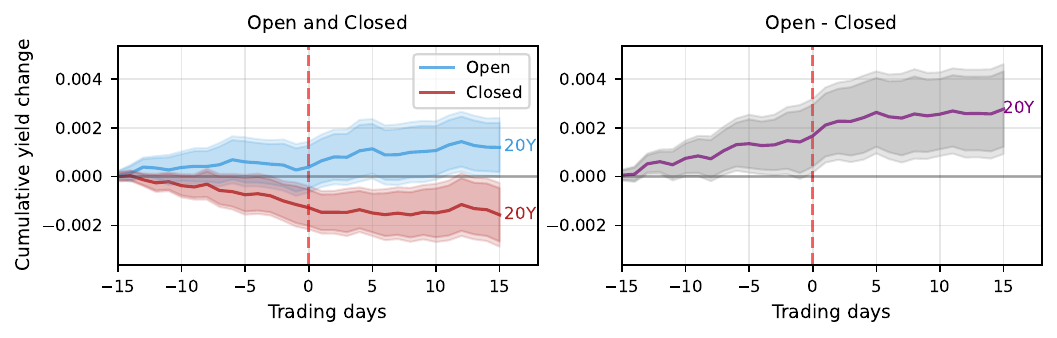}
     \end{subfigure}
     \\
        \begin{subfigure}[b]{\textwidth}
         \centering
         \includegraphics[width=0.9\textwidth]{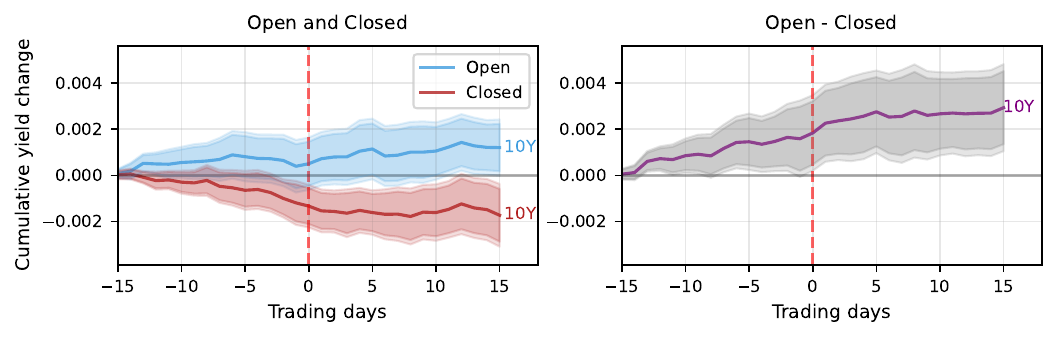}
     \end{subfigure}
     \\
        \begin{subfigure}[b]{\textwidth}
         \centering
         \includegraphics[width=0.9\textwidth]{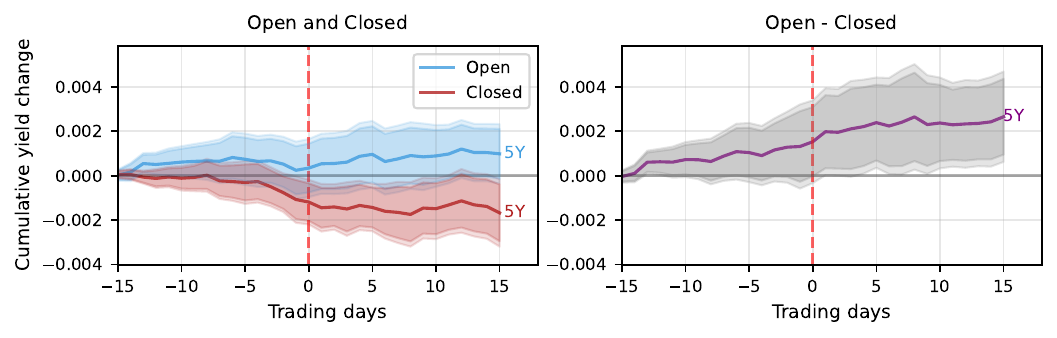}
     \end{subfigure}
     \\
        \begin{subfigure}[b]{\textwidth}
         \centering
         \includegraphics[width=0.9\textwidth]{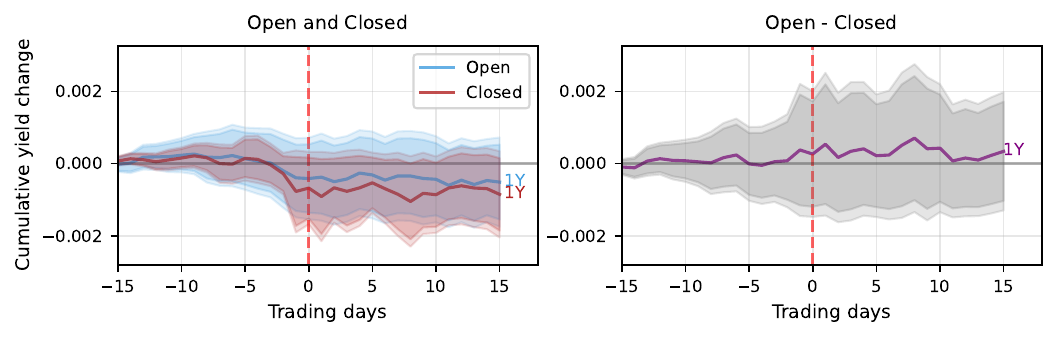}
     \end{subfigure}
     
     \caption{US treasury bond yield event study regression estimates: residualized (surprise)}
     \label{fig:regression_difference_residual_cesi}
    \note{Estimates of cumulative returns, residualized against the Citigroup US Economic Surprise Index, as described in Appendix~\ref{a:residualization}. The left panels show cumulative returns from 15 days prior to release, using equation~\eqref{eq:cum_return_paired}. The right panels show the estimated difference. Constant maturity duration noted in years. 90\% and 95\% confidence intervals are shaded.}
\end{figure}

\begin{figure}
     \centering
        \begin{subfigure}[b]{\textwidth}
         \centering
         \includegraphics[width=0.9\textwidth]{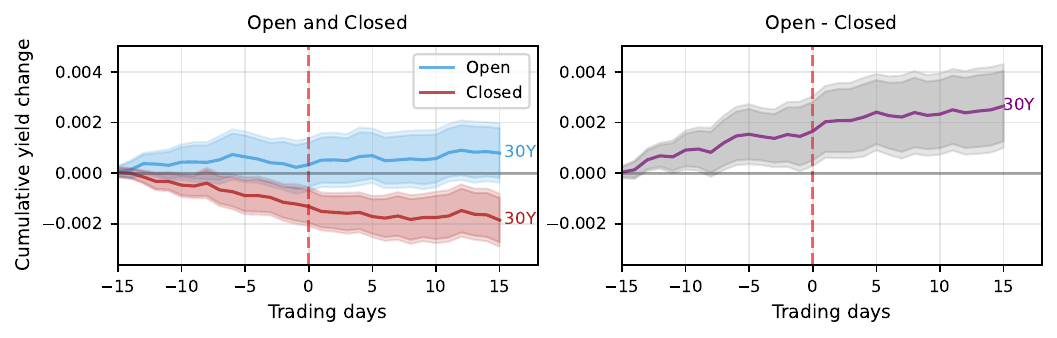}
     \end{subfigure}
     \\
        \begin{subfigure}[b]{\textwidth}
         \centering
         \includegraphics[width=0.9\textwidth]{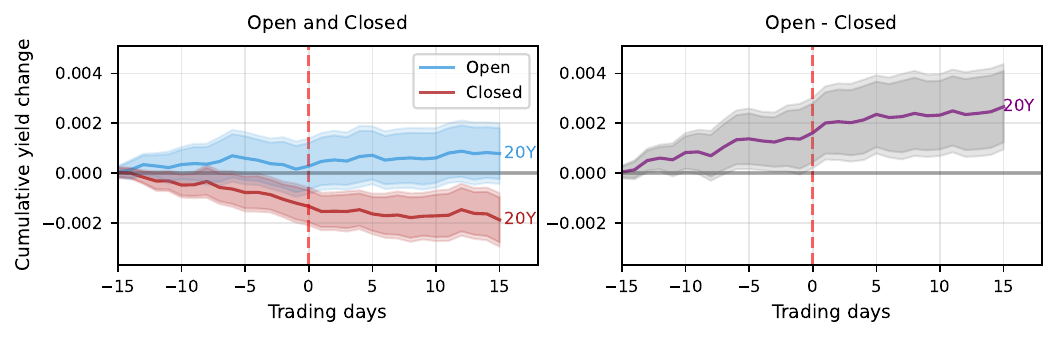}
     \end{subfigure}
     \\
        \begin{subfigure}[b]{\textwidth}
         \centering
         \includegraphics[width=0.9\textwidth]{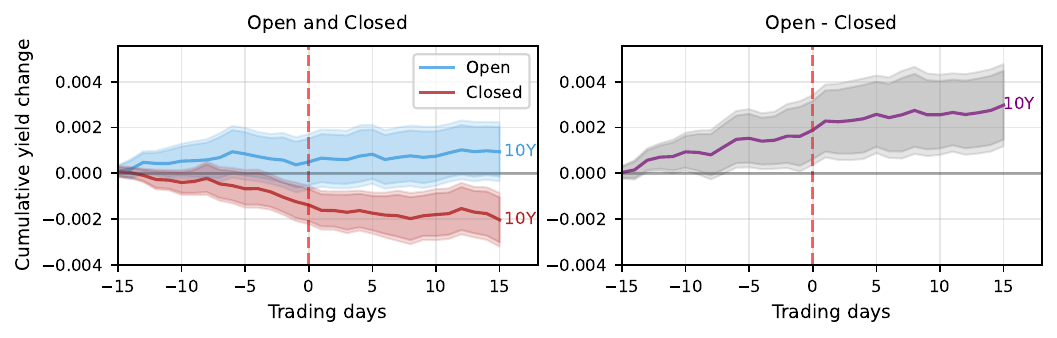}
     \end{subfigure}
     \\
        \begin{subfigure}[b]{\textwidth}
         \centering
         \includegraphics[width=0.9\textwidth]{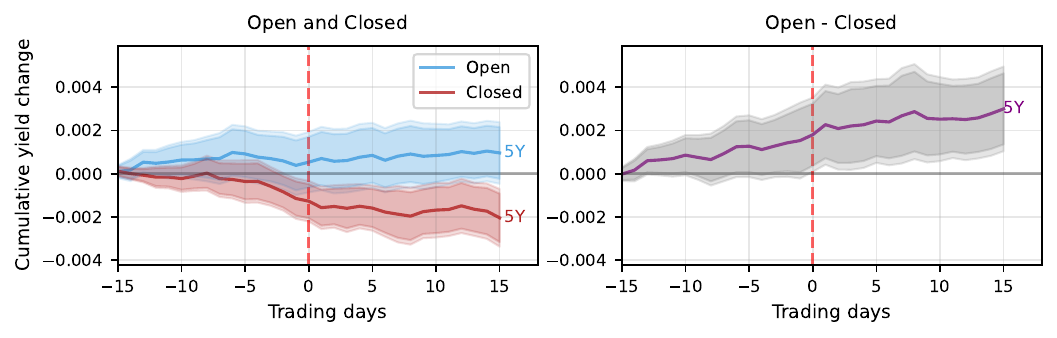}
     \end{subfigure}
     \\
        \begin{subfigure}[b]{\textwidth}
         \centering
         \includegraphics[width=0.9\textwidth]{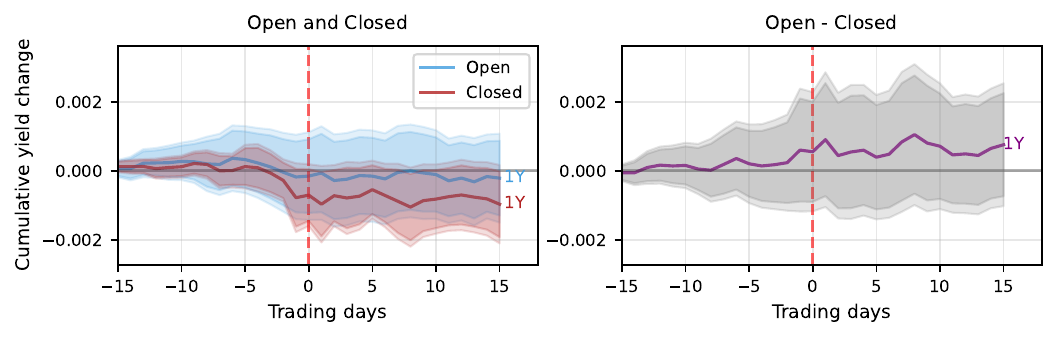}
     \end{subfigure}
     
     \caption{US treasury bond yield event study regression estimates: residualized (VIX)}
     \label{fig:regression_difference_residual_vix}
    \note{Estimates of cumulative returns, residualized against VIX, as described in Appendix~\ref{a:residualization}. The left panels show cumulative returns from 15 days prior to release, using equation~\eqref{eq:cum_return_paired}. The right panels show the estimated difference. Constant maturity duration noted in years. 90\% and 95\% confidence intervals are shaded.}
\end{figure}

\begin{figure}
     \centering
        \begin{subfigure}[b]{\textwidth}
         \centering
         \includegraphics[width=0.9\textwidth]{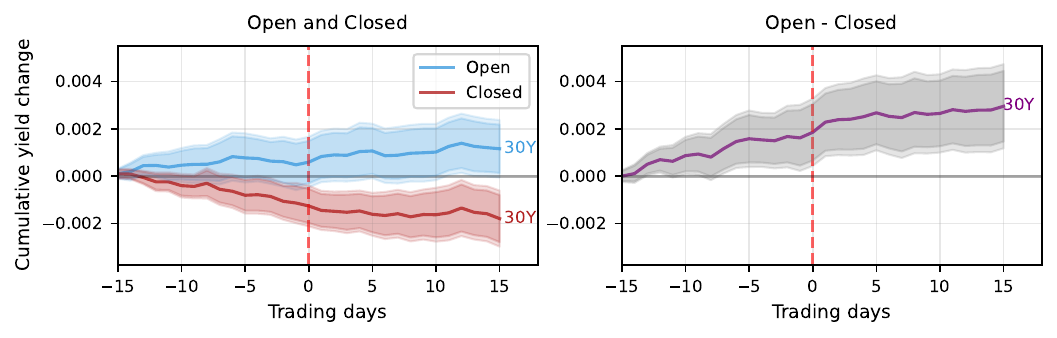}
     \end{subfigure}
     \\
        \begin{subfigure}[b]{\textwidth}
         \centering
         \includegraphics[width=0.9\textwidth]{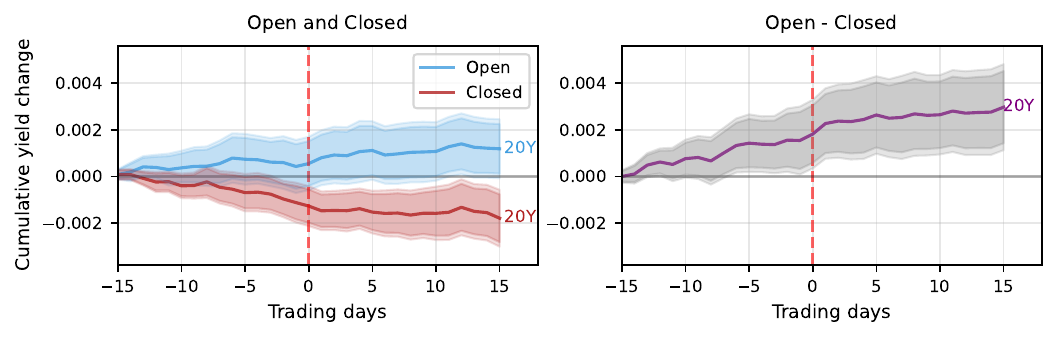}
     \end{subfigure}
     \\
        \begin{subfigure}[b]{\textwidth}
         \centering
         \includegraphics[width=0.9\textwidth]{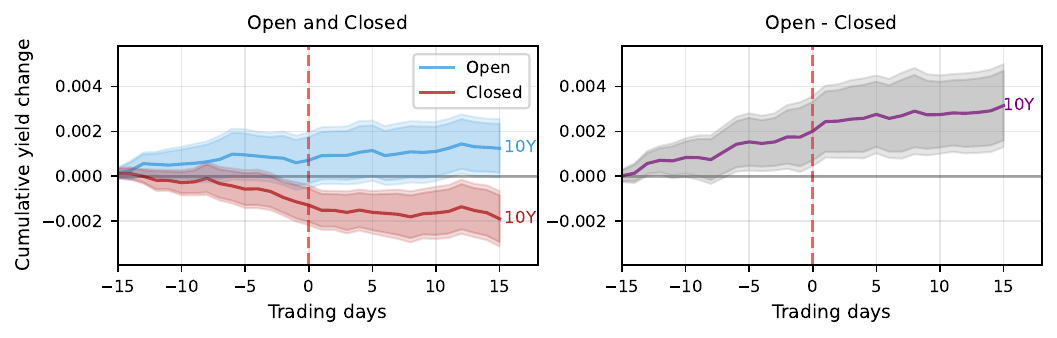}
     \end{subfigure}
     \\
        \begin{subfigure}[b]{\textwidth}
         \centering
         \includegraphics[width=0.9\textwidth]{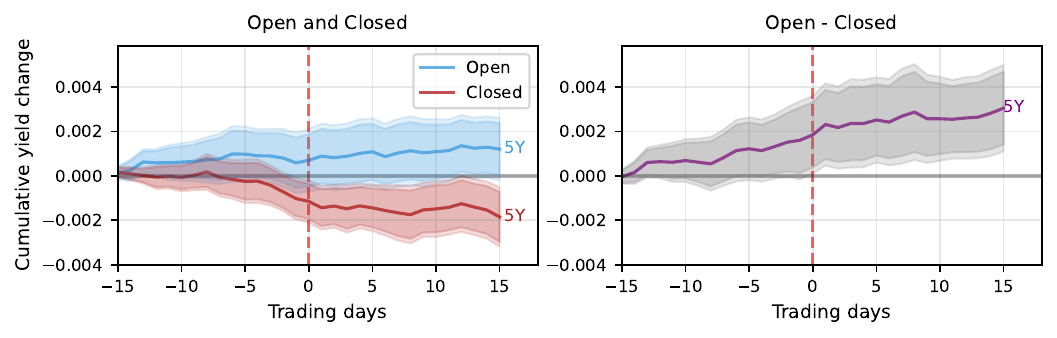}
     \end{subfigure}
     \\
        \begin{subfigure}[b]{\textwidth}
         \centering
         \includegraphics[width=0.9\textwidth]{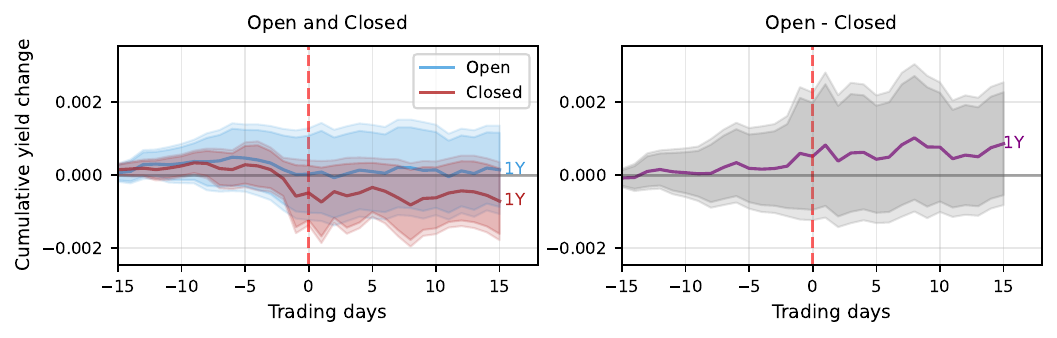}
     \end{subfigure}
     
     \caption{US treasury bond yield event study regression estimates: residualized (news sentiment)}
     \label{fig:regression_difference_residual_dnsi}
    \note{Estimates of cumulative returns, residualized against the Federal Reserve Bank of San Francisco Daily News Sentiment Index, as described in Appendix~\ref{a:residualization}. The left panels show cumulative returns from 15 days prior to release, using equation~\eqref{eq:cum_return_paired}. The right panels show the estimated difference. Constant maturity duration noted in years. 90\% and 95\% confidence intervals are shaded.}
\end{figure}

\begin{figure}
     \centering
        \begin{subfigure}[b]{\textwidth}
         \centering
         \includegraphics[width=0.9\textwidth]{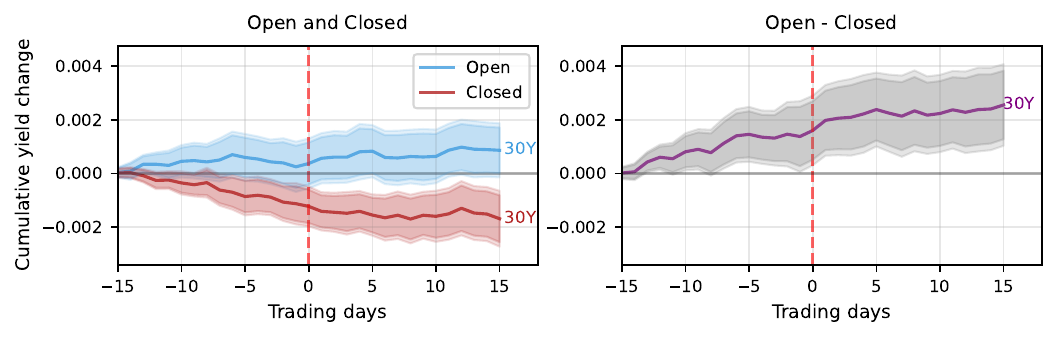}
     \end{subfigure}
     \\
        \begin{subfigure}[b]{\textwidth}
         \centering
         \includegraphics[width=0.9\textwidth]{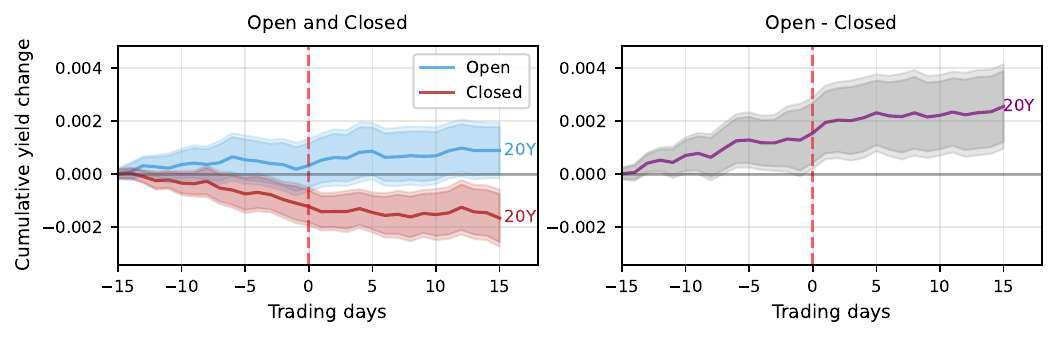}
     \end{subfigure}
     \\
        \begin{subfigure}[b]{\textwidth}
         \centering
         \includegraphics[width=0.9\textwidth]{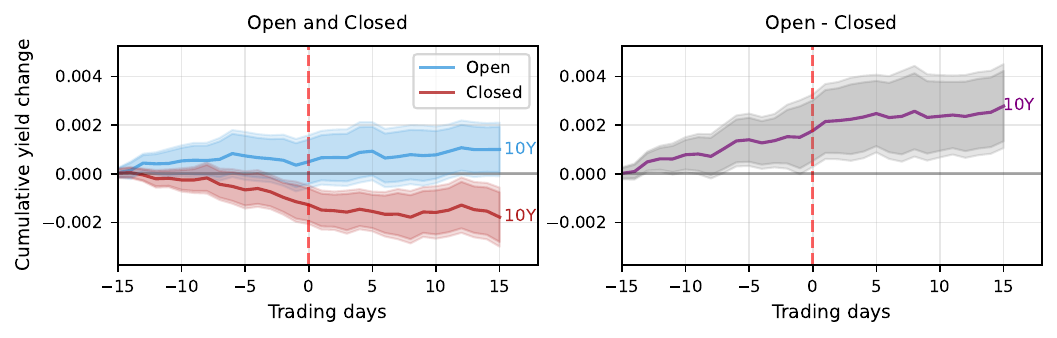}
     \end{subfigure}
     \\
        \begin{subfigure}[b]{\textwidth}
         \centering
         \includegraphics[width=0.9\textwidth]{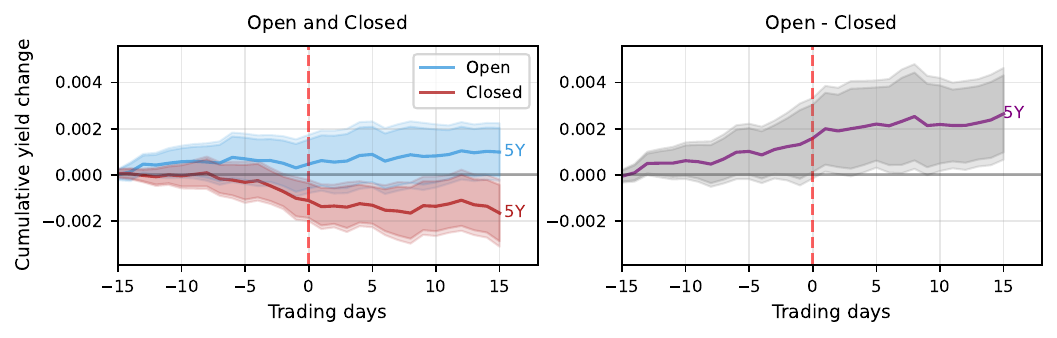}
     \end{subfigure}
     \\
        \begin{subfigure}[b]{\textwidth}
         \centering
         \includegraphics[width=0.9\textwidth]{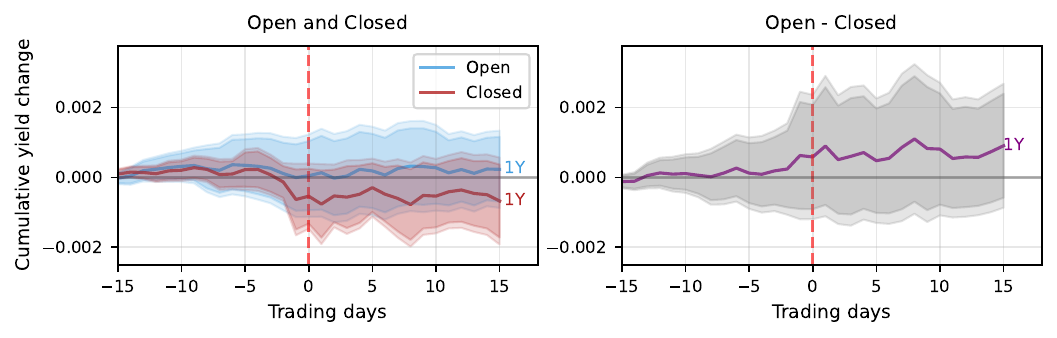}
     \end{subfigure}
     
     \caption{US treasury bond yield event study regression estimates: residualized (surprise + VIX + news sentiment)}
     \label{fig:regression_difference_residual_all}
    \note{Estimates of cumulative returns, residualized against the Citigroup US Economic Surprise Index, VIX, and the Federal Reserve Bank of San Francisco Daily
News Sentiment Index, as described in Appendix~\ref{a:residualization}. The left panels show cumulative returns from 15 days prior to release, using equation~\eqref{eq:cum_return_paired}. The right panels show the estimated difference. Constant maturity duration noted in years. 90\% and 95\% confidence intervals are shaded.}
\end{figure}

I first regress the yield outcome series on a constant plus 15 lags of each control series, and then run the regression procedure detailed in Section~\ref{s:analysis}. I conduct this for the level of surprise (Figure~\ref{fig:regression_difference_residual_cesi}), daily changes in VIX (Figure~\ref{fig:regression_difference_residual_vix}), level of news sentiment (Figure~\ref{fig:regression_difference_residual_dnsi}), and all three (Figure~\ref{fig:regression_difference_residual_all}). In some of these, the confidence bands for open include zero, but the magnitudes remain large and positive, and the difference between open and closed releases remains large and statistically significant.

%
%

\begin{figure}
     \centering
     \includegraphics[width=\textwidth]{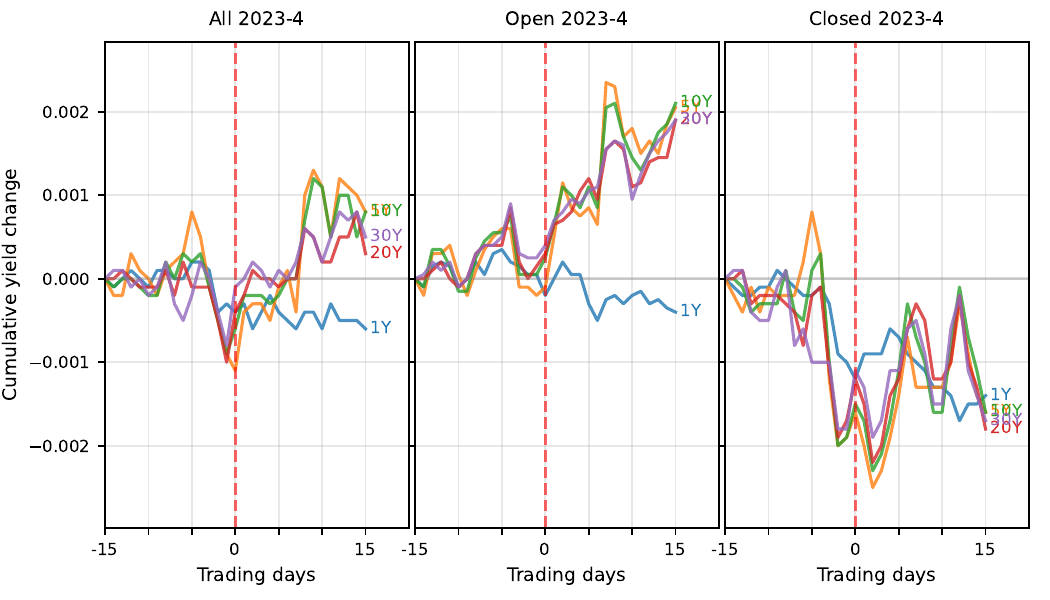}     
    \caption{US treasury bond movements around AI releases 2023-4}
        \label{fig:yield_event_study2023_4}
\note{Median change across release events relative to 15 business days before release. 0=event day. Includes AI releases in 2023-4. Constant maturity duration noted in years.}
\end{figure}

\begin{figure}[t]
\includegraphics[width=\columnwidth]{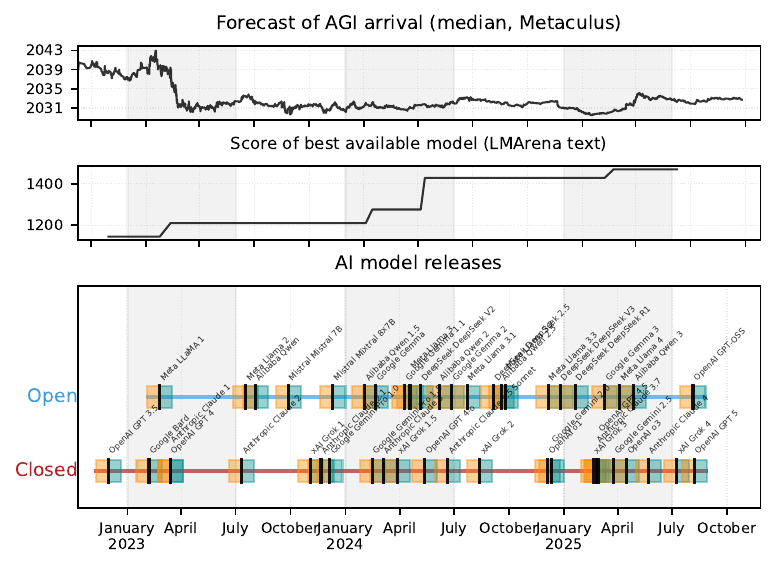}
\caption{Timeline}
\note{AI model releases and $\pm$15 business day window.  LMArena score of best model per ratings as of August 4, 2025 \citep{llmleaderboard2025}.}
\label{f:event_timeline}\end{figure}

\begin{figure}[ht]
     \begin{subfigure}[b]{\textwidth}
         \centering
         \includegraphics[width=0.8\textwidth]{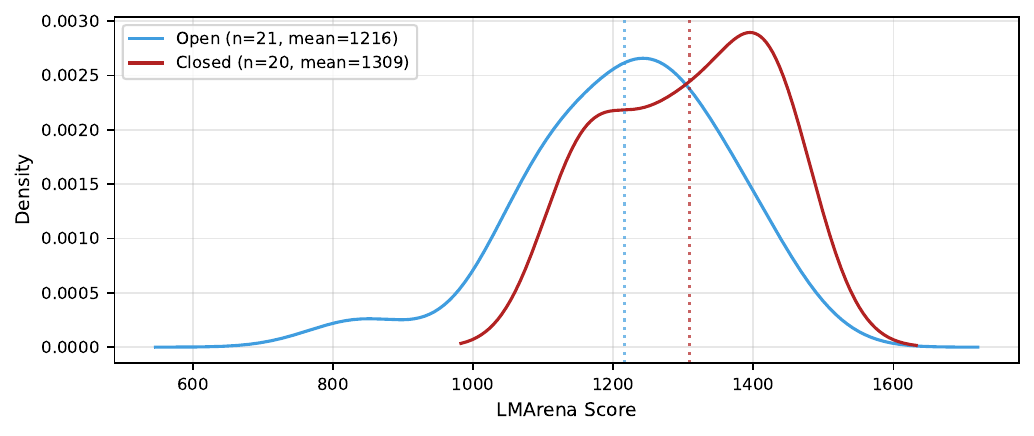}
         \caption{LMArena score}
     \end{subfigure}
     \\
     \begin{subfigure}[b]{\textwidth}
         \centering
         \includegraphics[width=0.8\textwidth]{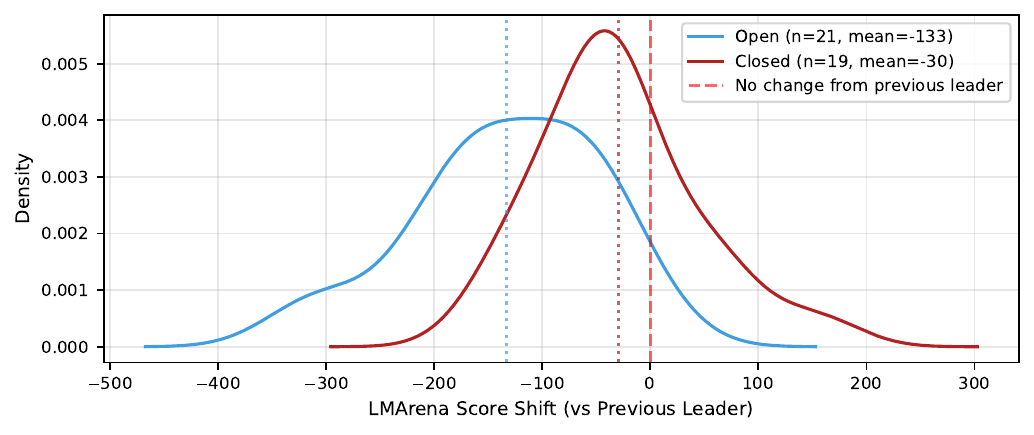}
         \caption{LMArena shift relative to frontier}
     \end{subfigure}
     \\
     \begin{subfigure}[b]{\textwidth}
         \centering
         \includegraphics[width=0.8\textwidth]{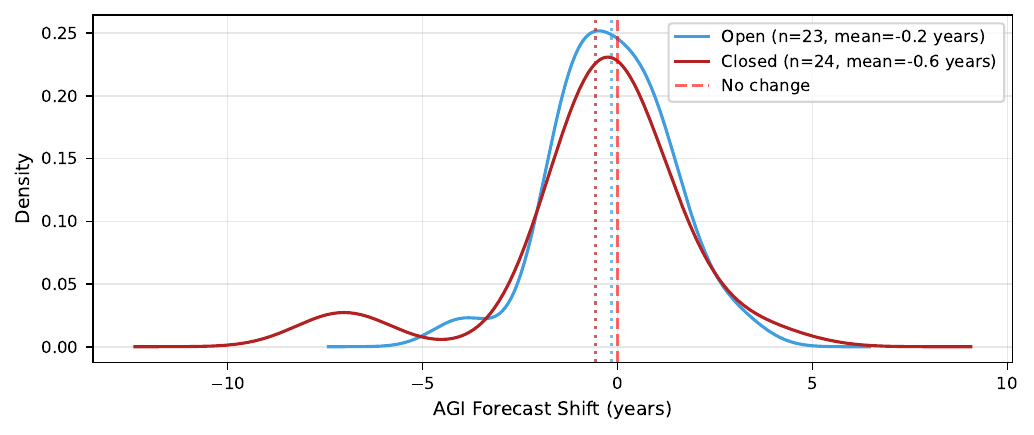}
         \caption{AGI forecast impact}
     \end{subfigure}
     
\caption{Comparison of releases}
\note{LMArena score of best model per ratings as of August 4, 2025 \citep{llmleaderboard2025}, for models that could be matched.}
\label{f:perf_comparison_open_closed}
\end{figure}

\begin{figure}[t]
\includegraphics[width=\columnwidth]{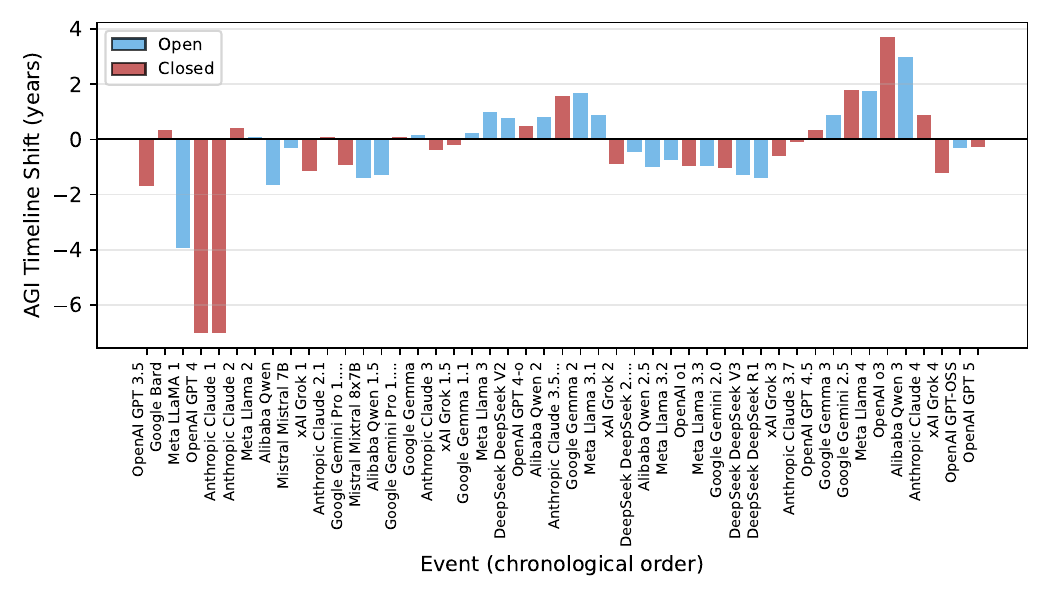}
\caption{AGI forecast shifts around AI model releases}
\note{Source: \citet{metaculus_when_2025} based on the difference in the median forecast from $W$=15 business days prior a release to 15 business days after.}
\label{f:agi_shift_histogram}\end{figure}

\begin{figure}
     \centering
        \begin{subfigure}[b]{\textwidth}
         \centering
         \includegraphics[width=\textwidth]{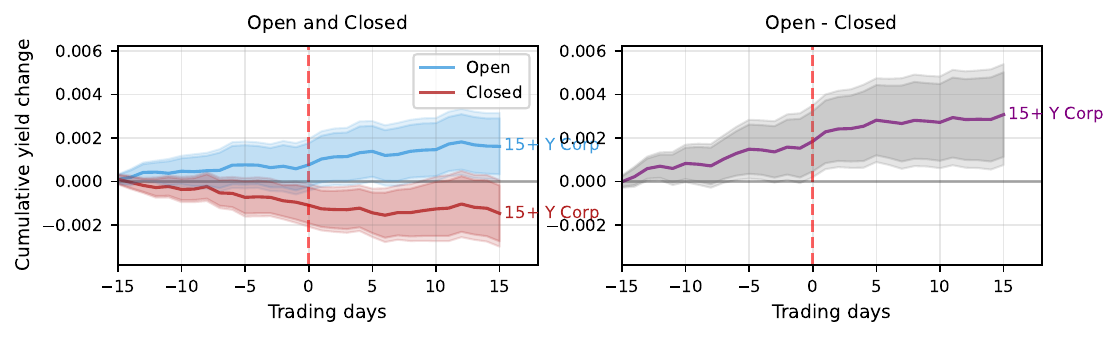}
     \end{subfigure}
     \\
        \begin{subfigure}[b]{\textwidth}
         \centering
         \includegraphics[width=\textwidth]{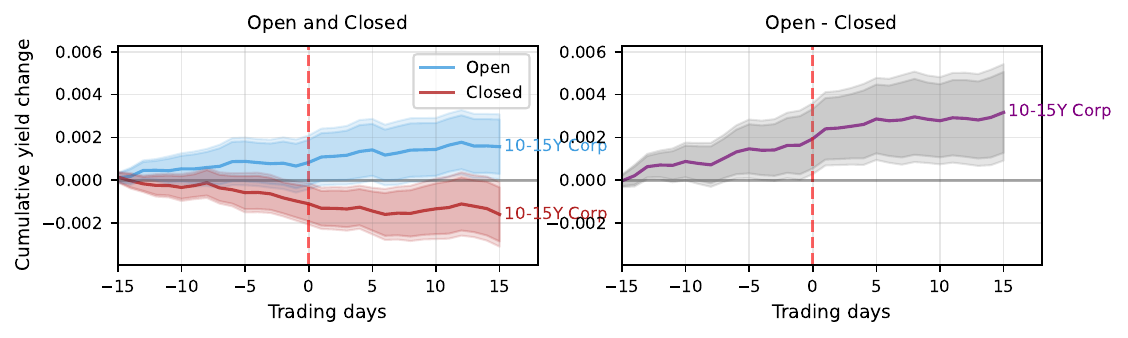}
     \end{subfigure}
     \\
        \begin{subfigure}[b]{\textwidth}
         \centering
         \includegraphics[width=\textwidth]{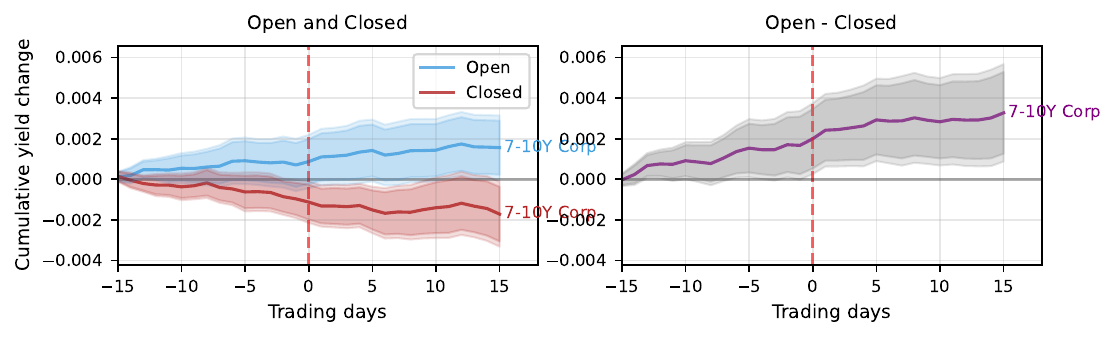}
     \end{subfigure}
     \\
        \begin{subfigure}[b]{\textwidth}
         \centering
         \includegraphics[width=\textwidth]{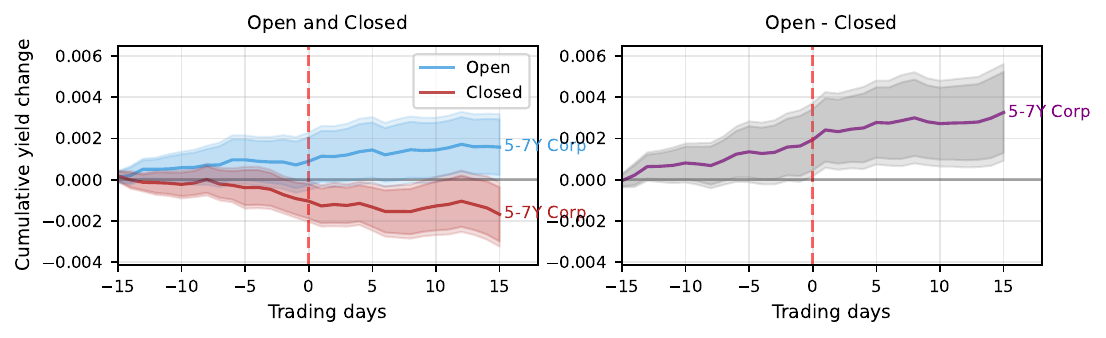}
     \end{subfigure}
     \\
        \begin{subfigure}[b]{\textwidth}
         \centering
         \includegraphics[width=\textwidth]{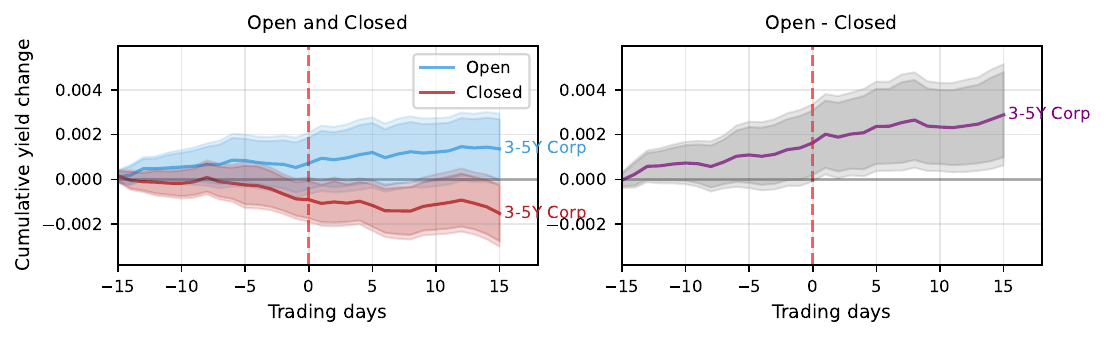}
     \end{subfigure}
     
     \caption{Corporate bond event study regression estimates}
     \label{fig:regression_difference_corporate}
    \note{Estimates of cumulative returns. The left panels show cumulative returns from 15 days prior to release, using equation~\eqref{eq:cum_return_paired}. The right panels show the estimated difference. 90\% and 95\% confidence intervals are shaded.}
\end{figure}

\begin{figure}
     \centering
        \begin{subfigure}[b]{\textwidth}
         \centering
         \includegraphics[width=\textwidth]{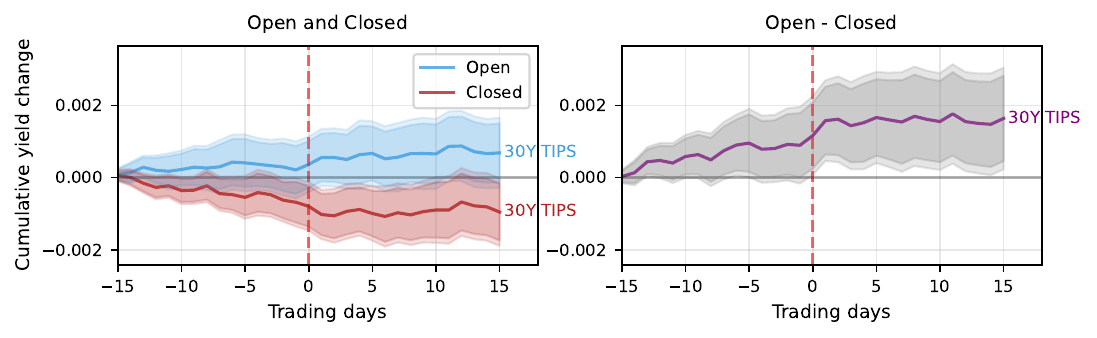}
     \end{subfigure}
     \\
        \begin{subfigure}[b]{\textwidth}
         \centering
         \includegraphics[width=\textwidth]{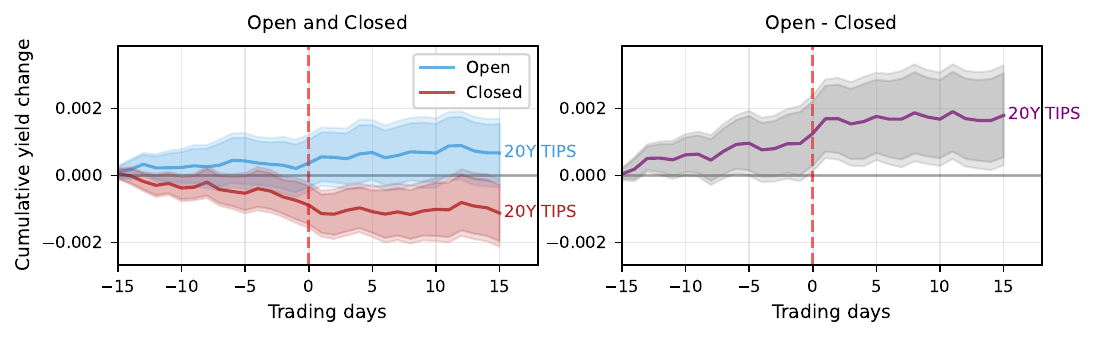}
     \end{subfigure}
     \\
        \begin{subfigure}[b]{\textwidth}
         \centering
         \includegraphics[width=\textwidth]{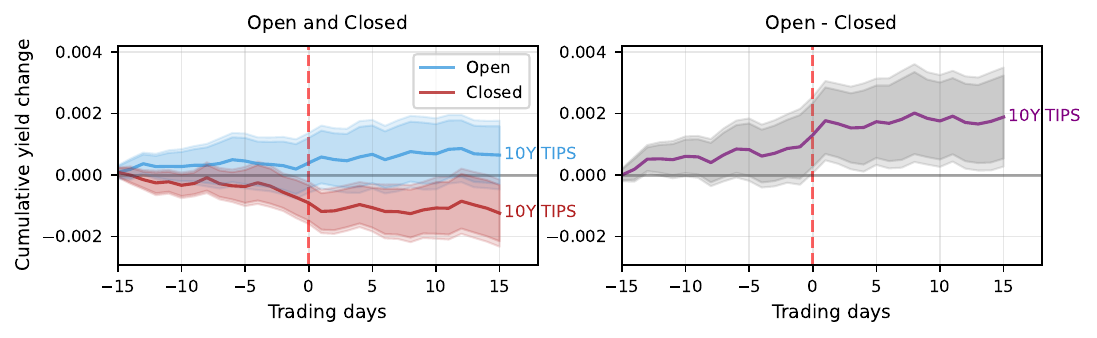}
     \end{subfigure}
     \\
        \begin{subfigure}[b]{\textwidth}
         \centering
         \includegraphics[width=\textwidth]{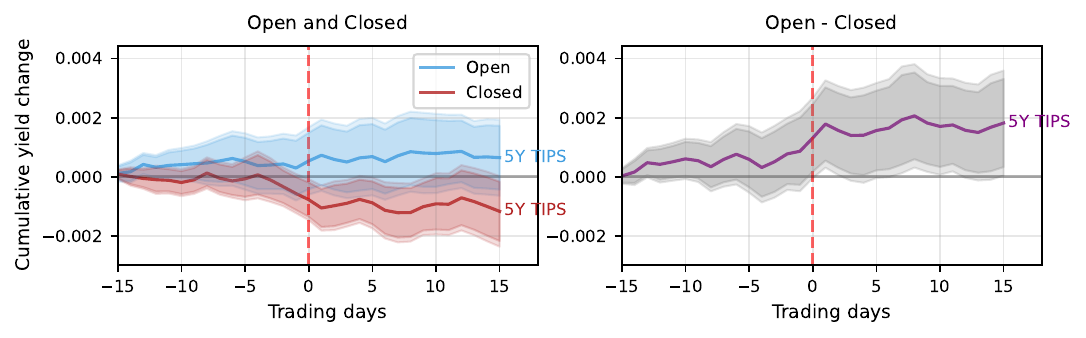}
     \end{subfigure}
     
     \caption{TIPS event study regression estimates}
     \label{fig:regression_difference_tips}
    \note{Estimates of cumulative returns. The left panels show cumulative returns from 15 days prior to release, using equation~\eqref{eq:cum_return_paired}. The right panels show the estimated difference. 90\% and 95\% confidence intervals are shaded.}
\end{figure}

\begin{figure}
     \centering
     \begin{subfigure}[b]{\textwidth}
         \centering
         \includegraphics[width=0.7\textwidth]{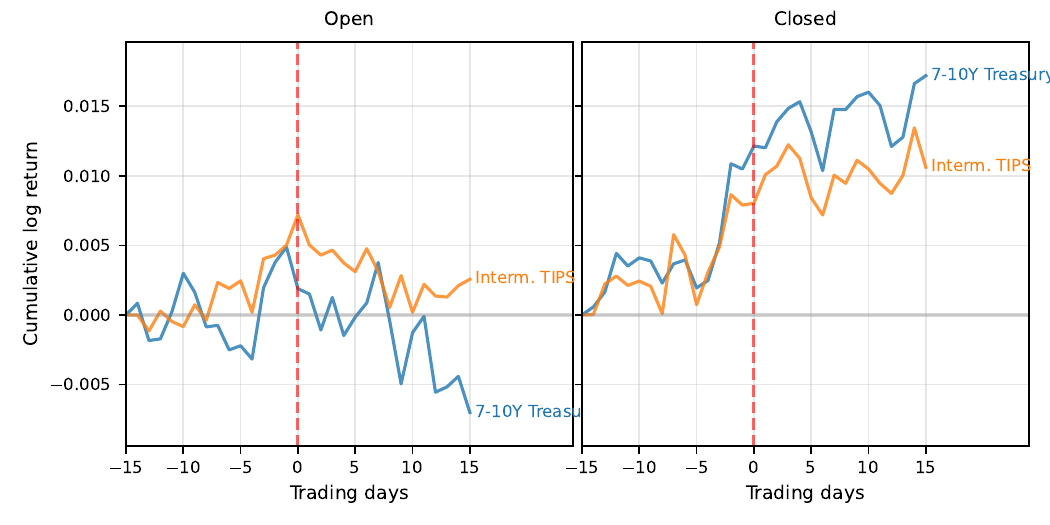}
         \caption{Breakeven intermediate}
         \label{fig:yield_event_study_open}
     \end{subfigure}
     \\
     \begin{subfigure}[b]{\textwidth}
         \centering
         \includegraphics[width=0.7\textwidth]{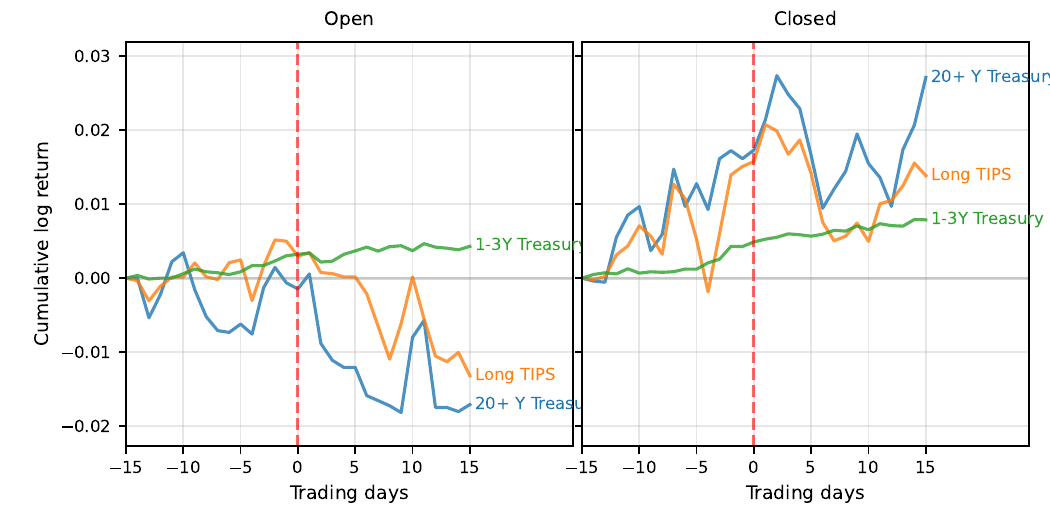}
         \caption{Breakeven long}
         \label{fig:yield_event_study_open}
     \end{subfigure}
        \caption{Inflation breakevens}
        \label{fig:otherindicators}
        \note{Median cumulative log return from 15 days before the release of AI models. Top panel shows IEF and TIP; bottom panel shows SHY, TLT, and LTPZ.}
\end{figure}

\begin{figure}
     \centering
     
        \begin{subfigure}[b]{\textwidth}
         \centering
         \includegraphics[width=0.9\textwidth]{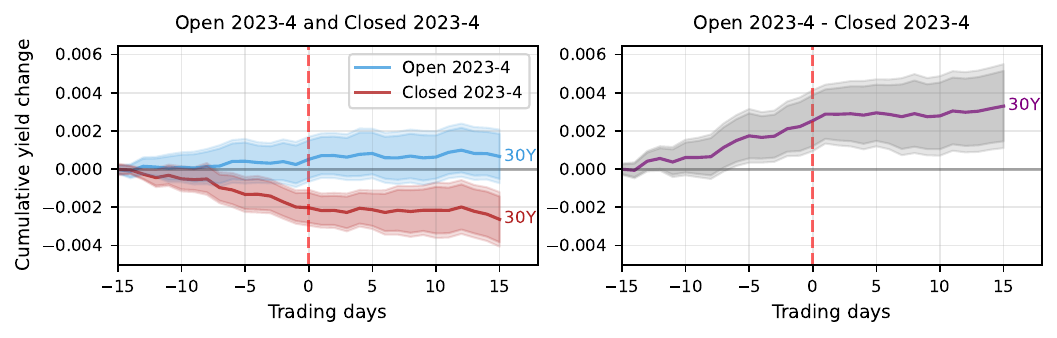}
     \end{subfigure}
     \\
        \begin{subfigure}[b]{\textwidth}
         \centering
         \includegraphics[width=0.9\textwidth]{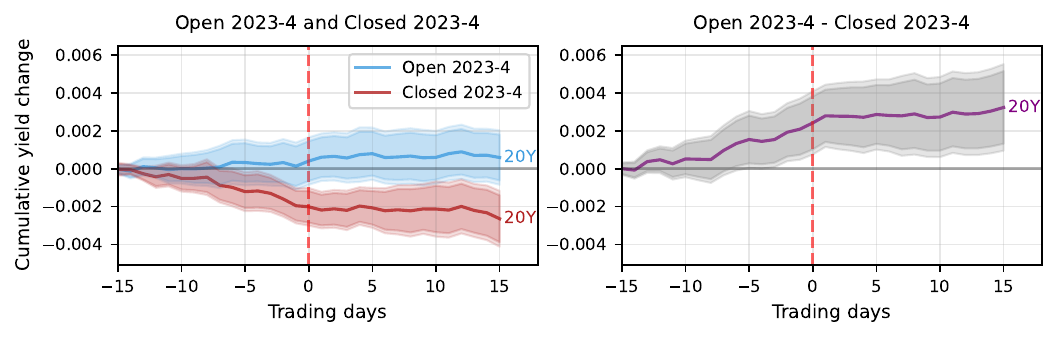}
     \end{subfigure}
     \\
        \begin{subfigure}[b]{\textwidth}
         \centering
         \includegraphics[width=0.9\textwidth]{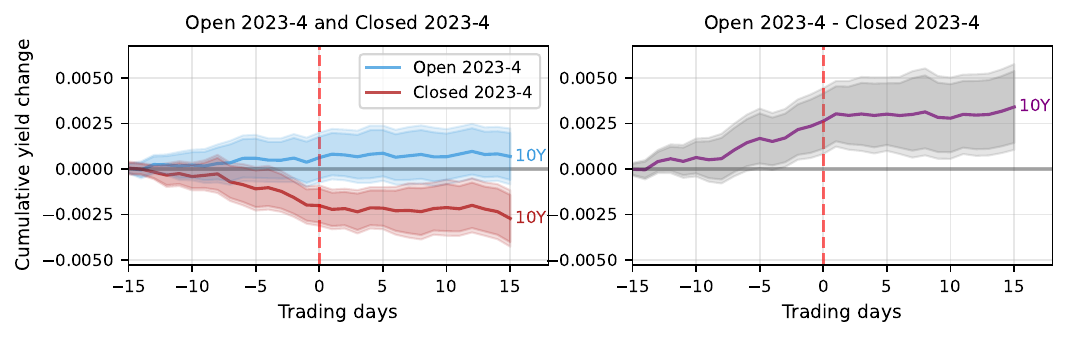}
     \end{subfigure}
     \\
        \begin{subfigure}[b]{\textwidth}
         \centering
         \includegraphics[width=0.9\textwidth]{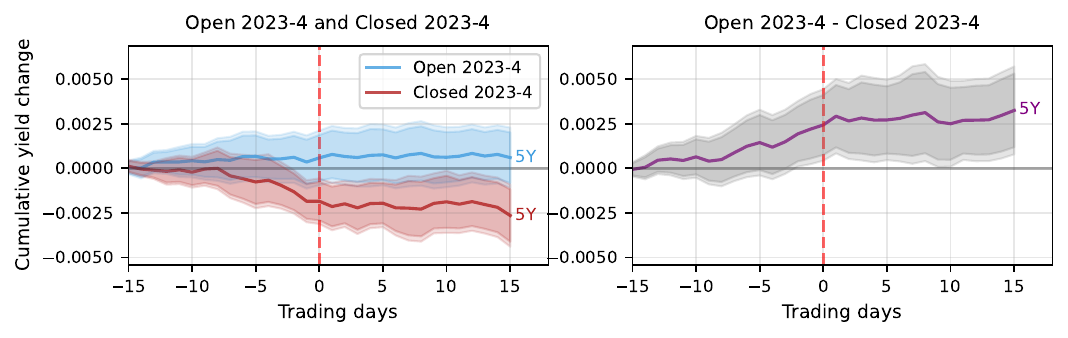}
     \end{subfigure}
     \\
        \begin{subfigure}[b]{\textwidth}
         \centering
         \includegraphics[width=0.9\textwidth]{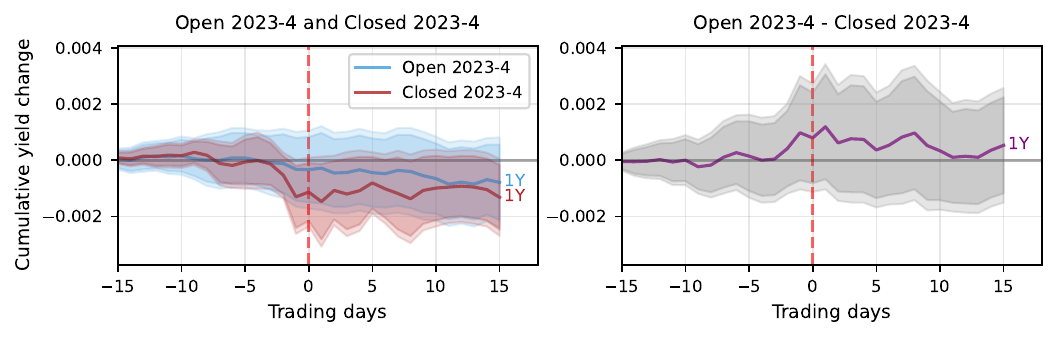}
     \end{subfigure}
     
     \caption{US treasury bond yield event study regression estimates (2023-24 releases)}
     \label{fig:regression_difference2023_24}
    \note{Estimates of cumulative yield changes. The left panels show cumulative returns from 15 days prior to release, using equation~\eqref{eq:cum_return_paired}. The right panels show the estimated difference. Figure~\ref{fig:regression_difference} shows the same plot for all years. Constant maturity duration noted in years. 90\% and 95\% confidence intervals are shaded.}
\end{figure}

\begin{figure}
     \centering
        \begin{subfigure}[b]{\textwidth}
         \centering
         \includegraphics[width=\textwidth]{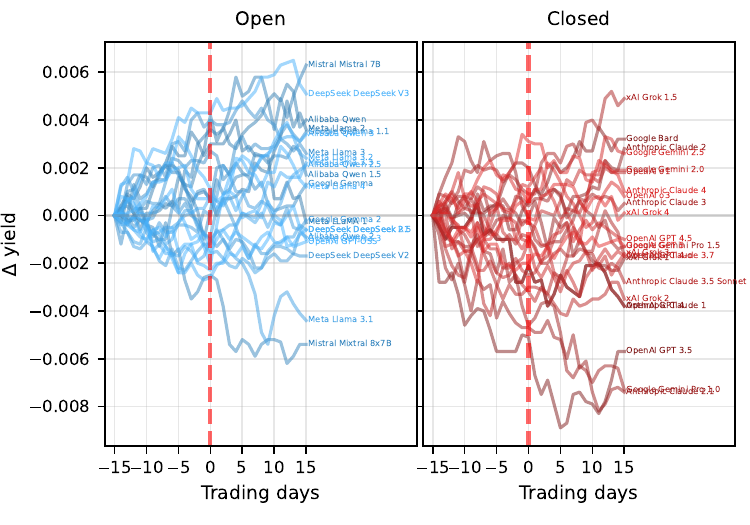}
         \caption{30Y}
     \end{subfigure}
     \\
        \begin{subfigure}[b]{\textwidth}
         \centering
         \includegraphics[width=\textwidth]{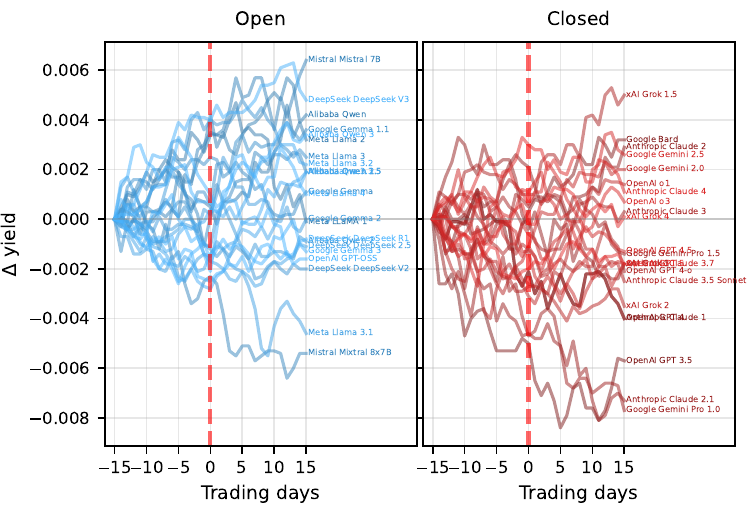}
         \caption{20Y}
     \end{subfigure}
     
     \caption{Yield change in US treasury bonds around all AI releases (1)}
     \label{fig:yields_by_event}
    \note{Differences in yields starting from 15 business days before each event.}
\end{figure}

\begin{figure}
     \centering
        \begin{subfigure}[b]{\textwidth}
         \centering
         \includegraphics[width=\textwidth]{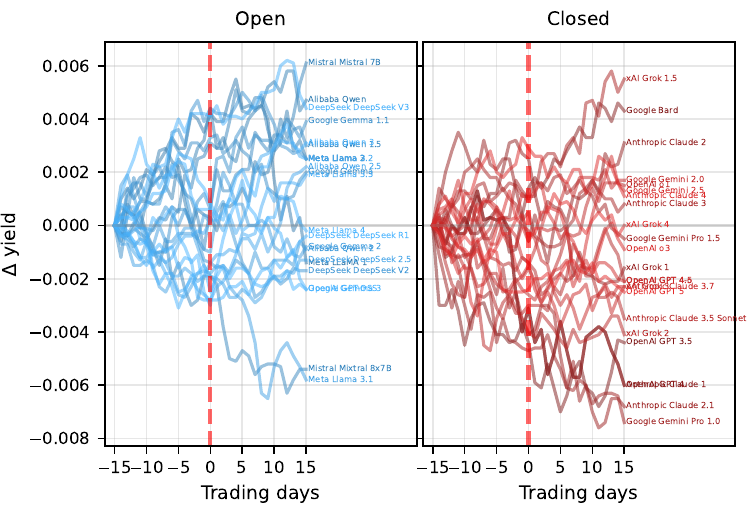}
         \caption{10Y}
     \end{subfigure}
     \\
        \begin{subfigure}[b]{\textwidth}
         \centering
         \includegraphics[width=\textwidth]{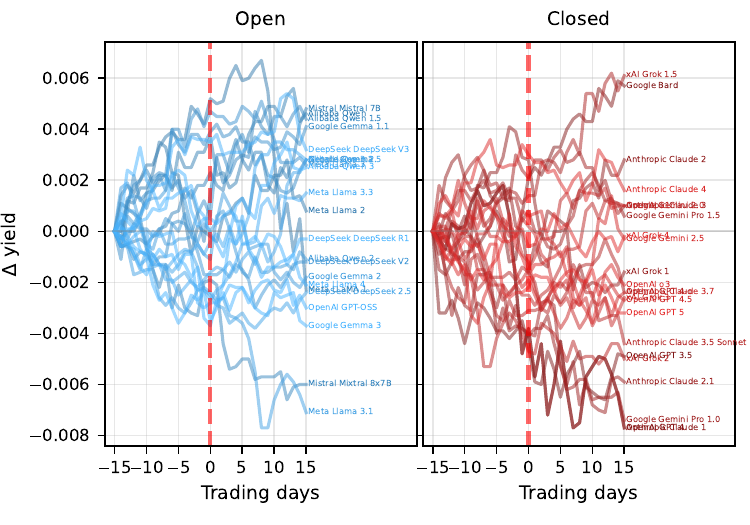}
         \caption{5Y}
     \end{subfigure}
     
     \caption{Yield change in US treasury bonds around all AI releases (2)}
     \label{fig:yields_by_event2}
    \note{Differences in yields starting from 15 business days before each event.}
\end{figure}

\begin{figure}
     \centering
        \begin{subfigure}[b]{\textwidth}
         \centering
         \includegraphics[width=\textwidth]{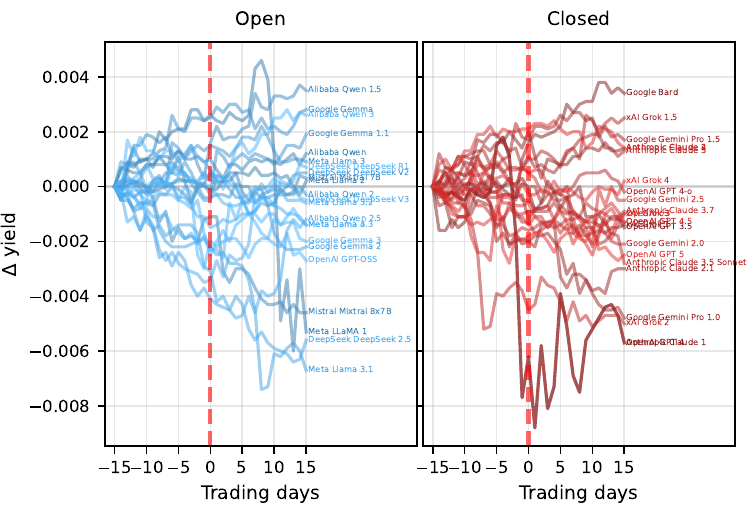}
         \caption{1Y}
     \end{subfigure}
     
     \caption{Yield change in US treasury bonds around all AI releases (3)}
     \label{fig:yields_by_event3}
    \note{Differences in yields starting from 15 business days before each event.}
\end{figure}

\begin{figure}[ht]
     \begin{subfigure}[b]{\textwidth}
         \centering
         \includegraphics[width=0.4\textwidth]{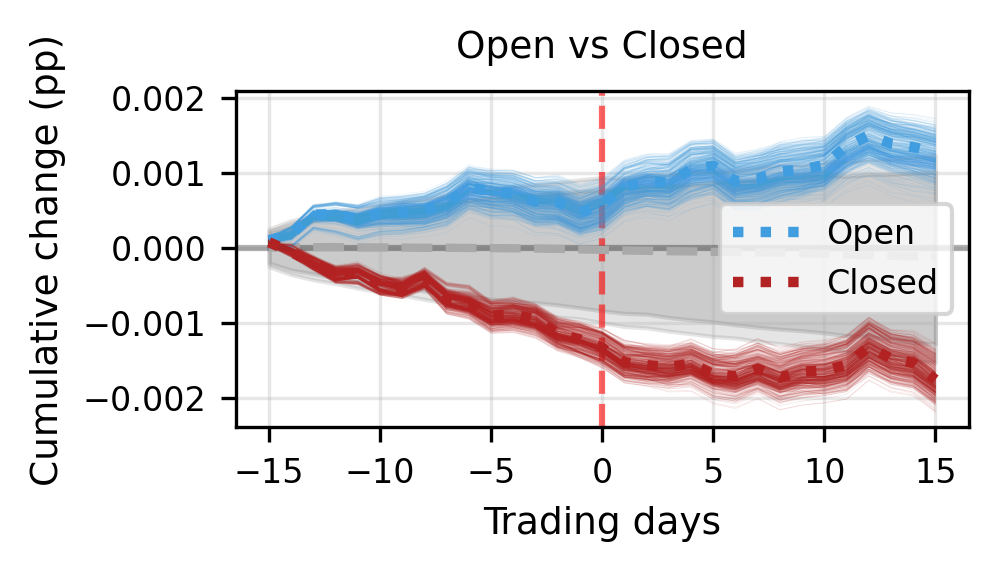}
         \includegraphics[width=0.4\textwidth]{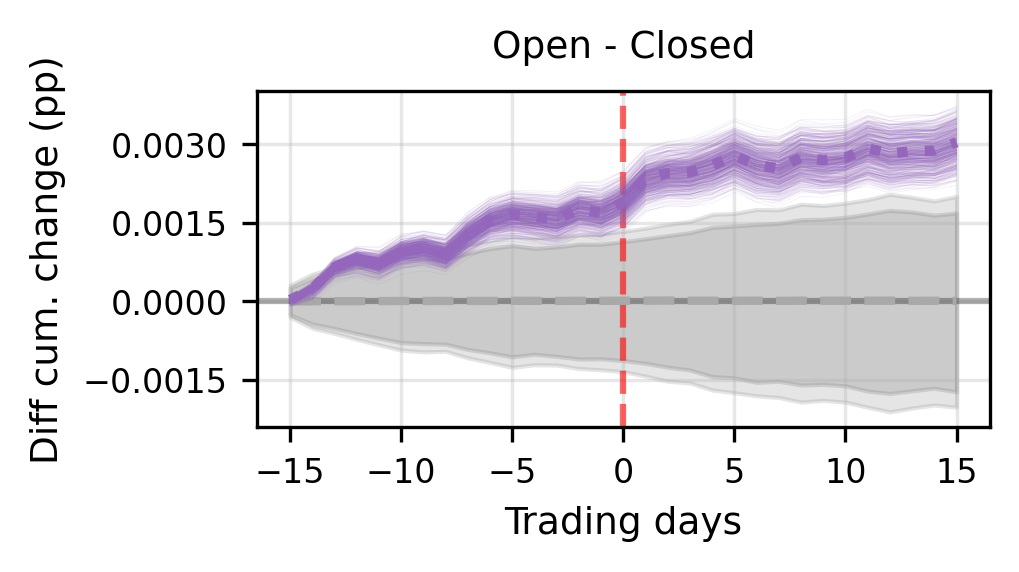}
         \caption{30 year}
     \end{subfigure}
     \\
     \begin{subfigure}[b]{\textwidth}
         \centering
         \includegraphics[width=0.4\textwidth]{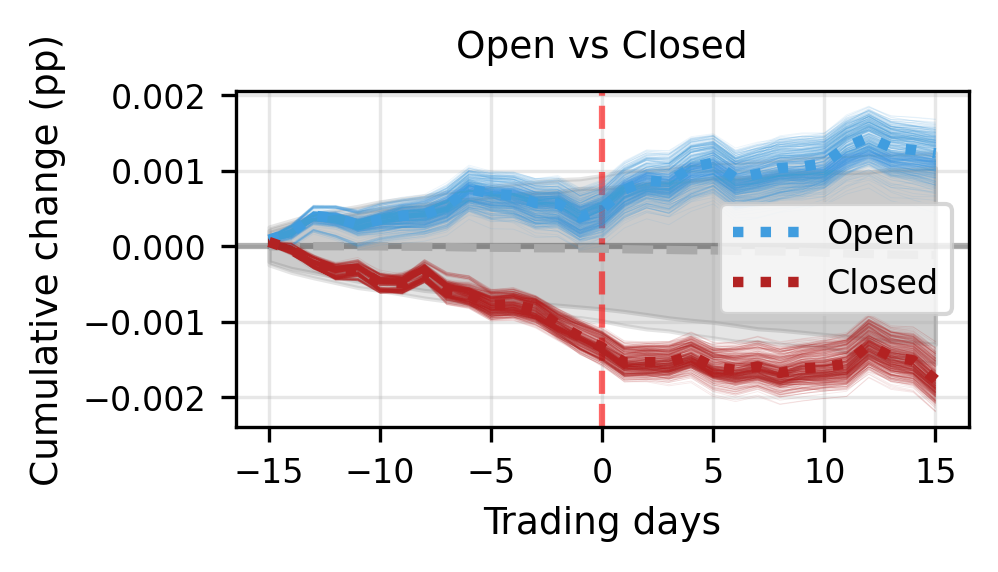}
         \includegraphics[width=0.4\textwidth]{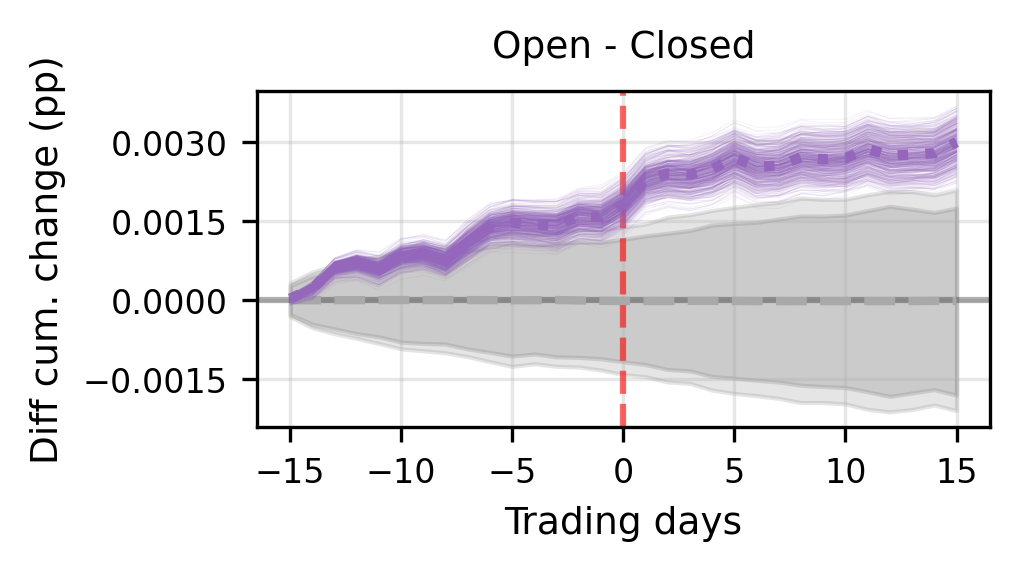}
         \caption{20 year}
     \end{subfigure}
     \\
     \begin{subfigure}[b]{\textwidth}
         \centering
         \includegraphics[width=0.4\textwidth]{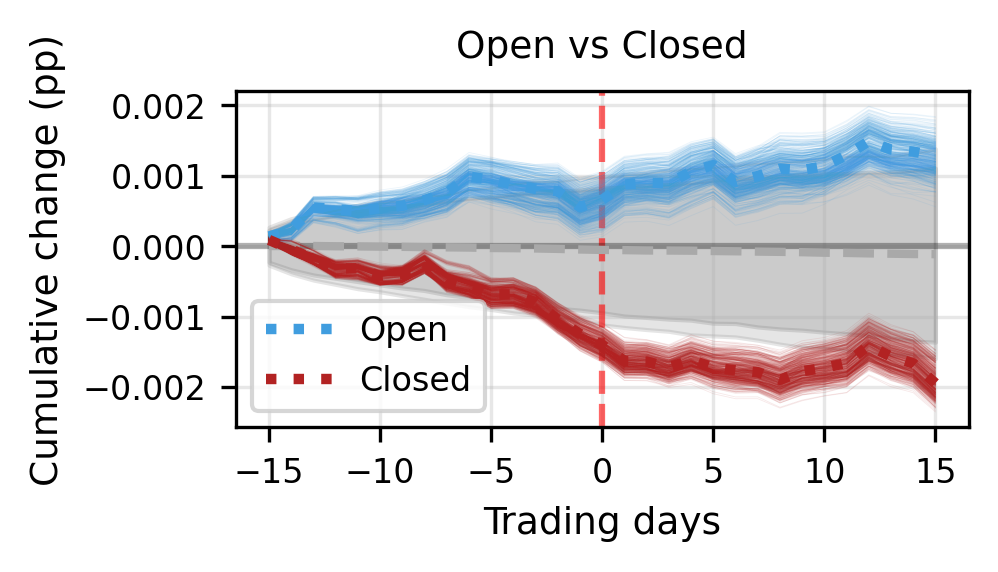}
         \includegraphics[width=0.4\textwidth]{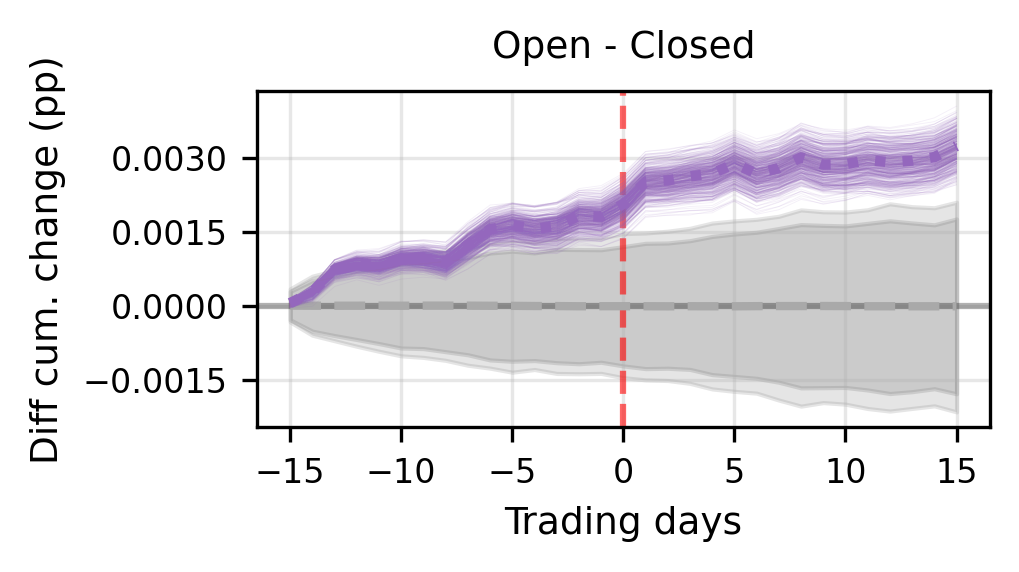}
         \caption{10 year}
     \end{subfigure}
     \\
     \begin{subfigure}[b]{\textwidth}
         \centering
         \includegraphics[width=0.4\textwidth]{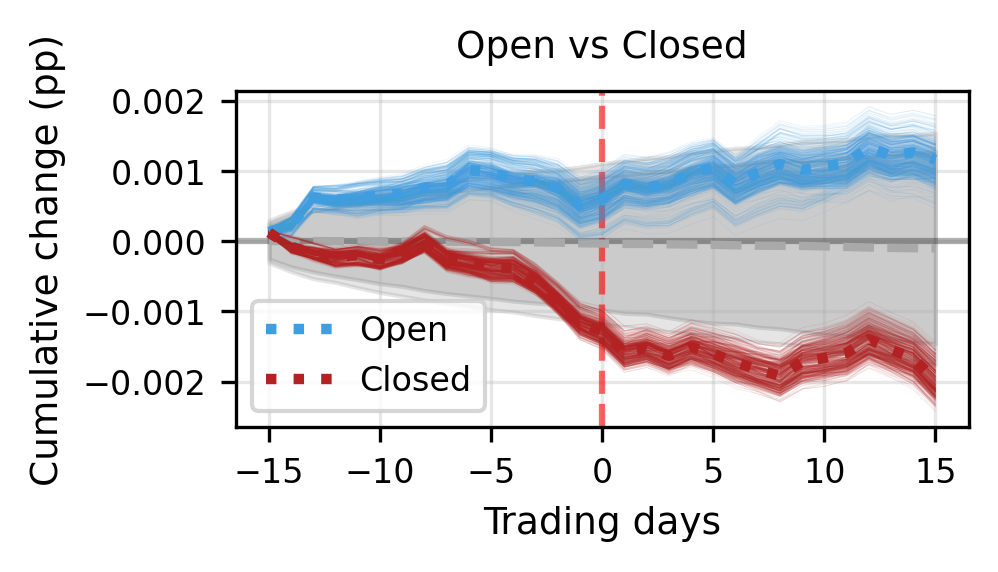}
         \includegraphics[width=0.4\textwidth]{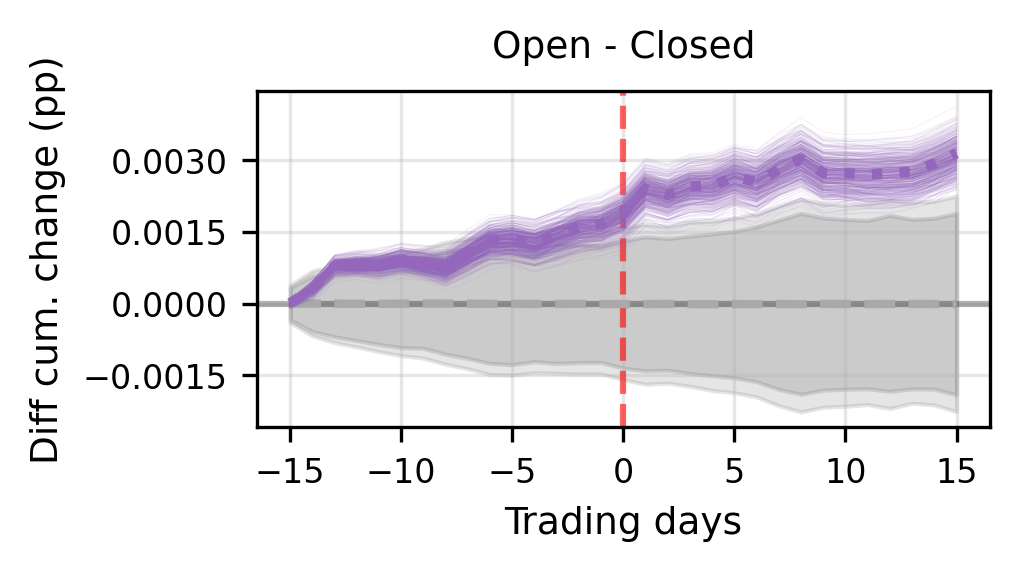}
         \caption{5 year}
     \end{subfigure}
     \\
     \begin{subfigure}[b]{\textwidth}
         \centering
         \includegraphics[width=0.4\textwidth]{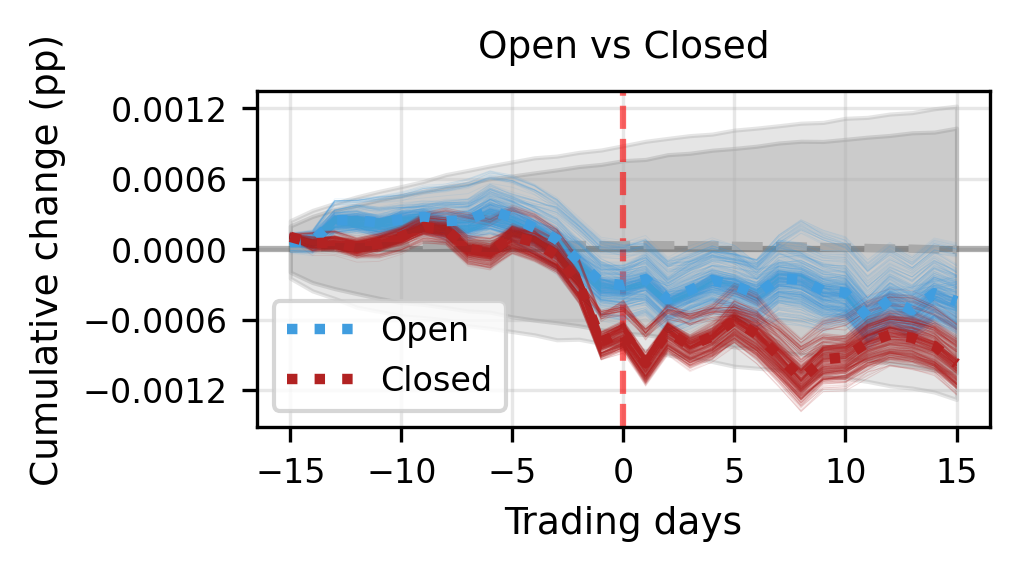}
         \includegraphics[width=0.4\textwidth]{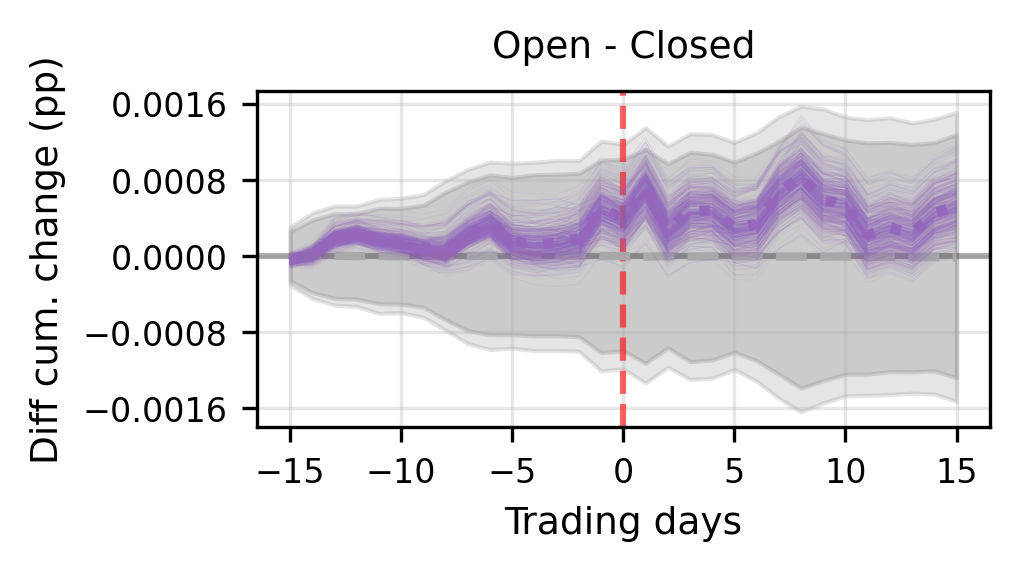}
         \caption{1 year}
     \end{subfigure}
     \\
\caption{Estimated yield change in US treasury bonds around major AI releases (OLS), permutation test, dropping events (1 pair)}
\note{OLS coefficient estimates. The shaded regions represent 90\% and 95\% of a placebo distribution, over 5,000 permutation draws. Dashed line indicates mean of the placebo distribution. Dotted line represents the main estimate. Thin lines represent the estimate resulting from dropping each pair of events (one from the open sample and one from the closed sample). For the first column, the placebo includes all dates in data. For last column, the placebo distribution includes the pooled event dates of closed and open AI model releases.}
\label{f:permutation_open_and_closed_ols_jackknife_drop1}\end{figure}

\begin{figure}[ht]
     \begin{subfigure}[b]{\textwidth}
         \centering
         \includegraphics[width=0.4\textwidth]{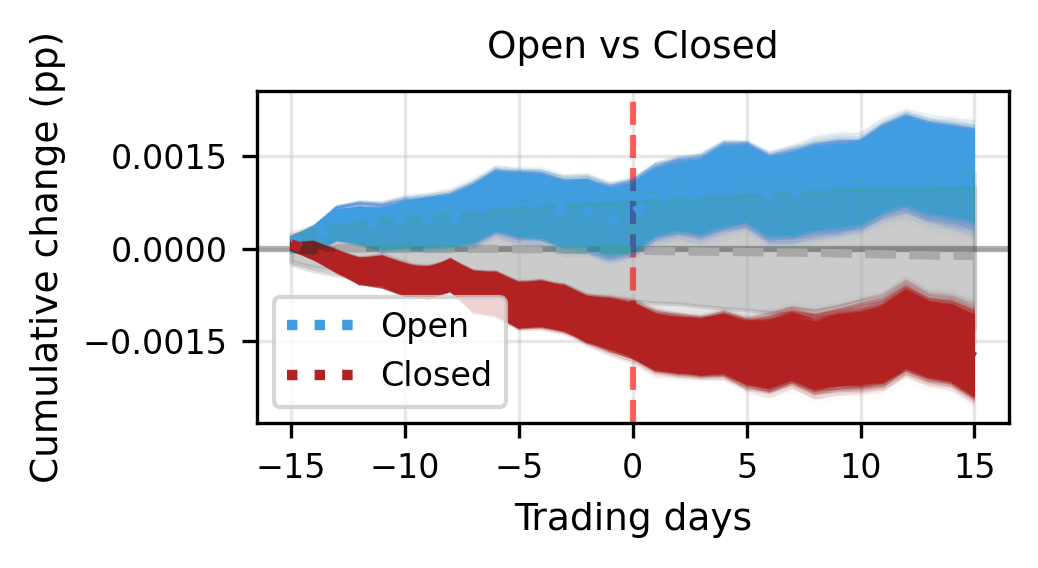}
         \includegraphics[width=0.4\textwidth]{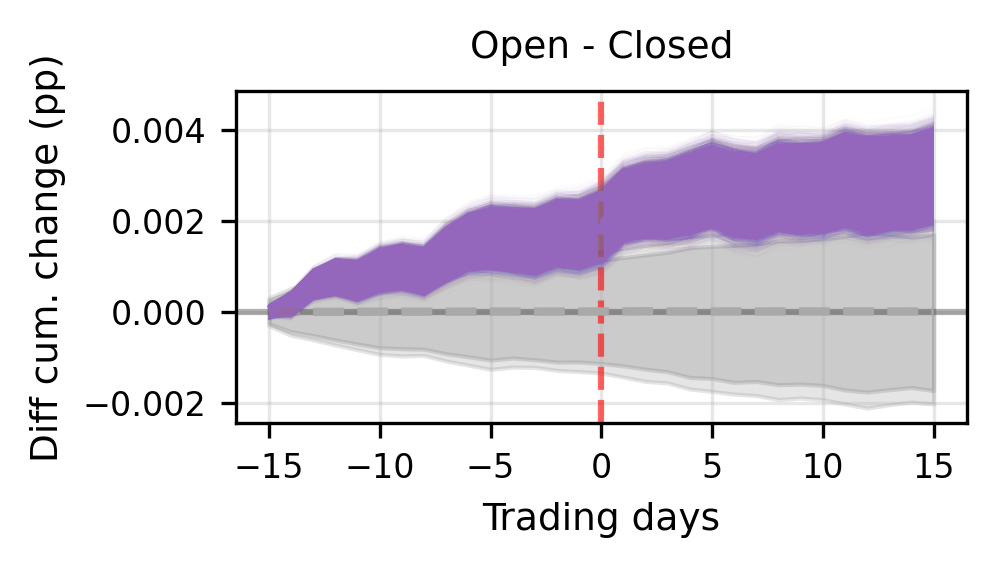}
         \caption{30 year}
     \end{subfigure}
     \\
     \begin{subfigure}[b]{\textwidth}
         \centering
         \includegraphics[width=0.4\textwidth]{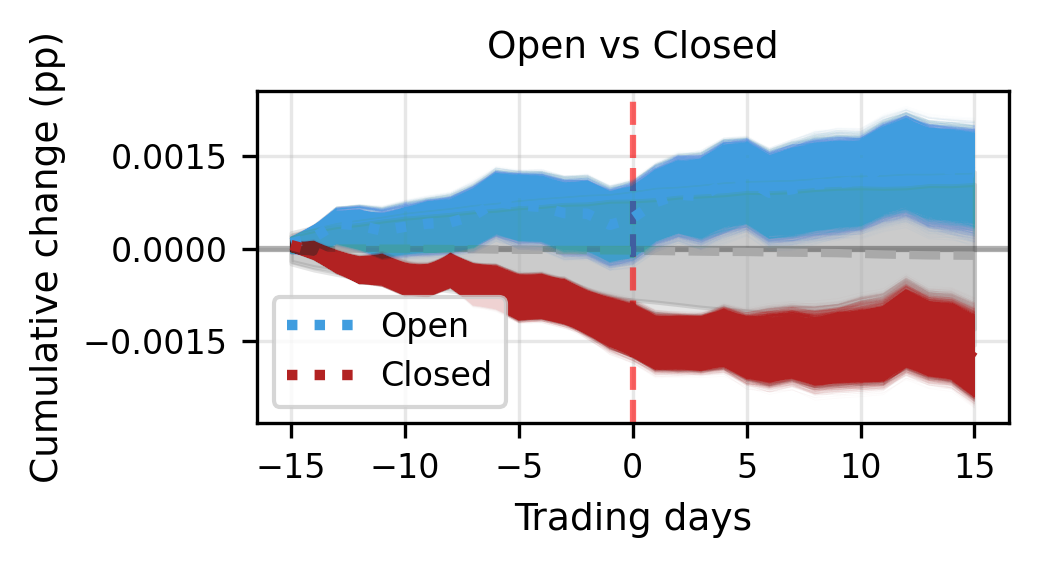}
         \includegraphics[width=0.4\textwidth]{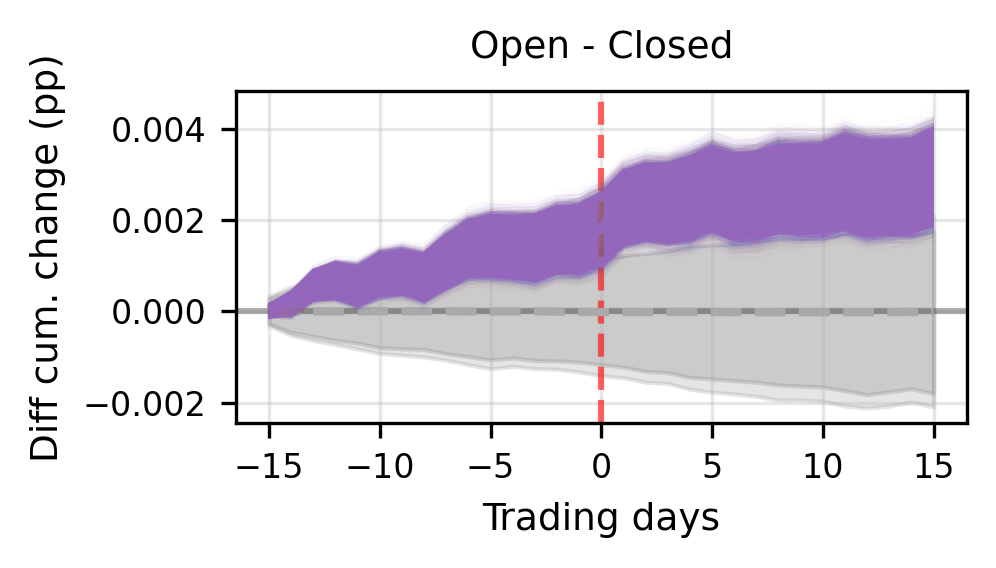}
         \caption{20 year}
     \end{subfigure}
     \\
     \begin{subfigure}[b]{\textwidth}
         \centering
         \includegraphics[width=0.4\textwidth]{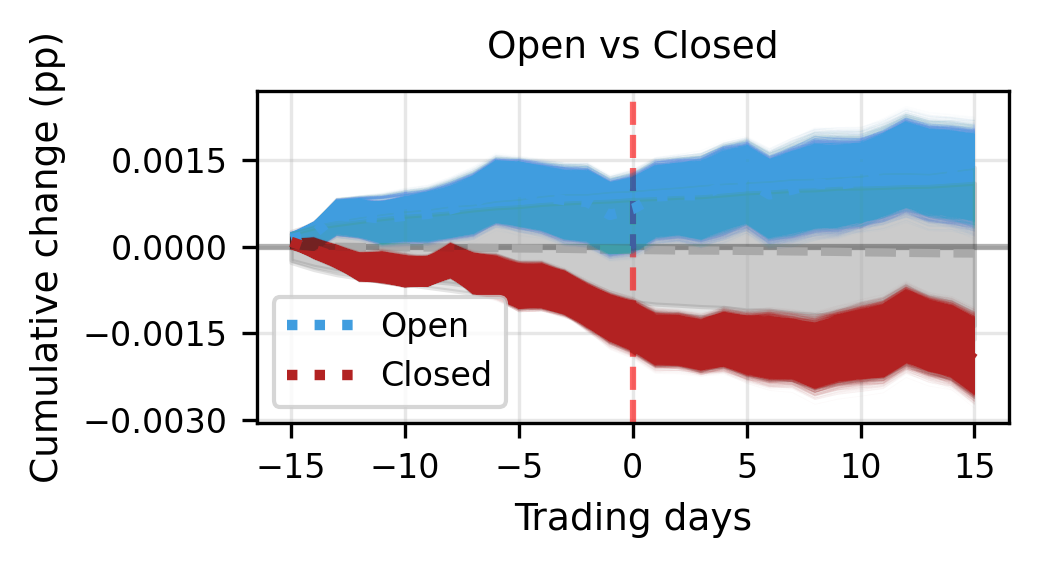}
         \includegraphics[width=0.4\textwidth]{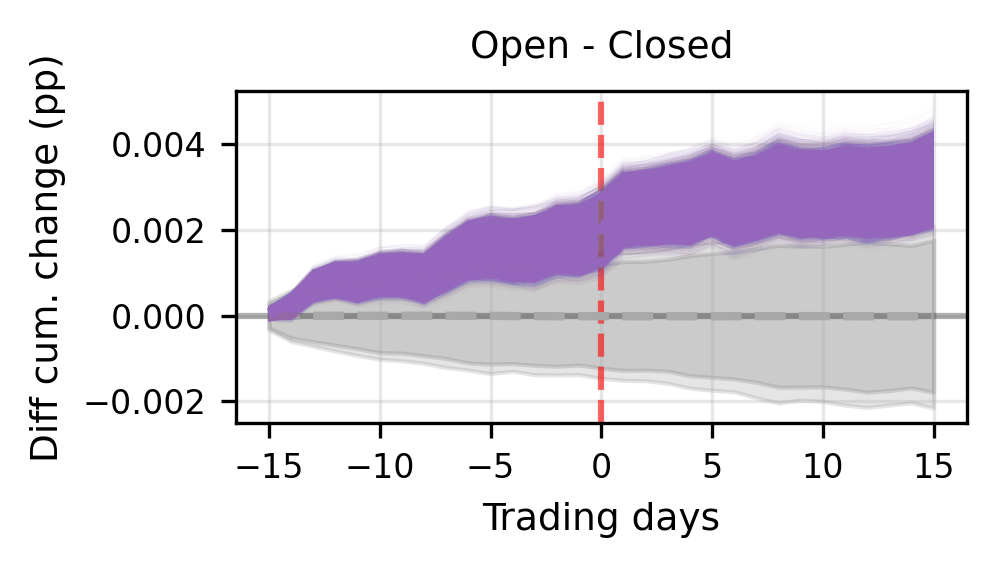}
         \caption{10 year}
     \end{subfigure}
     \\
     \begin{subfigure}[b]{\textwidth}
         \centering
         \includegraphics[width=0.4\textwidth]{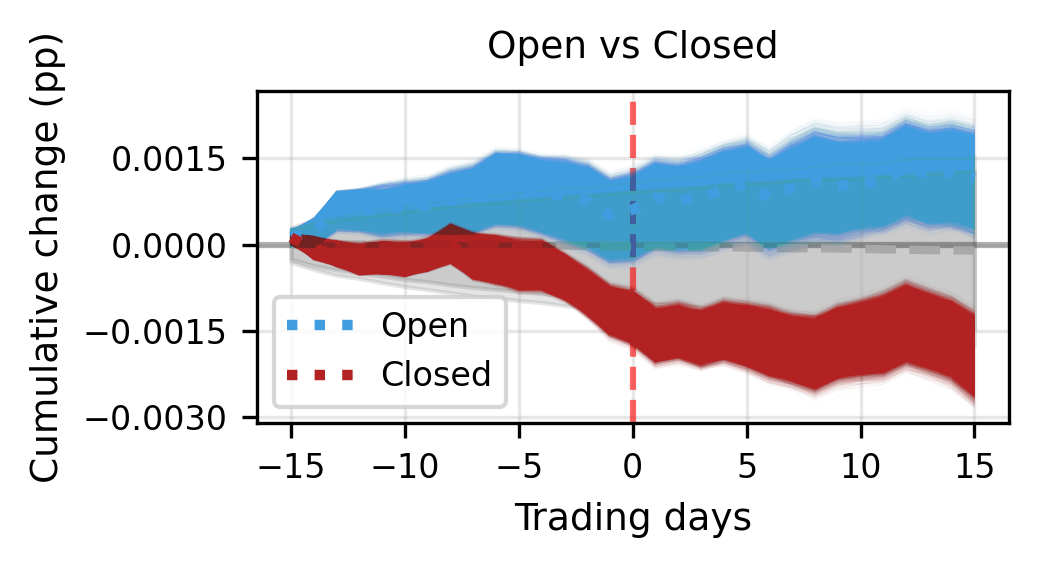}
         \includegraphics[width=0.4\textwidth]{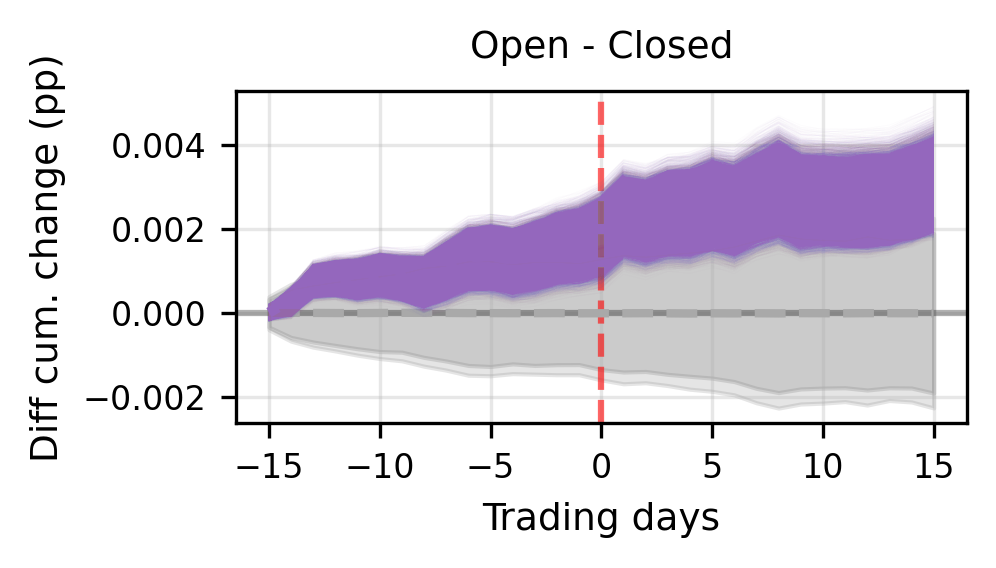}
         \caption{5 year}
     \end{subfigure}
     \\
     \begin{subfigure}[b]{\textwidth}
         \centering
         \includegraphics[width=0.4\textwidth]{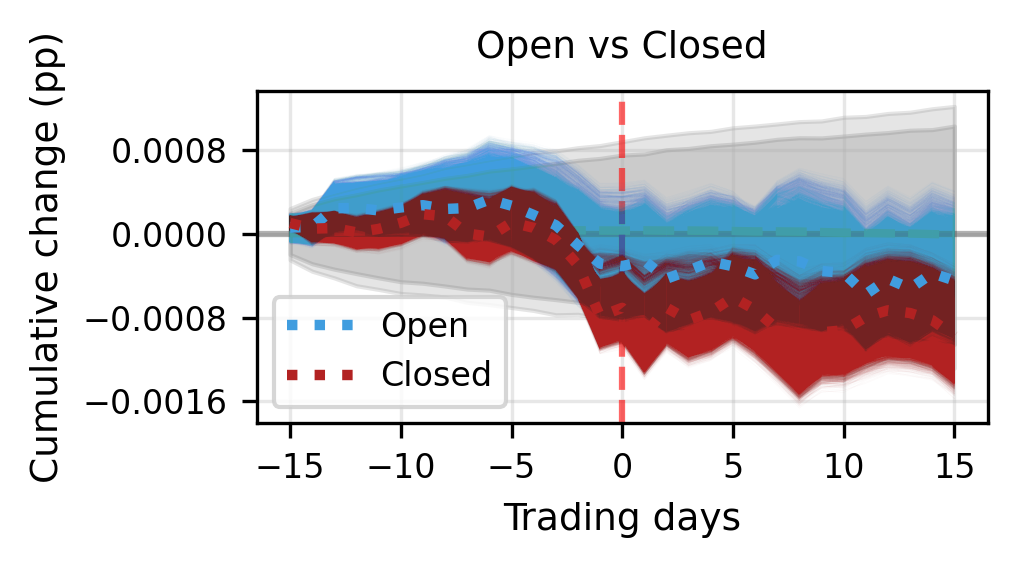}
         \includegraphics[width=0.4\textwidth]{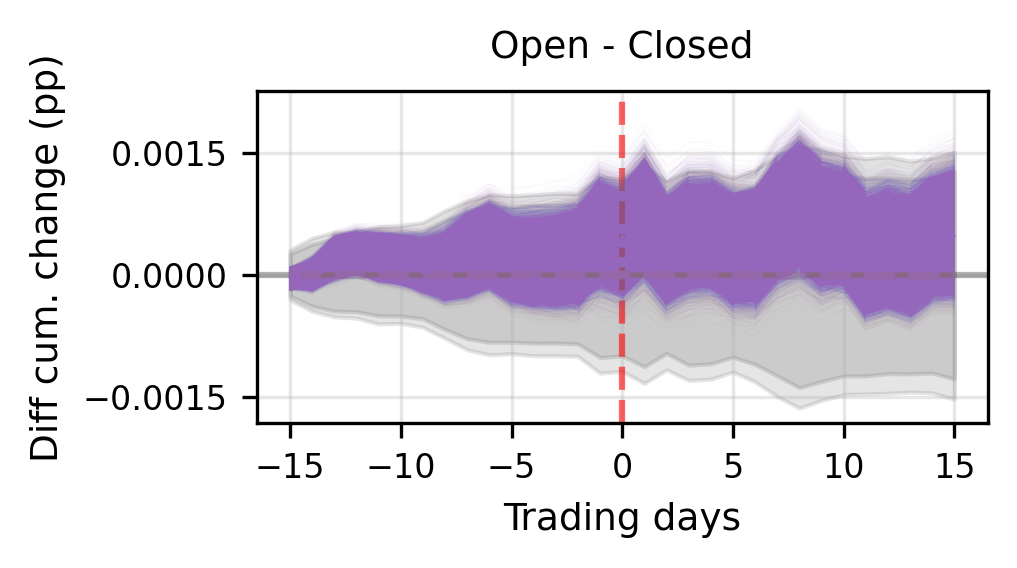}
         \caption{1 year}
     \end{subfigure}
     \\
\caption{Estimated yield change in US treasury bonds around major AI releases (OLS), permutation test, dropping events (2 pairs)}
\note{OLS coefficient estimates. The shaded regions represent 90\% and 95\% of a placebo distribution, over 5,000 permutation draws. Dashed line indicates mean of the placebo distribution. Dotted line represents the main estimate. Thin lines represent the estimate resulting from dropping two pairs of events (two from the open sample and two from the closed sample). For the first column, the placebo includes all dates in data. For last column, the placebo distribution includes the pooled event dates of closed and open AI model releases.}
\label{f:permutation_open_and_closed_ols_jackknife_drop2}\end{figure}

\begin{figure}[ht]
     \begin{subfigure}[b]{\textwidth}
         \centering
         \includegraphics[width=0.4\textwidth]{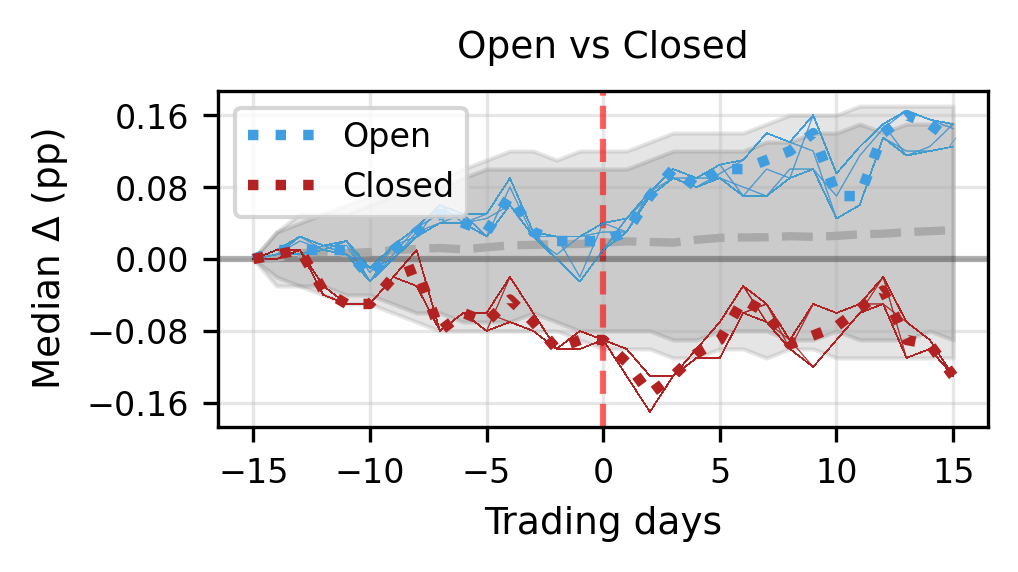}
         \includegraphics[width=0.4\textwidth]{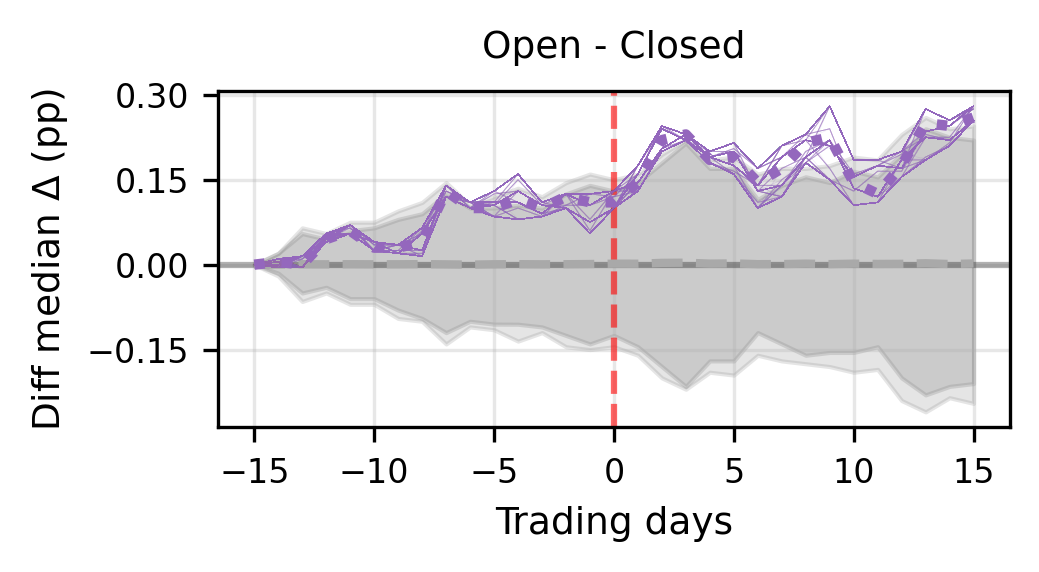}
         \caption{30 year}
     \end{subfigure}
     \\
     \begin{subfigure}[b]{\textwidth}
         \centering
         \includegraphics[width=0.4\textwidth]{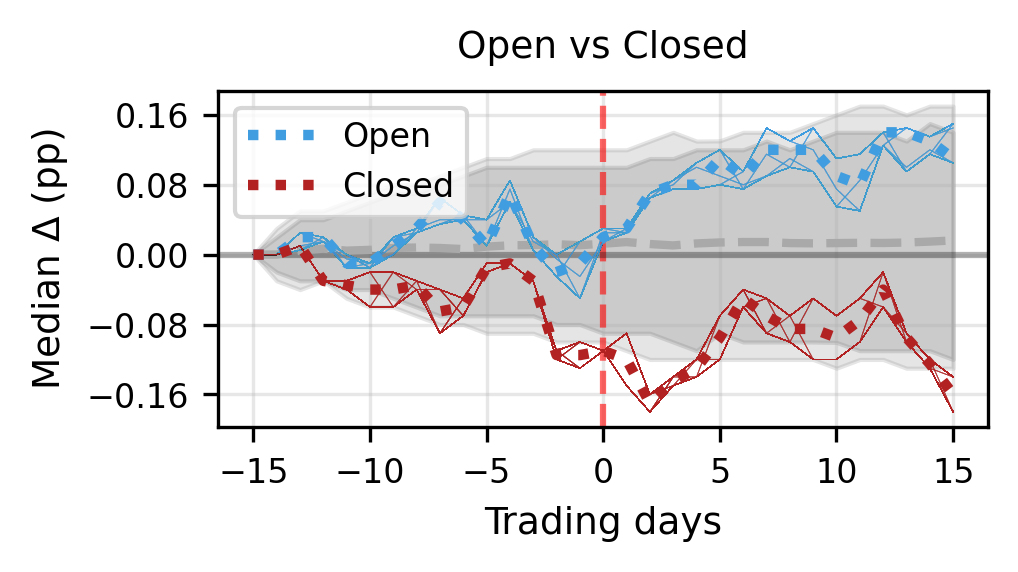}
         \includegraphics[width=0.4\textwidth]{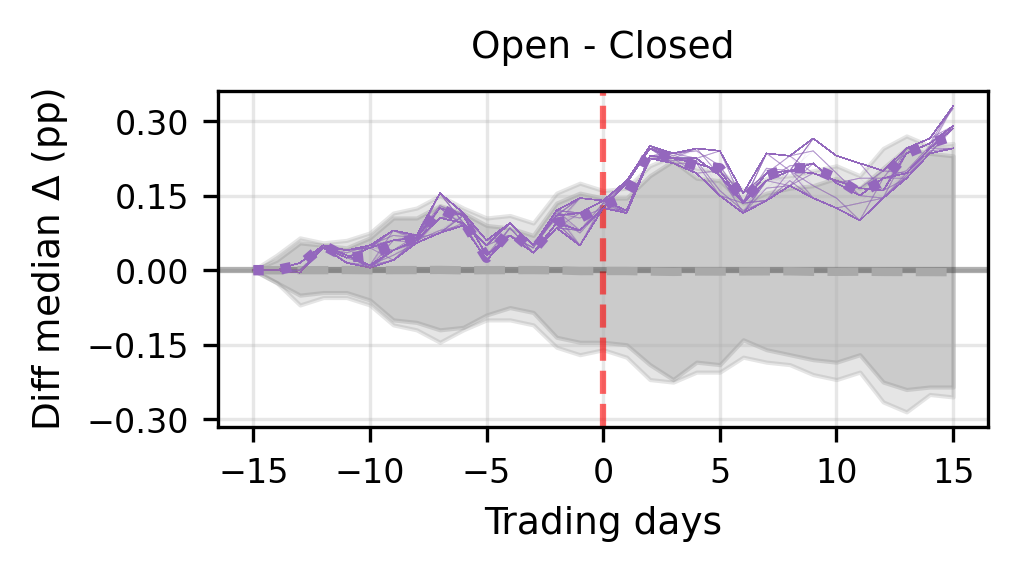}
         \caption{20 year}
     \end{subfigure}
     \\
     \begin{subfigure}[b]{\textwidth}
         \centering
         \includegraphics[width=0.4\textwidth]{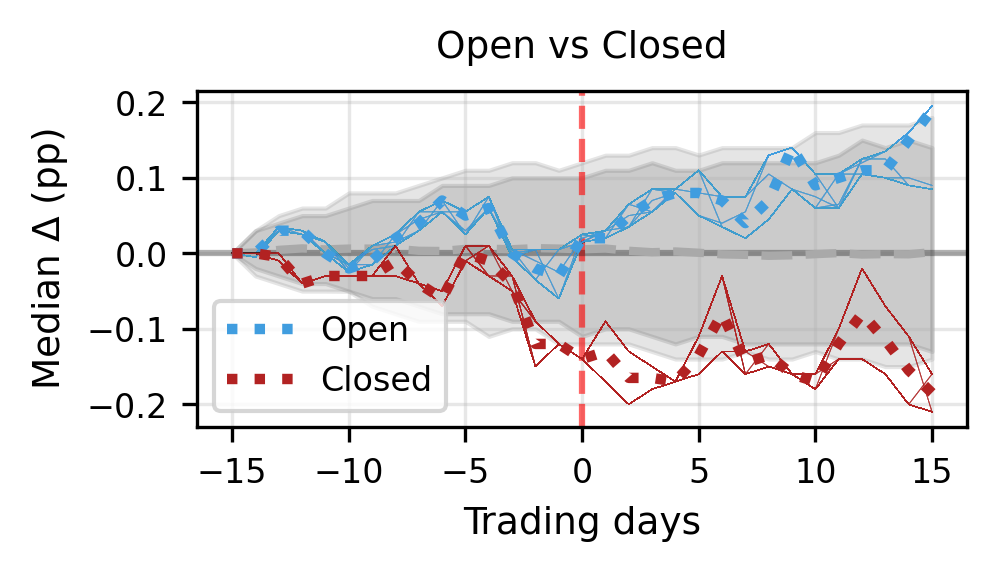}
         \includegraphics[width=0.4\textwidth]{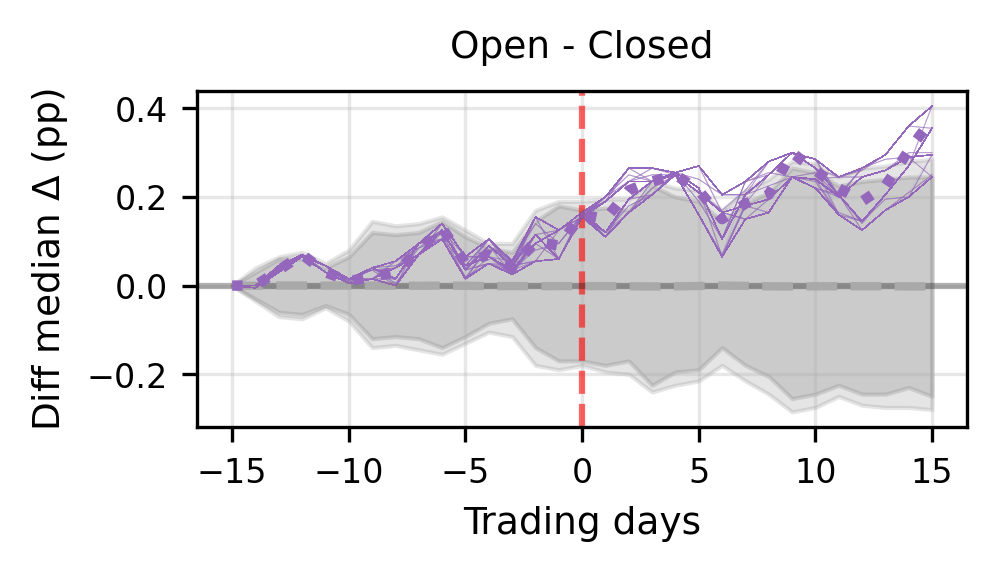}
         \caption{10 year}
     \end{subfigure}
     \\
     \begin{subfigure}[b]{\textwidth}
         \centering
         \includegraphics[width=0.4\textwidth]{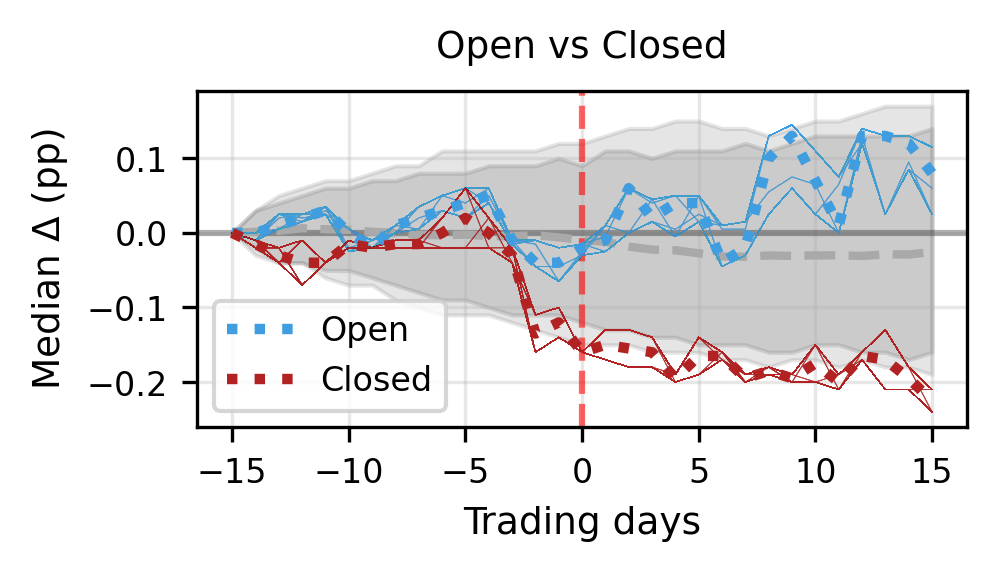}
         \includegraphics[width=0.4\textwidth]{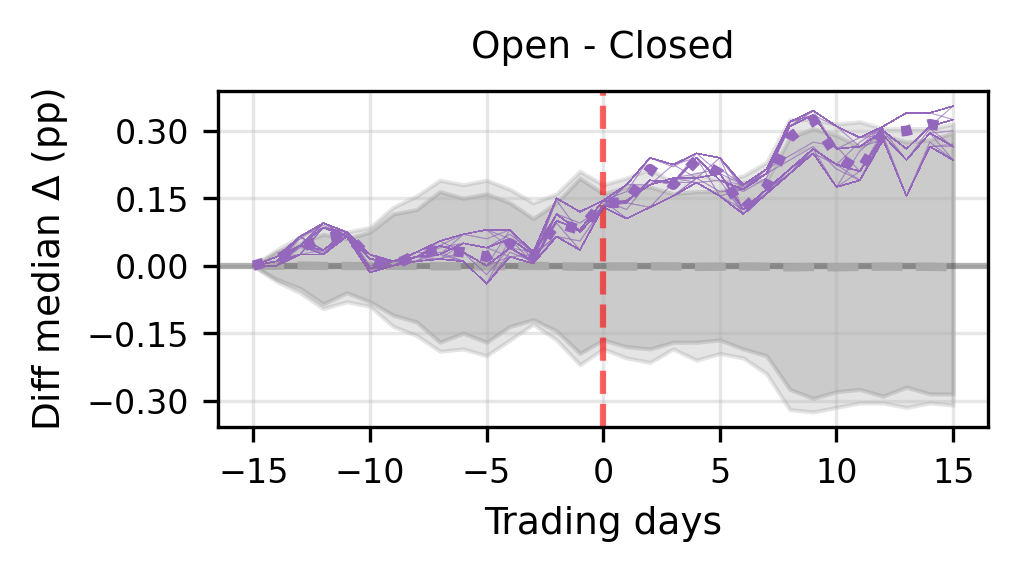}
         \caption{5 year}
     \end{subfigure}
     \\
     \begin{subfigure}[b]{\textwidth}
         \centering
         \includegraphics[width=0.4\textwidth]{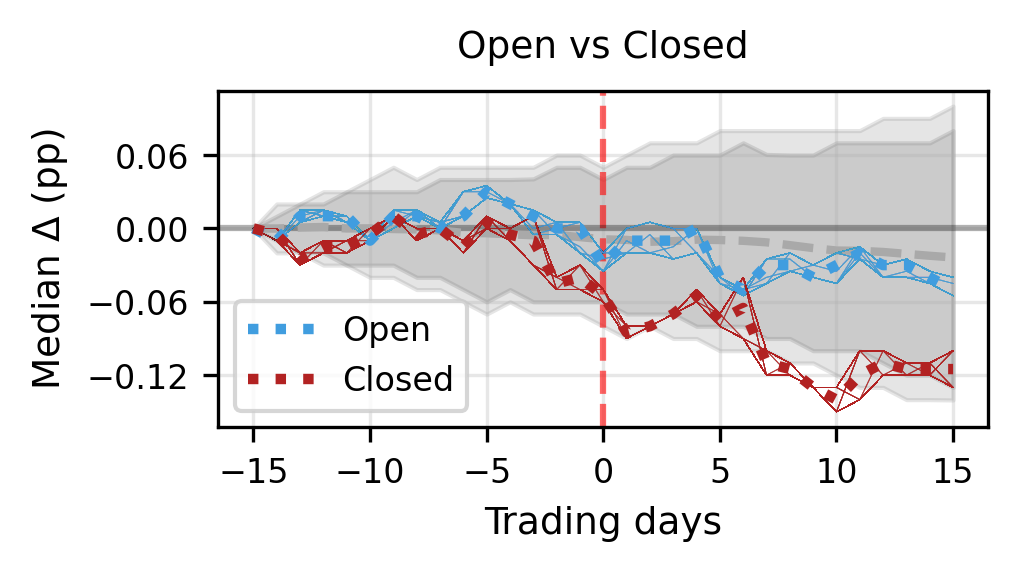}
         \includegraphics[width=0.4\textwidth]{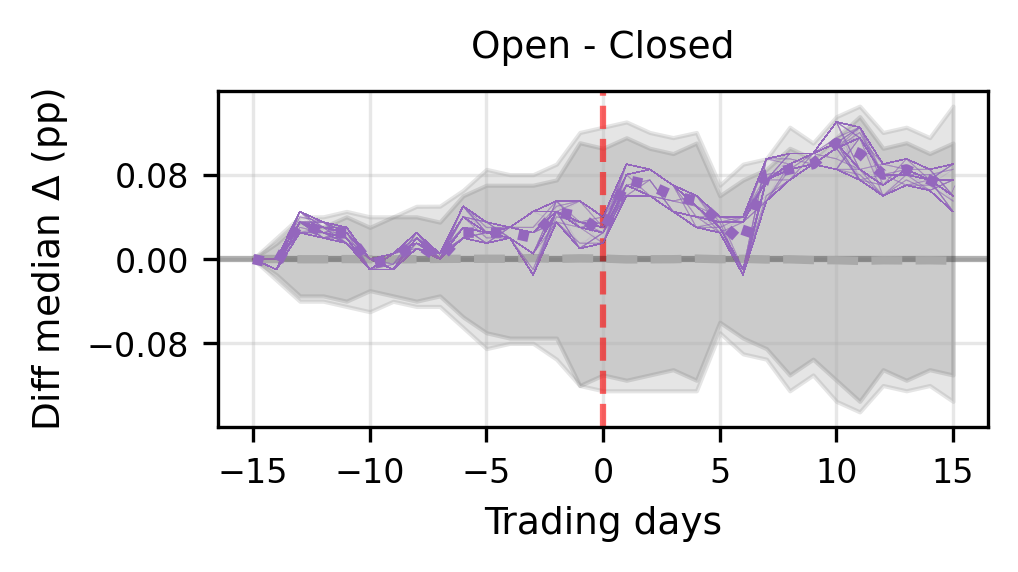}
         \caption{1 year}
     \end{subfigure}
     \\
\caption{Median yield change in US treasury bonds around major AI releases, permutation test, dropping events (1 pair)}
\note{Dotted line shows the estimated change for the median event, relative to 15 days before the event. The shaded regions represent 90\% and 95\% of a placebo distribution, over 5,000 permutation draws. Dashed line indicates mean of the placebo distribution. Thin lines represent the estimate resulting from dropping each pair of events (one from the open sample and one from the closed sample). For the first column, the placebo includes all dates in data. For last column, the placebo distribution includes the pooled event dates of closed and open AI model releases.}
\label{f:permutation_open_and_closed_medianchange_jackknife_drop1}\end{figure}

\begin{figure}[ht]
     \begin{subfigure}[b]{\textwidth}
         \centering
         \includegraphics[width=0.4\textwidth]{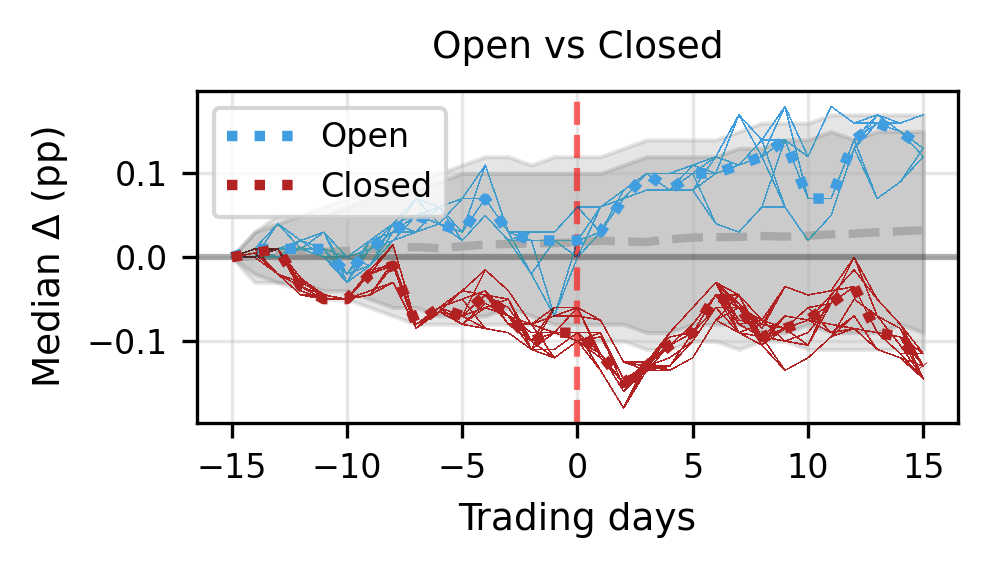}
         \includegraphics[width=0.4\textwidth]{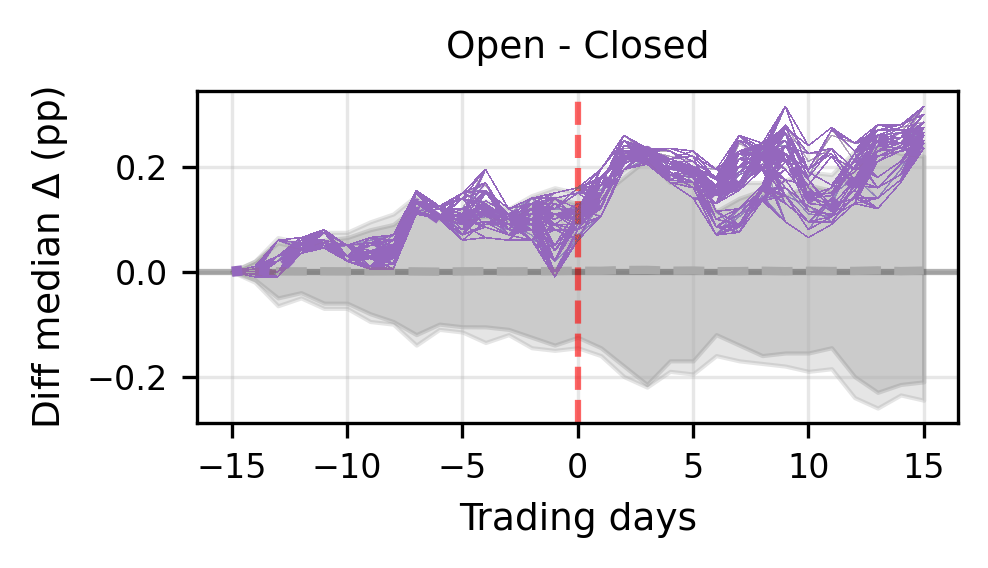}
         \caption{30 year}
     \end{subfigure}
     \\
     \begin{subfigure}[b]{\textwidth}
         \centering
         \includegraphics[width=0.4\textwidth]{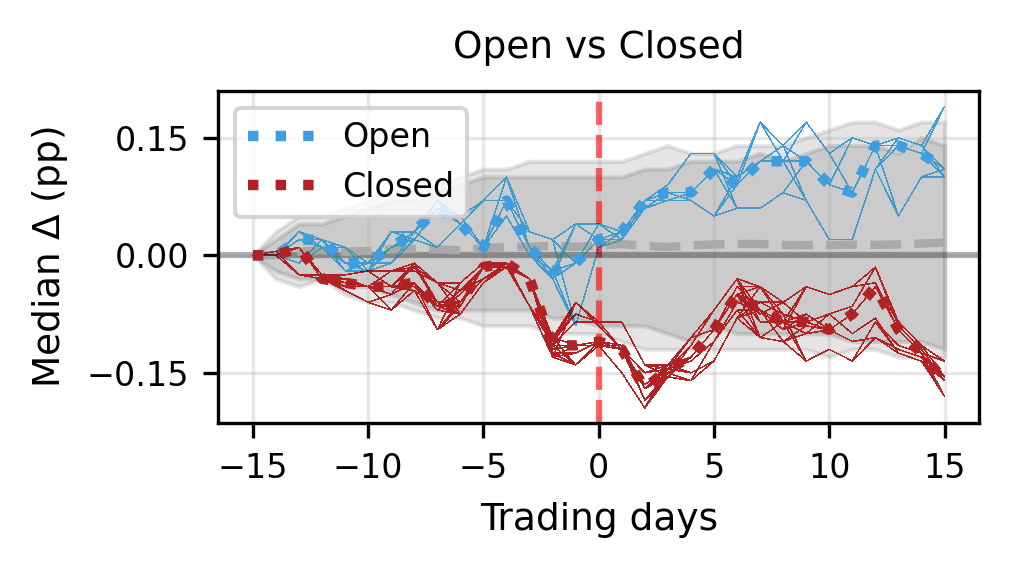}
         \includegraphics[width=0.4\textwidth]{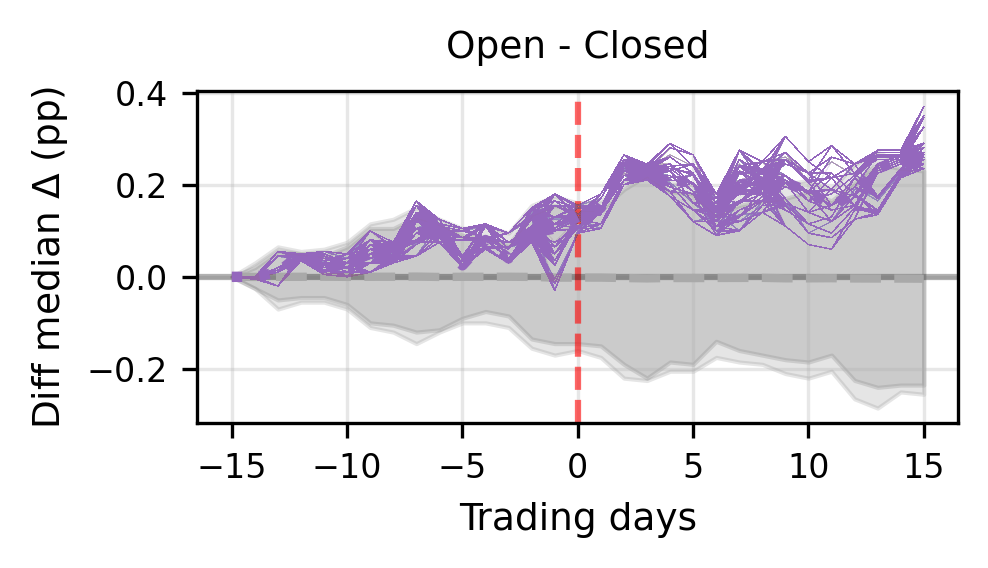}
         \caption{20 year}
     \end{subfigure}
     \\
     \begin{subfigure}[b]{\textwidth}
         \centering
         \includegraphics[width=0.4\textwidth]{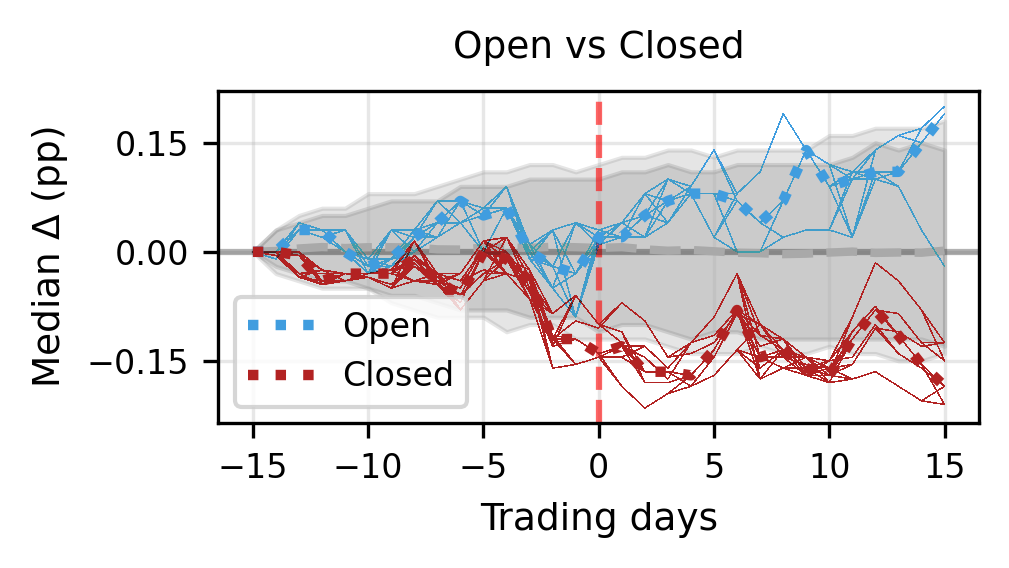}
         \includegraphics[width=0.4\textwidth]{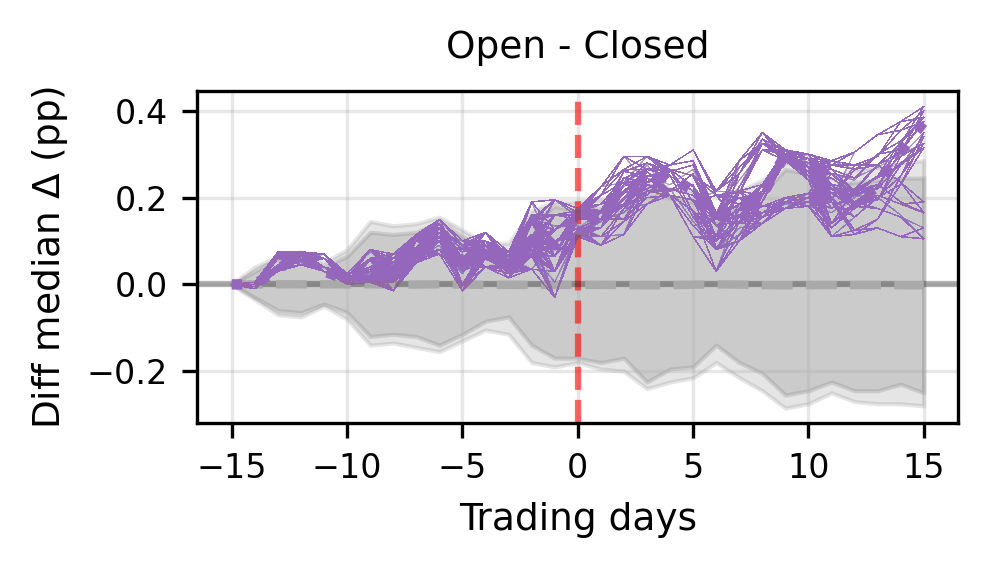}
         \caption{10 year}
     \end{subfigure}
     \\
     \begin{subfigure}[b]{\textwidth}
         \centering
         \includegraphics[width=0.4\textwidth]{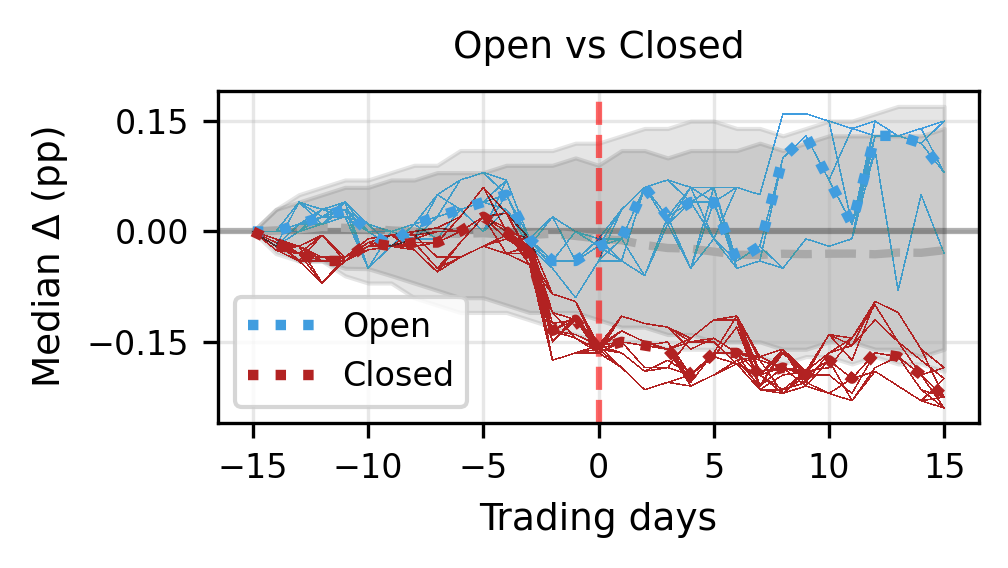}
         \includegraphics[width=0.4\textwidth]{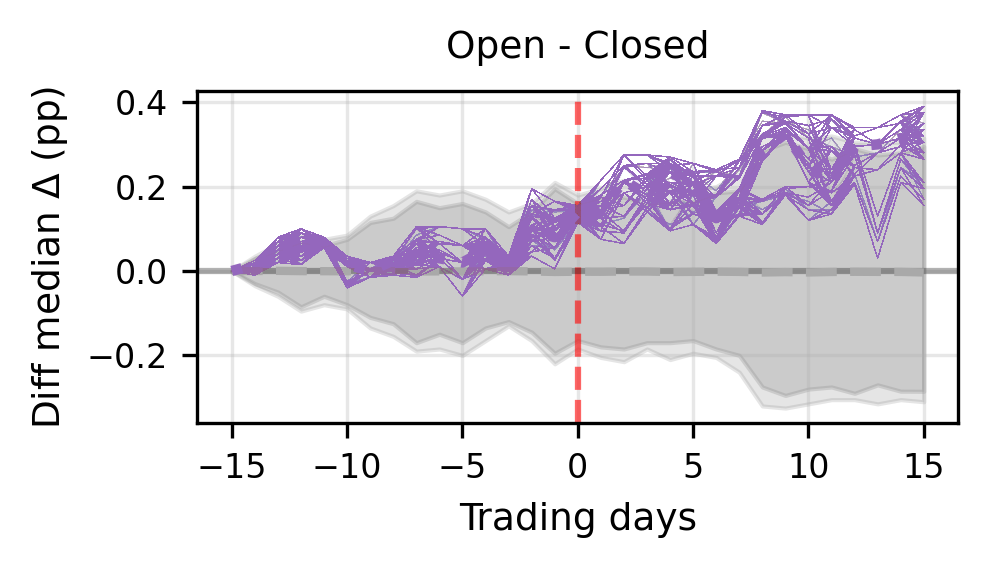}
         \caption{5 year}
     \end{subfigure}
     \\
     \begin{subfigure}[b]{\textwidth}
         \centering
         \includegraphics[width=0.4\textwidth]{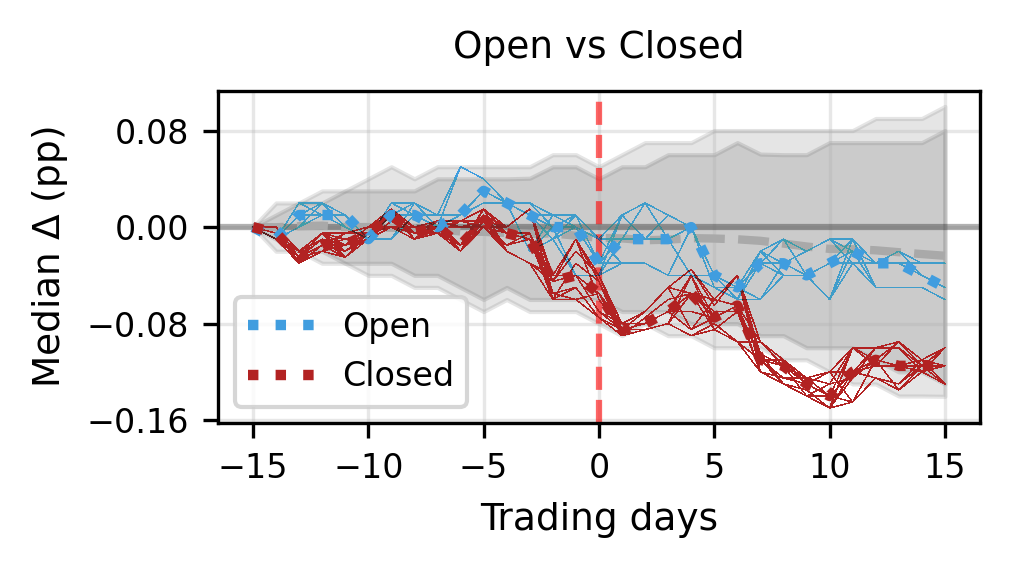}
         \includegraphics[width=0.4\textwidth]{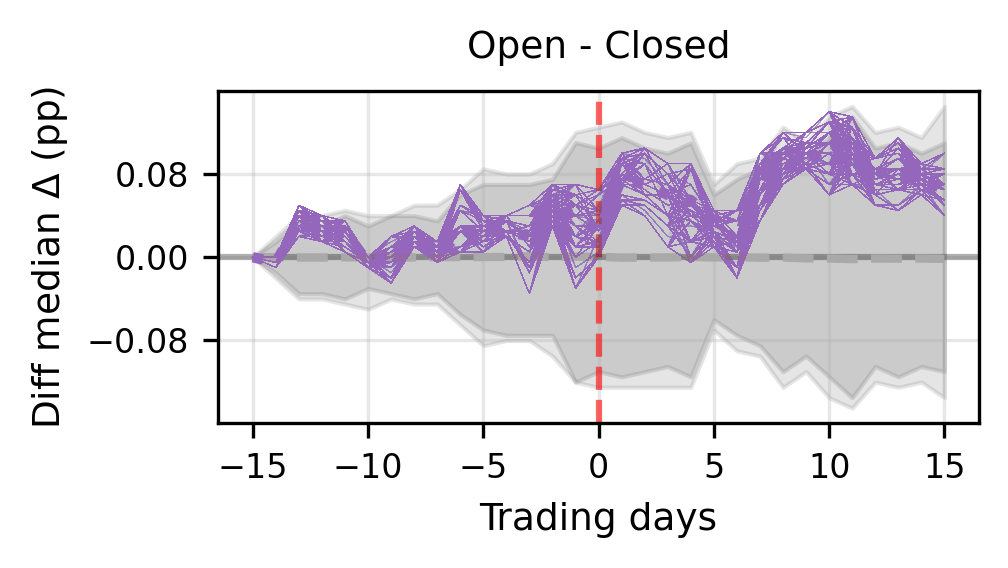}
         \caption{1 year}
     \end{subfigure}
     \\
\caption{Median yield change in US treasury bonds around major AI releases, permutation test, dropping events (2 pairs)}
\note{Dotted line shows the estimated change for the median event, relative to 15 days before the event. The shaded regions represent 90\% and 95\% of a placebo distribution, over 5,000 permutation draws. Dashed line indicates mean of the placebo distribution. Thin lines represent the estimate resulting from dropping two pairs of events (two from the open sample and two from the closed sample). For the first column, the placebo includes all dates in data. For last column, the placebo distribution includes the pooled event dates of closed and open AI model releases.}
\label{f:permutation_open_and_closed_medianchange_jackknife_drop2}\end{figure}

\begin{figure}[ht]
     \begin{subfigure}[b]{\textwidth}
         \centering
         \includegraphics[width=0.4\textwidth]{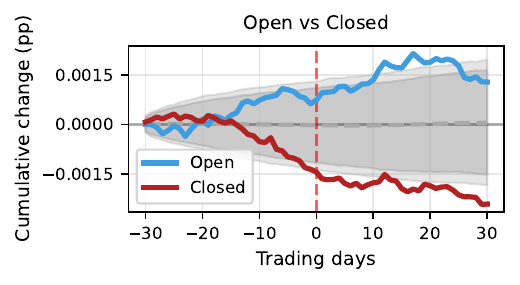}
         \includegraphics[width=0.4\textwidth]{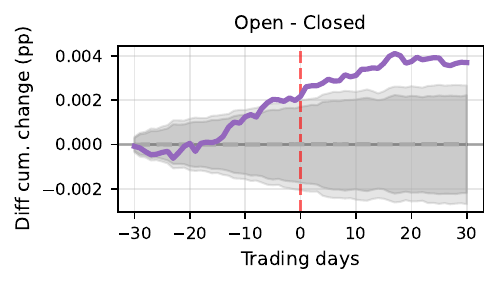}
         \caption{30 year}
     \end{subfigure}
     \\
     \begin{subfigure}[b]{\textwidth}
         \centering
         \includegraphics[width=0.4\textwidth]{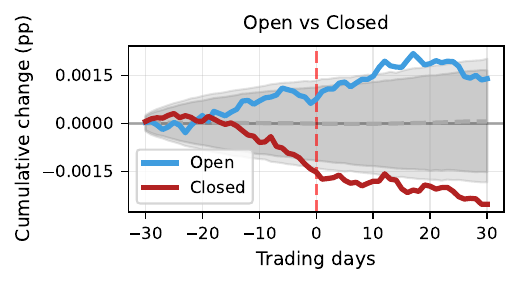}
         \includegraphics[width=0.4\textwidth]{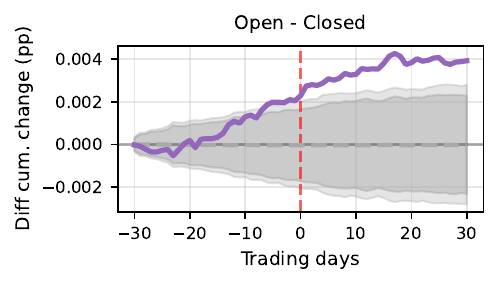}
         \caption{20 year}
     \end{subfigure}
     \\
     \begin{subfigure}[b]{\textwidth}
         \centering
         \includegraphics[width=0.4\textwidth]{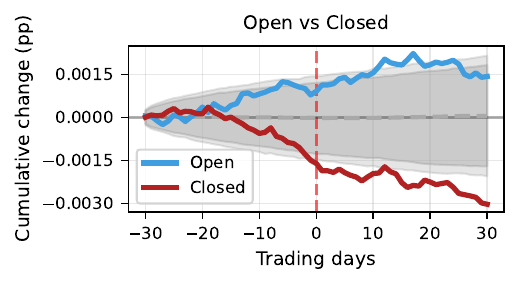}
         \includegraphics[width=0.4\textwidth]{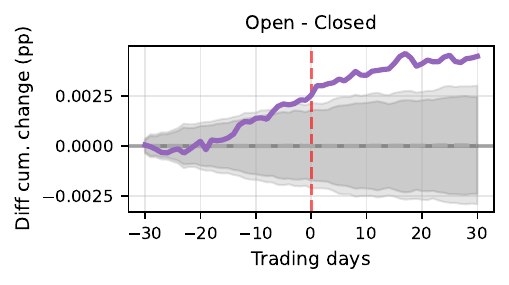}
         \caption{10 year}
     \end{subfigure}
     \\
     \begin{subfigure}[b]{\textwidth}
         \centering
         \includegraphics[width=0.4\textwidth]{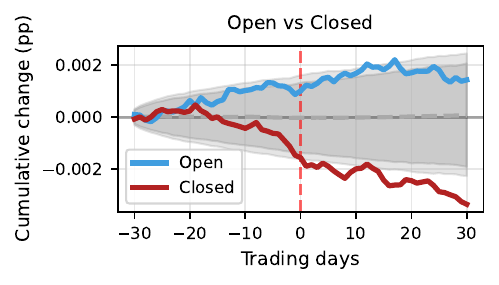}
         \includegraphics[width=0.4\textwidth]{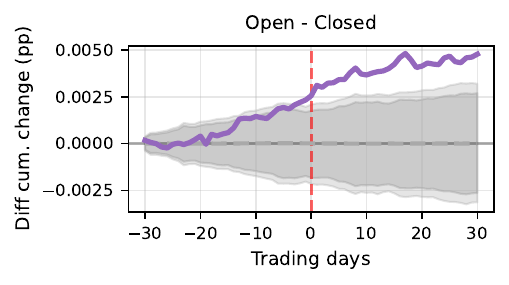}
         \caption{5 year}
     \end{subfigure}
     \\
     \begin{subfigure}[b]{\textwidth}
         \centering
         \includegraphics[width=0.4\textwidth]{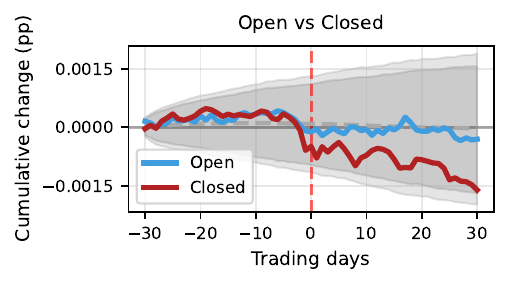}
         \includegraphics[width=0.4\textwidth]{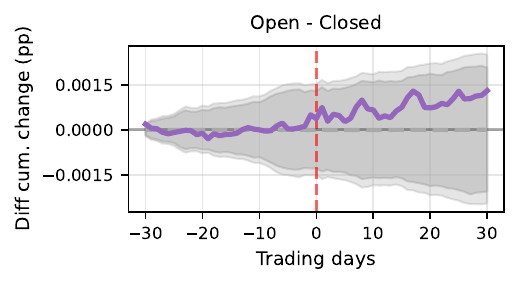}
         \caption{1 year}
     \end{subfigure}
     \\
\caption{Estimated yield change in US treasury bonds around major AI releases (OLS) and permutation test ($\pm$30 day window)}
\note{OLS coefficient estimates. The shaded regions represent 90\% and 95\% of a placebo distribution, over 5,000 permutation draws. Dashed line indicates mean of the placebo distribution. For the first column, the placebo includes all dates in data. For last column, the placebo distribution includes the pooled event dates of closed and open AI model releases.}
\label{f:permutation_open_and_closed_ols_30day}\end{figure}

\begin{figure}[ht]
     \begin{subfigure}[b]{\textwidth}
         \centering
         \includegraphics[width=0.4\textwidth]{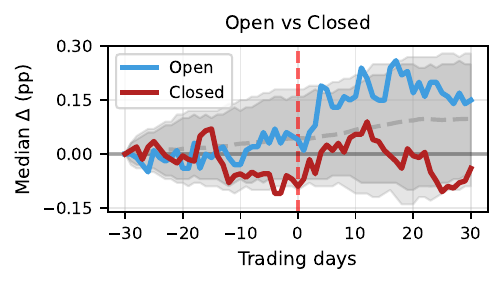}
         \includegraphics[width=0.4\textwidth]{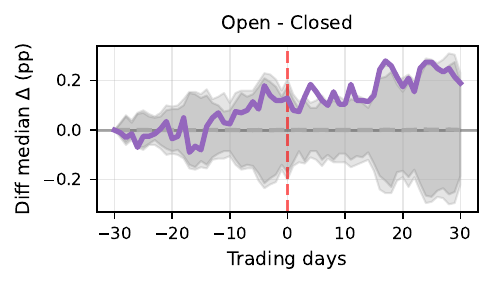}
         \caption{30 year}
     \end{subfigure}
     \\
     \begin{subfigure}[b]{\textwidth}
         \centering
         \includegraphics[width=0.4\textwidth]{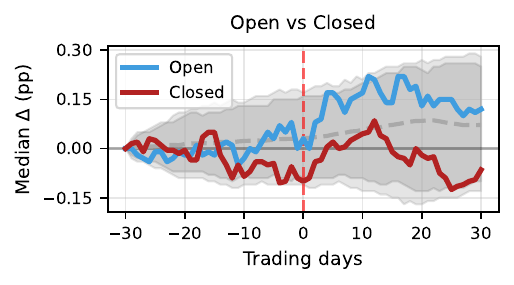}
         \includegraphics[width=0.4\textwidth]{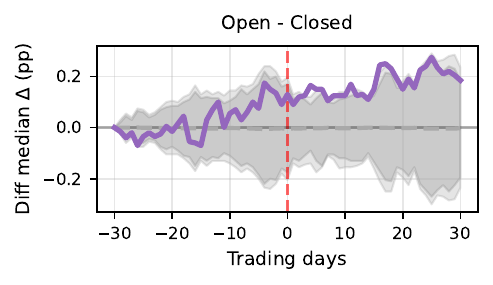}
         \caption{20 year}
     \end{subfigure}
     \\
     \begin{subfigure}[b]{\textwidth}
         \centering
         \includegraphics[width=0.4\textwidth]{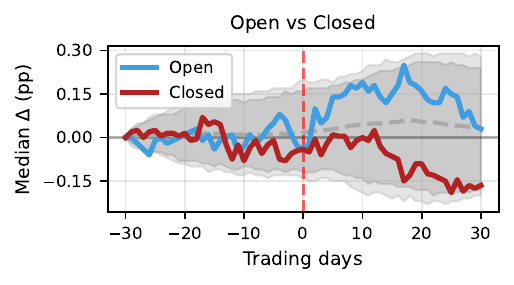}
         \includegraphics[width=0.4\textwidth]{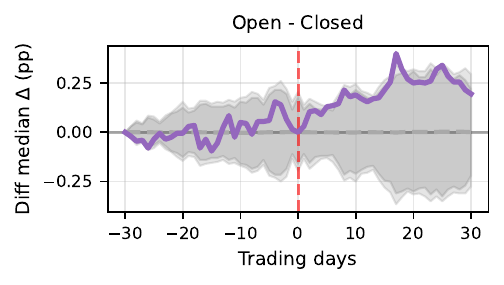}
         \caption{10 year}
     \end{subfigure}
     \\
     \begin{subfigure}[b]{\textwidth}
         \centering
         \includegraphics[width=0.4\textwidth]{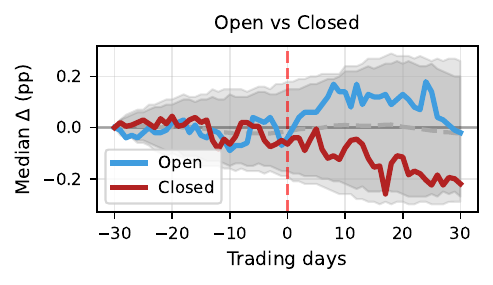}
         \includegraphics[width=0.4\textwidth]{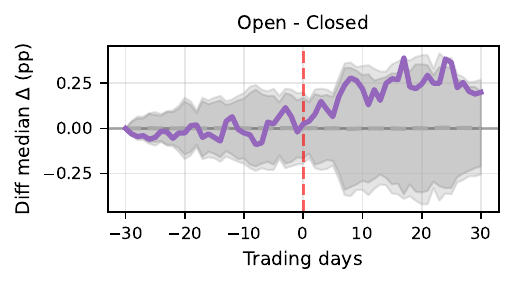}
         \caption{5 year}
     \end{subfigure}
     \\
     \begin{subfigure}[b]{\textwidth}
         \centering
         \includegraphics[width=0.4\textwidth]{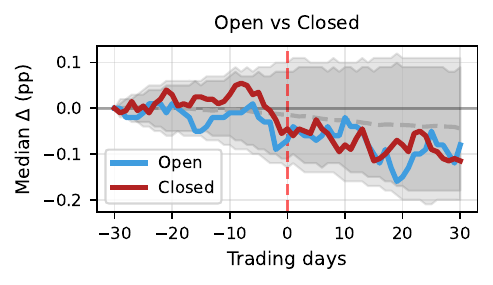}
         \includegraphics[width=0.4\textwidth]{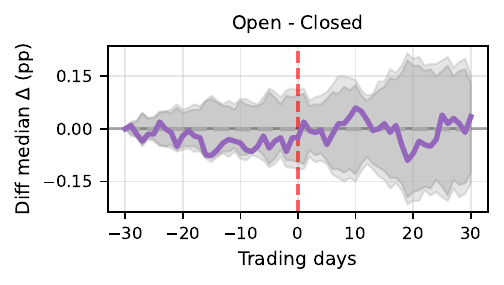}
         \caption{1 year}
     \end{subfigure}
     \\
\caption{Median yield change in US treasury bonds around major AI releases and permutation test ($\pm$30 day window)}
\note{Solid line shows the estimated change for the median event, relative to 30 days before the event. The shaded regions represent 90\% and 95\% of a placebo distribution, over 5,000 permutation draws. Dashed line indicates mean of the placebo distribution. For left column, the placebo includes all dates in data. For right column, the placebo distribution includes the pooled event dates of closed and open AI model releases.}
\label{f:permutation_open_and_closed_medianchange_30day}\end{figure}

\begin{figure}[ht]
     \begin{subfigure}[b]{\textwidth}
         \centering
         \includegraphics[width=0.4\textwidth]{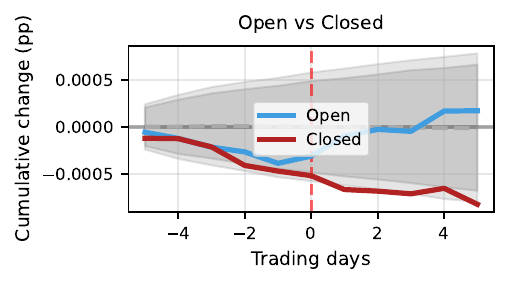}
         \includegraphics[width=0.4\textwidth]{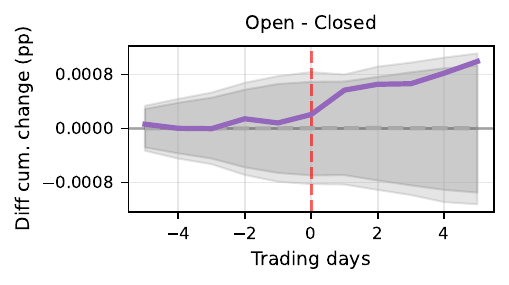}
         \caption{30 year}
     \end{subfigure}
     \\
     \begin{subfigure}[b]{\textwidth}
         \centering
         \includegraphics[width=0.4\textwidth]{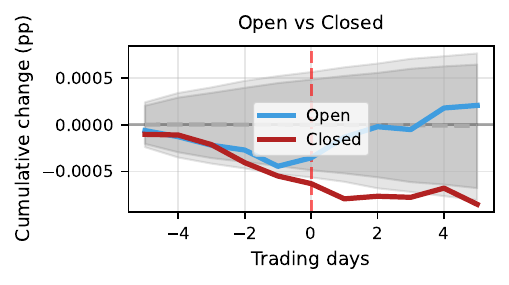}
         \includegraphics[width=0.4\textwidth]{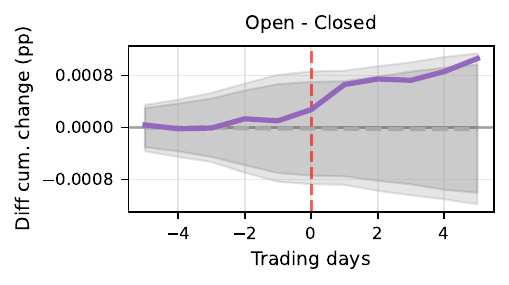}
         \caption{20 year}
     \end{subfigure}
     \\
     \begin{subfigure}[b]{\textwidth}
         \centering
         \includegraphics[width=0.4\textwidth]{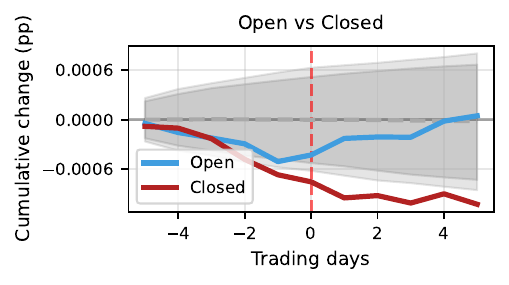}
         \includegraphics[width=0.4\textwidth]{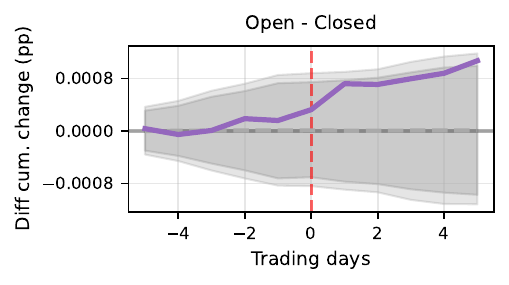}
         \caption{10 year}
     \end{subfigure}
     \\
     \begin{subfigure}[b]{\textwidth}
         \centering
         \includegraphics[width=0.4\textwidth]{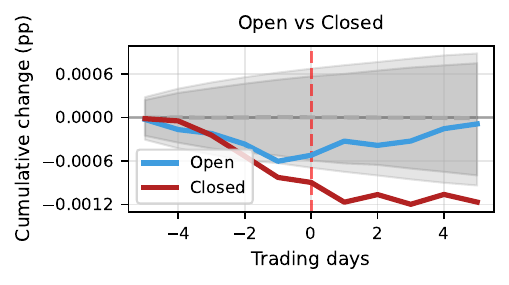}
         \includegraphics[width=0.4\textwidth]{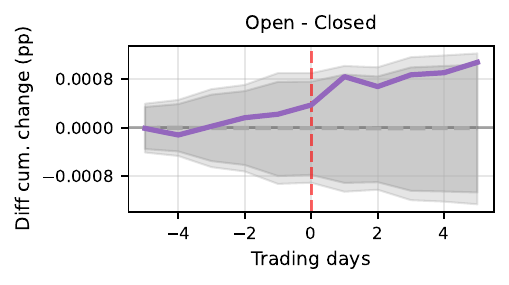}
         \caption{5 year}
     \end{subfigure}
     \\
     \begin{subfigure}[b]{\textwidth}
         \centering
         \includegraphics[width=0.4\textwidth]{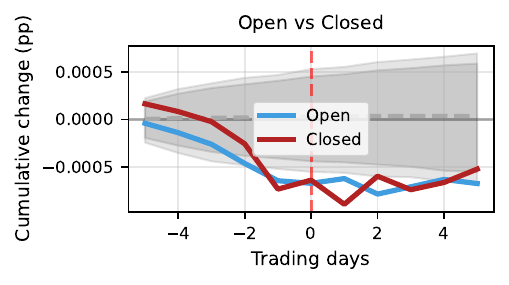}
         \includegraphics[width=0.4\textwidth]{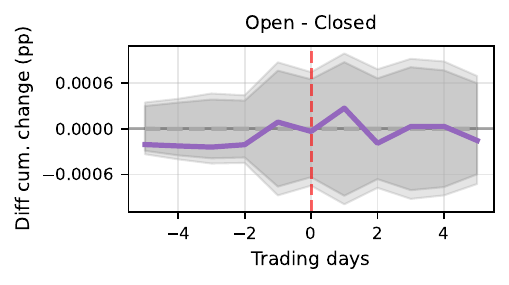}
         \caption{1 year}
     \end{subfigure}
     \\
\caption{Estimated yield change in US treasury bonds around major AI releases (OLS) and permutation test ($\pm$5 day window)}
\note{OLS coefficient estimates. The shaded regions represent 90\% and 95\% of a placebo distribution, over 5,000 permutation draws. Dashed line indicates mean of the placebo distribution. For the first column, the placebo includes all dates in data. For last column, the placebo distribution includes the pooled event dates of closed and open AI model releases.}
\label{f:permutation_open_and_closed_ols_5day}\end{figure}

\begin{figure}[ht]
     \begin{subfigure}[b]{\textwidth}
         \centering
         \includegraphics[width=0.4\textwidth]{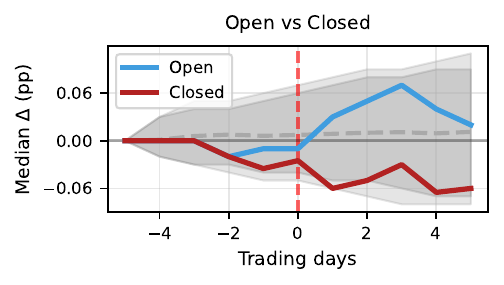}
         \includegraphics[width=0.4\textwidth]{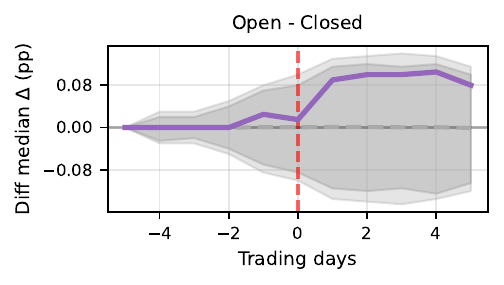}
         \caption{30 year}
     \end{subfigure}
     \\
     \begin{subfigure}[b]{\textwidth}
         \centering
         \includegraphics[width=0.4\textwidth]{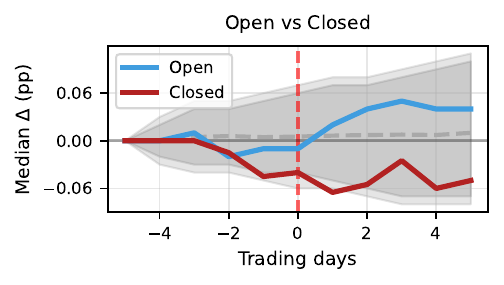}
         \includegraphics[width=0.4\textwidth]{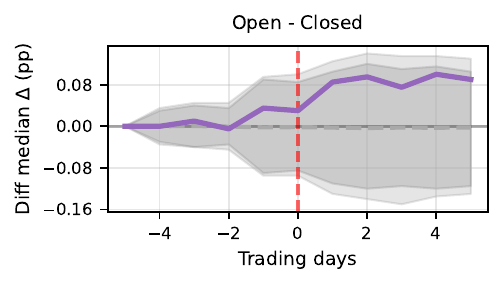}
         \caption{20 year}
     \end{subfigure}
     \\
     \begin{subfigure}[b]{\textwidth}
         \centering
         \includegraphics[width=0.4\textwidth]{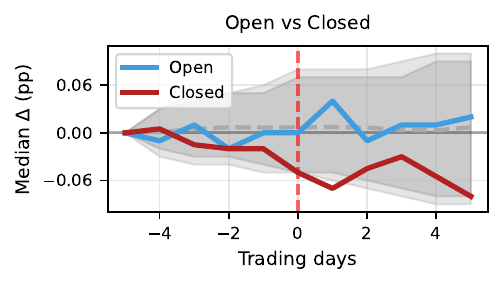}
         \includegraphics[width=0.4\textwidth]{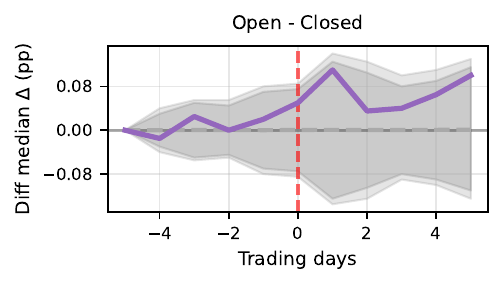}
         \caption{10 year}
     \end{subfigure}
     \\
     \begin{subfigure}[b]{\textwidth}
         \centering
         \includegraphics[width=0.4\textwidth]{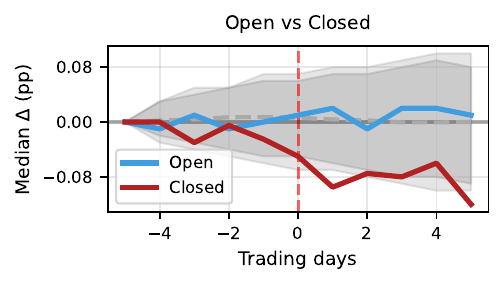}
         \includegraphics[width=0.4\textwidth]{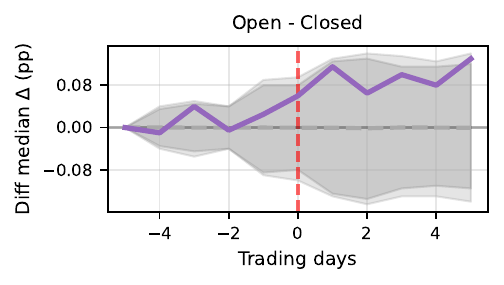}
         \caption{5 year}
     \end{subfigure}
     \\
     \begin{subfigure}[b]{\textwidth}
         \centering
         \includegraphics[width=0.4\textwidth]{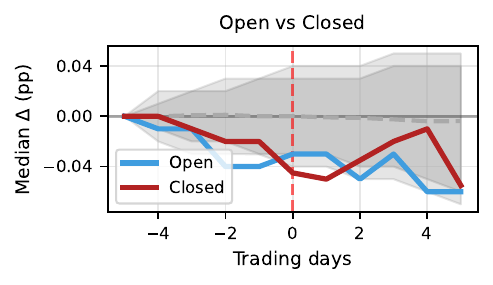}
         \includegraphics[width=0.4\textwidth]{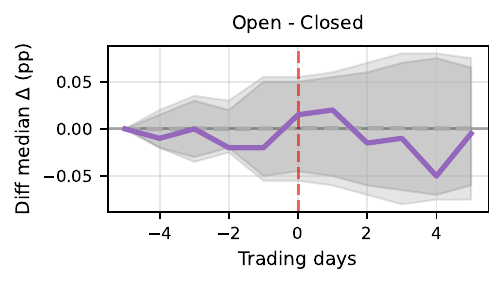}
         \caption{1 year}
     \end{subfigure}
     \\
\caption{Median yield change in US treasury bonds around major AI releases and permutation test ($\pm$5 day window)}
\note{Solid line shows the estimated change for the median event, relative to 5 days before the event. The shaded regions represent 90\% and 95\% of a placebo distribution, over 5,000 permutation draws. Dashed line indicates mean of the placebo distribution. For left column, the placebo includes all dates in data. For right column, the placebo distribution includes the pooled event dates of closed and open AI model releases.}
\label{f:permutation_open_and_closed_medianchange_5day}\end{figure}

\begin{figure}
    \centering
    \begin{subfigure}[b]{\textwidth}
         \centering
         \includegraphics[width=0.9\textwidth]{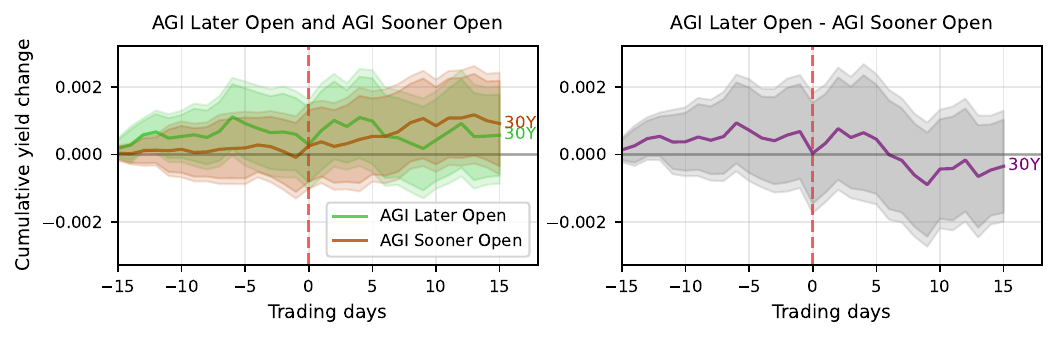}
         \caption{AGI forecast impact among open models}
     \end{subfigure}
     \\
     \begin{subfigure}[b]{\textwidth}
         \centering
         \includegraphics[width=0.9\textwidth]{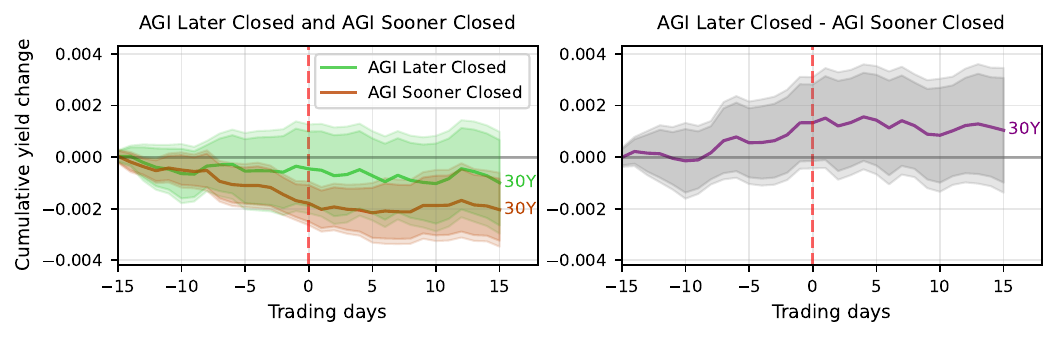}
         \caption{AGI forecast impact among closed models}
     \end{subfigure}
     \caption{US treasury bond (30Y) yield event study regression: interacted splits}
     \label{fig:regression_interacted_splits} 
    \note{Estimates of cumulative returns. The left panels show cumulative returns from 15 days prior to release, using equation~\eqref{eq:cum_return_paired} and subsetting by the designated characteristic. The right panels show the estimated difference between the designated subsets. 30 year maturity duration. 90\% and 95\% confidence intervals are shaded.}
\end{figure}

\begin{figure}
     \centering
        \begin{subfigure}[b]{\textwidth}
         \centering
         \includegraphics[width=\textwidth]{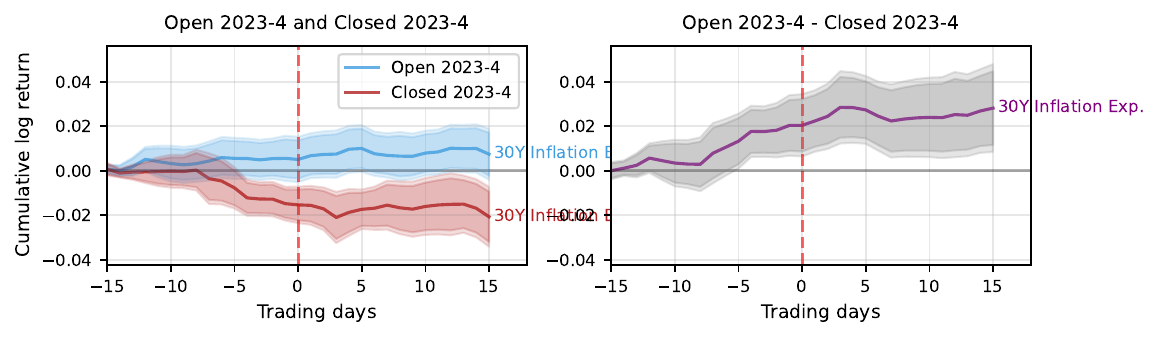}
         \caption{Inflation expectations (RINF)}
     \end{subfigure}
     \\
        \begin{subfigure}[b]{\textwidth}
         \centering
         \includegraphics[width=\textwidth]{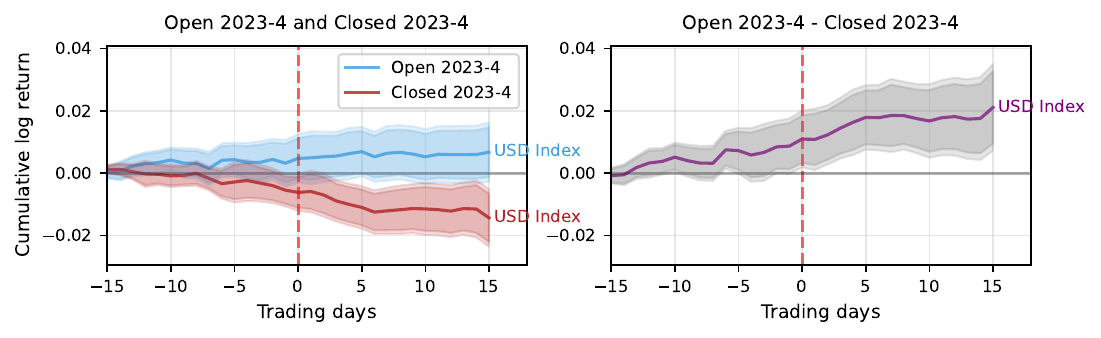}
         \caption{US dollar currency index (UUP)}
     \end{subfigure}
     \\
        \begin{subfigure}[b]{\textwidth}
         \centering
         \includegraphics[width=\textwidth]{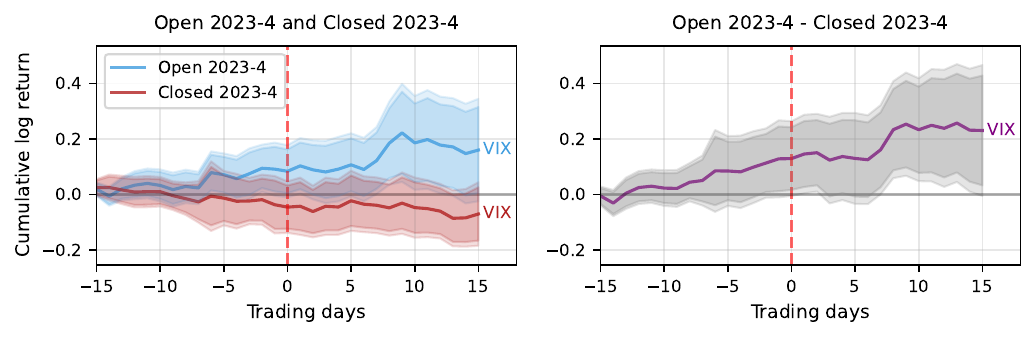}
         \caption{Uncertainty index (VIX)}
     \end{subfigure}
     \caption{Other outcomes event study regression estimates, 2023-4 releases}
     \label{fig:regression_difference_other_2023_4}
    \note{Estimates of cumulative returns. The left panels show cumulative returns from 15 days prior to release, using equation~\eqref{eq:cum_return_paired}. The right panels show the estimated difference. 90\% and 95\% confidence intervals are shaded.}
\end{figure}

\begin{figure}
     \centering
        \begin{subfigure}[b]{\textwidth}
         \centering
         \includegraphics[width=0.7\textwidth]{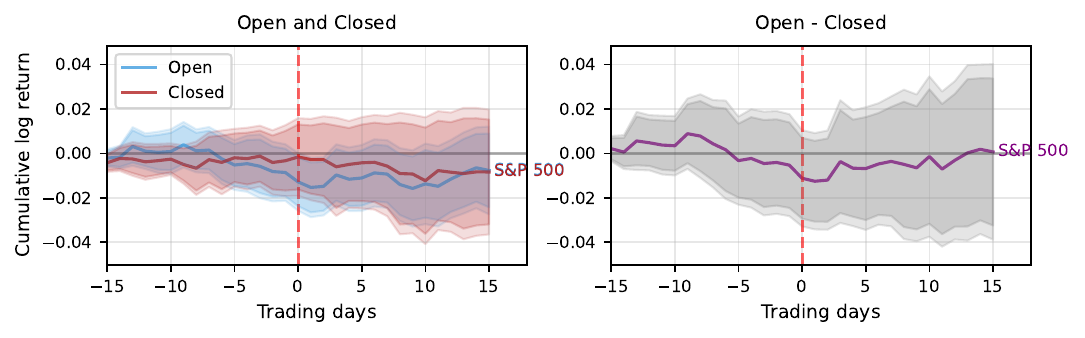}
     \end{subfigure}
     \\
        \begin{subfigure}[b]{\textwidth}
         \centering
         \includegraphics[width=0.7\textwidth]{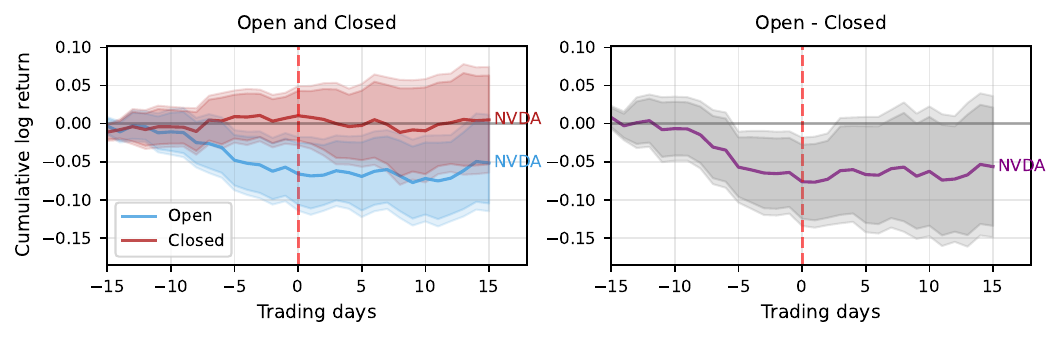}
     \end{subfigure}
     \\
        \begin{subfigure}[b]{\textwidth}
         \centering
         \includegraphics[width=0.7\textwidth]{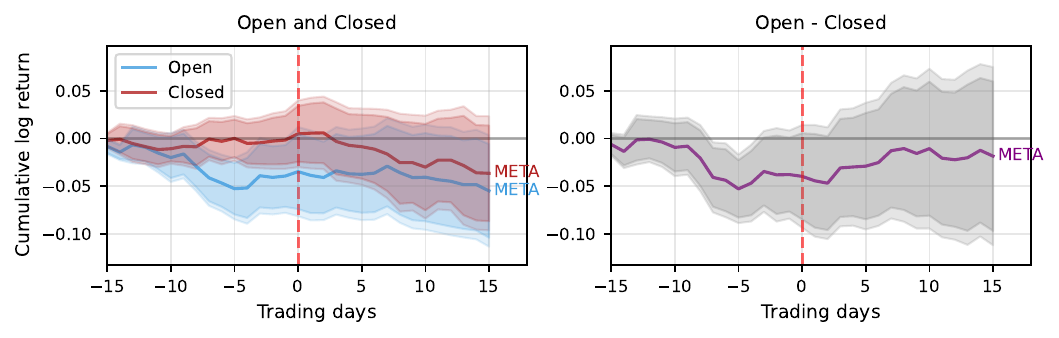}
     \end{subfigure}
     \\
        \begin{subfigure}[b]{\textwidth}
         \centering
         \includegraphics[width=0.7\textwidth]{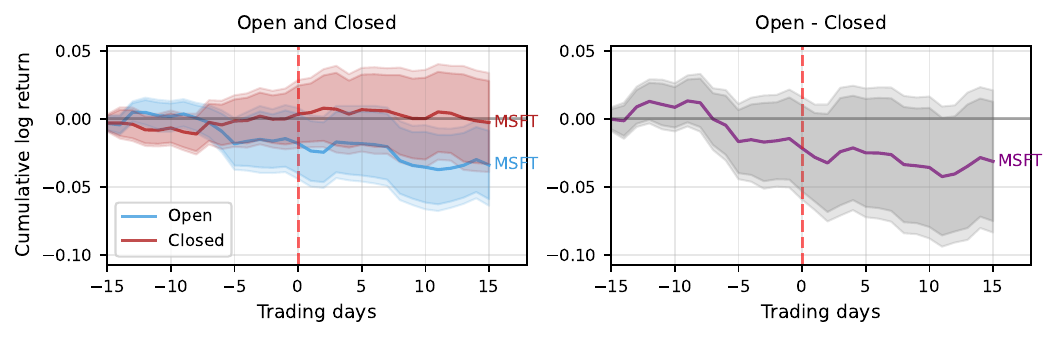}
     \end{subfigure}
     \\
        \begin{subfigure}[b]{\textwidth}
         \centering
         \includegraphics[width=0.7\textwidth]{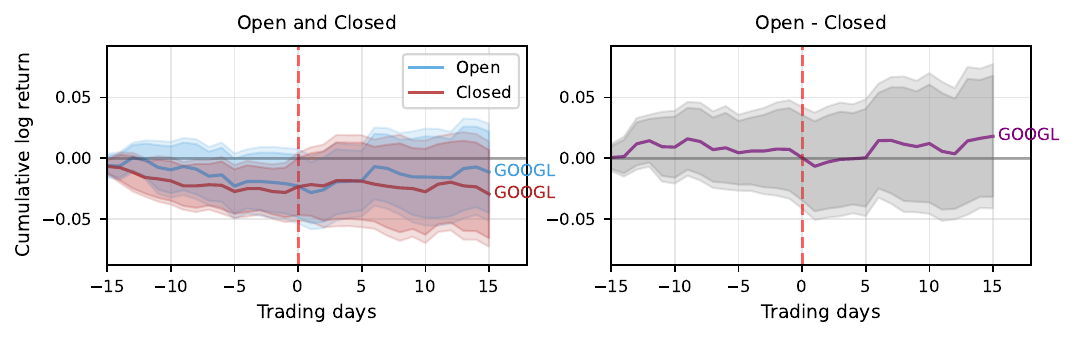}
     \end{subfigure}
     \\
        \begin{subfigure}[b]{\textwidth}
         \centering
         \includegraphics[width=0.7\textwidth]{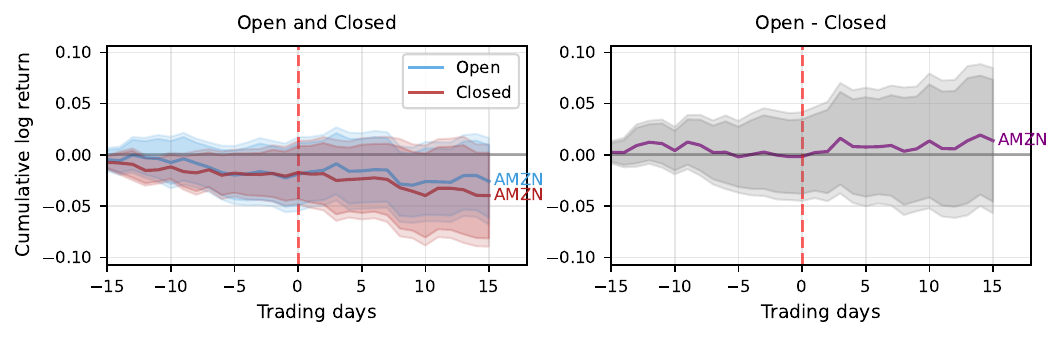}
     \end{subfigure}
     \caption{Equity return event study regression estimates}
     \label{fig:regression_equity}
    \note{Estimates of cumulative returns. The left panels show cumulative returns from 15 days prior to release, using equation~\eqref{eq:cum_return_paired}. The right panels show the estimated difference. Constant maturity duration noted in years. 90\% and 95\% confidence intervals are shaded.}
\end{figure}

\begin{figure}
     \centering
        \begin{subfigure}[b]{\textwidth}
         \centering
         \includegraphics[width=0.7\textwidth]{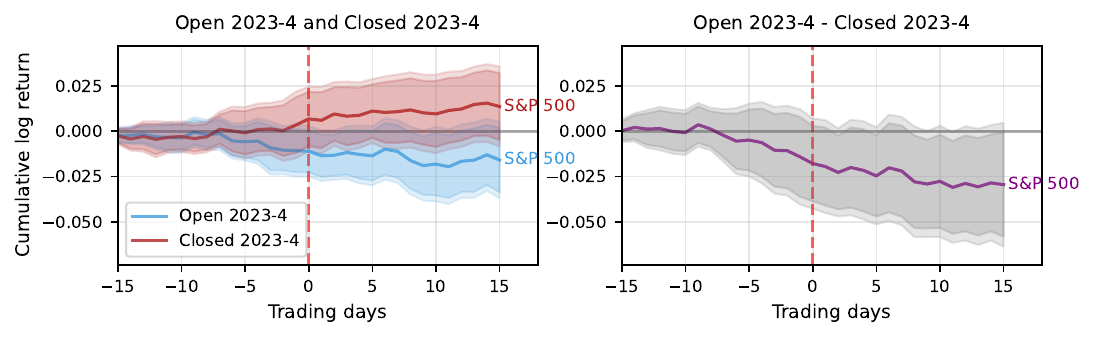}
     \end{subfigure}
     \\
        \begin{subfigure}[b]{\textwidth}
         \centering
         \includegraphics[width=0.7\textwidth]{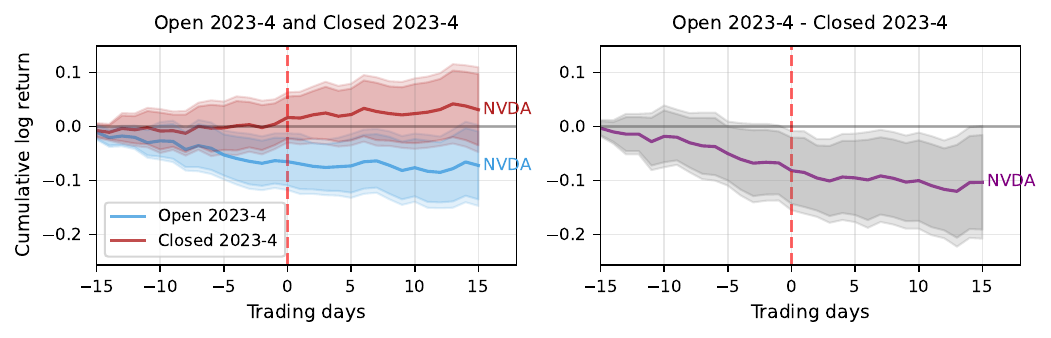}
     \end{subfigure}
     \\
        \begin{subfigure}[b]{\textwidth}
         \centering
         \includegraphics[width=0.7\textwidth]{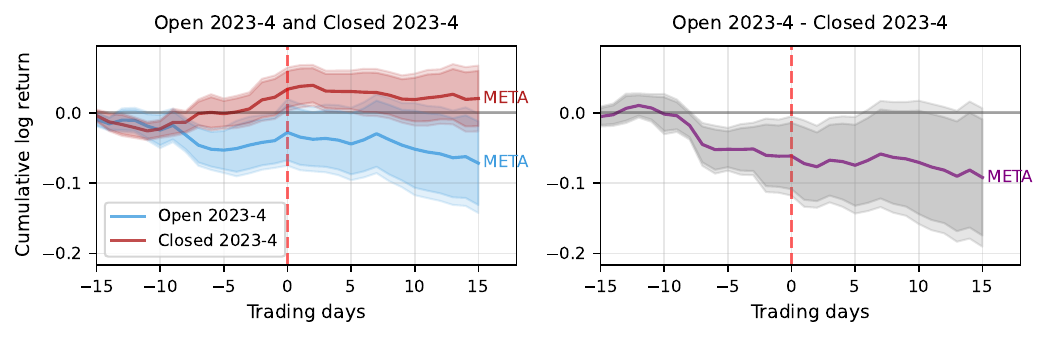}
     \end{subfigure}
     \\
        \begin{subfigure}[b]{\textwidth}
         \centering
         \includegraphics[width=0.7\textwidth]{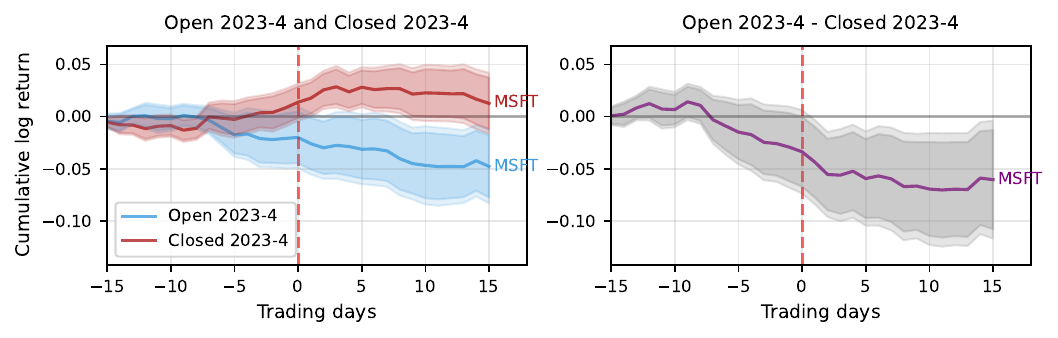}
     \end{subfigure}
     \\
        \begin{subfigure}[b]{\textwidth}
         \centering
         \includegraphics[width=0.7\textwidth]{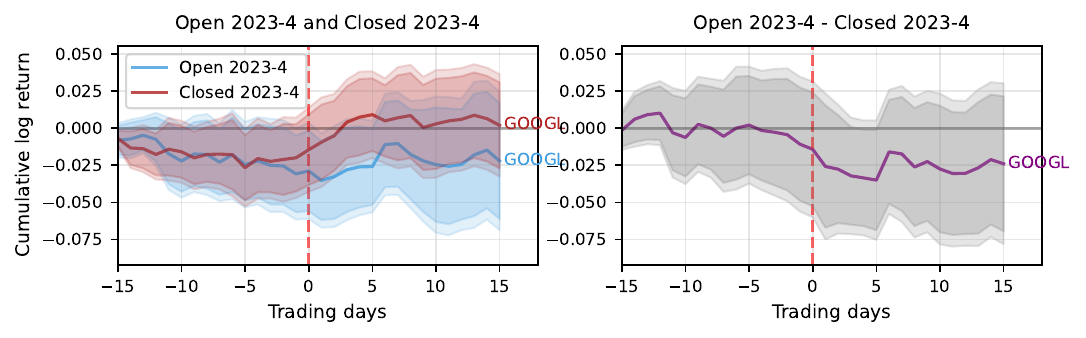}
     \end{subfigure}
     \\
        \begin{subfigure}[b]{\textwidth}
         \centering
         \includegraphics[width=0.7\textwidth]{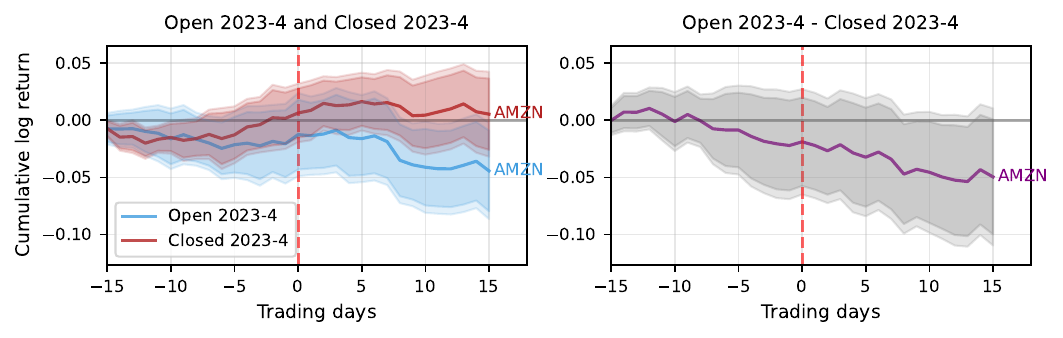}
     \end{subfigure}
     \caption{Equity return event study regression estimates, 2023-4 releases}
     \label{fig:regression_equity_2023_4}
    \note{Estimates of cumulative returns. The left panels show cumulative returns from 15 days prior to release, using equation~\eqref{eq:cum_return_paired}. The right panels show the estimated difference. Constant maturity duration noted in years. 90\% and 95\% confidence intervals are shaded.}
\end{figure}

\clearpage
%
%

\begin{table}[htbp]
\centering
\caption{Model Releases}
\label{tab:model_release_dates}
\resizebox{!}{0.5\textheight}{%
\begin{tabular}{lll}
\toprule
Date & Model & Open \\
\midrule
11/30/2022 & OpenAI GPT 3.5 &  \\
02/06/2023 & Google Bard &  \\
02/24/2023 & Meta LLaMA 1 & x \\
03/14/2023 & OpenAI GPT 4 &  \\
03/14/2023 & Anthropic Claude 1 &  \\
07/11/2023 & Anthropic Claude 2 &  \\
07/18/2023 & Meta Llama 2 & x \\
08/03/2023 & Alibaba Qwen & x \\
09/27/2023 & Mistral Mistral 7B & x \\
11/03/2023 & xAI Grok 1 &  \\
11/21/2023 & Anthropic Claude 2.1 &  \\
12/06/2023 & Google Gemini Pro 1.0 &  \\
12/11/2023 & Mistral Mixtral 8x7B & x \\
02/04/2024 & Alibaba Qwen 1.5 & x \\
02/15/2024 & Google Gemini Pro 1.5 &  \\
02/21/2024 & Google Gemma & x \\
03/04/2024 & Anthropic Claude 3 &  \\
03/28/2024 & xAI Grok 1.5 &  \\
04/09/2024 & Google Gemma 1.1 & x \\
04/18/2024 & Meta Llama 3 & x \\
05/06/2024 & DeepSeek V2 & x \\
05/13/2024 & OpenAI GPT 4-o &  \\
06/06/2024 & Alibaba Qwen 2 & x \\
06/20/2024 & Anthropic Claude 3.5 Sonnet &  \\
06/27/2024 & Google Gemma 2 & x \\
07/23/2024 & Meta Llama 3.1 & x \\
08/13/2024 & xAI Grok 2 &  \\
09/05/2024 & DeepSeek 2.5 & x \\
09/18/2024 & Alibaba Qwen 2.5 & x \\
09/25/2024 & Meta Llama 3.2 & x \\
12/05/2024 & OpenAI o1 &  \\
12/06/2024 & Meta Llama 3.3 & x \\
12/11/2024 & Google Gemini 2.0 &  \\
12/26/2024 & DeepSeek V3 & x \\
01/20/2025 & DeepSeek R1 & x \\
02/19/2025 & xAI Grok 3 &  \\
02/24/2025 & Anthropic Claude 3.7 &  \\
02/27/2025 & OpenAI GPT 4.5 &  \\
03/10/2025 & Google Gemma 3 & x \\
03/25/2025 & Google Gemini 2.5 &  \\
04/05/2025 & Meta Llama 4 & x \\
04/16/2025 & OpenAI o3 &  \\
04/28/2025 & Alibaba Qwen 3 & x \\
05/22/2025 & Anthropic Claude 4 &  \\
07/09/2025 & xAI Grok 4 &  \\
08/05/2025 & OpenAI GPT-OSS & x \\
08/07/2025 & OpenAI GPT 5 &  \\
\bottomrule
\end{tabular}
}
\end{table}

\begin{landscape}
    \begin{table}
        \resizebox{9in}{!}{            \begin{tabular}{lccccccccccccccc}
\hline\hline
Day & \multicolumn{3}{c}{1Y} & \multicolumn{3}{c}{5Y} & \multicolumn{3}{c}{10Y} & \multicolumn{3}{c}{20Y} & \multicolumn{3}{c}{30Y} \\
     & Open & Closed & Diff. & Open & Closed & Diff. & Open & Closed & Diff. & Open & Closed & Diff. & Open & Closed & Diff. \\
\hline
-15 & 0.6 (0.65) & 1.0 (0.18) & -0.4 (0.73) & 1.3 (0.39) & 1.2 (0.29) & 0.1 (0.97) & 1.4 (0.25) & 0.9 (0.46) & 0.5 (0.72) & 0.9 (0.42) & 0.6 (0.60) & 0.3 (0.81) & 1.0 (0.35) & 0.7 (0.48) & 0.3 (0.82) \\
-14 & 0.7 (0.70) & 0.5 (0.70) & 0.2 (0.94) & 2.3 (0.35) & -0.8 (0.71) & 3.1 (0.31) & 2.3 (0.29) & -0.5 (0.80) & 2.8 (0.29) & 1.8 (0.37) & -0.4 (0.77) & 2.2 (0.32) & 1.8 (0.34) & -0.6 (0.70) & 2.4 (0.29) \\
-13 & 2.5 (0.25) & 0.6 (0.70) & 1.9 (0.47) & 6.2 (0.03)** & -1.7 (0.47) & 7.9 (0.03)** & 5.5 (0.03)** & -1.8 (0.37) & 7.3 (0.02)** & 4.1 (0.08)* & -2.1 (0.21) & 6.2 (0.02)** & 4.4 (0.05)** & -2.1 (0.20) & 6.5 (0.01)** \\
-12 & 2.5 (0.24) & 0.2 (0.93) & 2.3 (0.40) & 5.9 (0.06)* & -2.2 (0.44) & 8.2 (0.04)** & 5.2 (0.06)* & -3.2 (0.21) & 8.4 (0.01)** & 3.7 (0.18) & -3.3 (0.14) & 7.0 (0.02)** & 4.5 (0.11) & -3.5 (0.08)* & 8.0 (0.01)*** \\
-11 & 2.3 (0.38) & 0.5 (0.78) & 1.8 (0.58) & 6.2 (0.12) & -1.9 (0.53) & 8.1 (0.08)* & 5.0 (0.18) & -3.2 (0.23) & 8.1 (0.05)** & 2.9 (0.41) & -3.0 (0.21) & 5.9 (0.12) & 3.8 (0.26) & -3.3 (0.14) & 7.2 (0.06)* \\
-10 & 2.5 (0.37) & 1.1 (0.53) & 1.5 (0.67) & 6.5 (0.13) & -2.6 (0.46) & 9.1 (0.07)* & 5.4 (0.17) & -4.3 (0.17) & 9.7 (0.03)** & 3.5 (0.35) & -4.9 (0.09)* & 8.4 (0.05)* & 4.6 (0.23) & -4.9 (0.07)* & 9.5 (0.03)** \\
 -9 & 2.8 (0.39) & 1.9 (0.29) & 0.9 (0.82) & 6.8 (0.15) & -1.6 (0.65) & 8.4 (0.13) & 5.7 (0.20) & -4.0 (0.21) & 9.7 (0.06)* & 4.0 (0.34) & -4.9 (0.10)* & 8.9 (0.07)* & 4.8 (0.25) & -5.4 (0.05)** & 10.2 (0.03)** \\
 -8 & 2.4 (0.56) & 1.7 (0.44) & 0.7 (0.88) & 7.6 (0.15) & -0.0 (0.99) & 7.7 (0.23) & 6.5 (0.16) & -2.4 (0.54) & 8.9 (0.12) & 4.2 (0.33) & -3.4 (0.35) & 7.6 (0.15) & 5.0 (0.23) & -4.1 (0.22) & 9.1 (0.08)* \\
 -7 & 2.5 (0.58) & 0.0 (0.99) & 2.4 (0.65) & 8.0 (0.18) & -2.5 (0.60) & 10.5 (0.12) & 7.6 (0.16) & -4.8 (0.23) & 12.4 (0.04)** & 5.3 (0.29) & -5.7 (0.11) & 11.1 (0.05)** & 5.9 (0.23) & -6.8 (0.05)* & 12.7 (0.02)** \\
 -6 & 3.3 (0.49) & -0.3 (0.93) & 3.5 (0.53) & 10.1 (0.12) & -3.1 (0.52) & 13.2 (0.07)* & 9.9 (0.09)* & -5.6 (0.18) & 15.5 (0.02)** & 7.6 (0.16) & -6.3 (0.10) & 13.9 (0.02)** & 8.0 (0.13) & -7.4 (0.05)** & 15.4 (0.01)** \\
 -5 & 2.7 (0.57) & 1.1 (0.74) & 1.6 (0.78) & 9.8 (0.13) & -3.7 (0.48) & 13.6 (0.07)* & 9.4 (0.11) & -6.9 (0.13) & 16.3 (0.02)** & 7.0 (0.21) & -7.8 (0.07)* & 14.7 (0.03)** & 7.5 (0.16) & -9.0 (0.03)** & 16.6 (0.01)** \\
 -4 & 1.8 (0.71) & 0.6 (0.87) & 1.2 (0.84) & 8.8 (0.16) & -3.8 (0.48) & 12.6 (0.09)* & 8.7 (0.12) & -6.8 (0.14) & 15.5 (0.02)** & 6.7 (0.20) & -7.7 (0.08)* & 14.4 (0.03)** & 7.2 (0.15) & -8.9 (0.04)** & 16.1 (0.01)** \\
 -3 & 0.8 (0.87) & -0.6 (0.86) & 1.4 (0.82) & 8.6 (0.19) & -5.7 (0.24) & 14.2 (0.07)* & 8.1 (0.16) & -7.9 (0.07)* & 16.1 (0.02)** & 5.7 (0.29) & -8.4 (0.04)** & 14.2 (0.03)** & 6.1 (0.22) & -9.5 (0.02)** & 15.7 (0.02)** \\
 -2 & -1.1 (0.85) & -3.0 (0.32) & 1.9 (0.78) & 7.6 (0.27) & -8.4 (0.07)* & 16.0 (0.05)* & 7.8 (0.21) & -10.4 (0.01)** & 18.2 (0.02)** & 5.6 (0.33) & -10.4 (0.01)** & 16.0 (0.03)** & 6.1 (0.27) & -11.5 (0.01)*** & 17.5 (0.01)** \\
 -1 & -2.8 (0.64) & -7.9 (0.14) & 5.0 (0.58) & 5.1 (0.47) & -11.5 (0.02)** & 16.7 (0.07)* & 5.5 (0.38) & -12.5 (0.01)*** & 18.0 (0.03)** & 3.8 (0.51) & -12.0 (0.01)*** & 15.8 (0.04)** & 4.7 (0.40) & -12.2 (0.01)*** & 16.9 (0.02)** \\
  0 & -3.1 (0.63) & -7.1 (0.13) & 4.0 (0.65) & 6.1 (0.40) & -12.9 (0.01)** & 19.0 (0.04)** & 6.6 (0.30) & -14.0 (<0.01)*** & 20.6 (0.01)** & 4.9 (0.40) & -13.4 (<0.01)*** & 18.3 (0.02)** & 5.7 (0.31) & -13.4 (<0.01)*** & 19.0 (0.01)** \\
  1 & -2.6 (0.71) & -9.7 (0.10)* & 7.1 (0.47) & 8.1 (0.28) & -15.8 (<0.01)*** & 24.0 (0.01)** & 8.7 (0.18) & -16.3 (<0.01)*** & 25.1 (<0.01)*** & 7.5 (0.22) & -15.4 (<0.01)*** & 22.9 (<0.01)*** & 8.1 (0.15) & -15.3 (<0.01)*** & 23.4 (<0.01)*** \\
  2 & -4.3 (0.52) & -7.0 (0.12) & 2.8 (0.76) & 7.6 (0.32) & -15.2 (<0.01)*** & 22.8 (0.02)** & 9.0 (0.18) & -16.4 (<0.01)*** & 25.4 (<0.01)*** & 8.7 (0.16) & -15.3 (<0.01)*** & 24.1 (<0.01)*** & 8.8 (0.13) & -15.7 (<0.01)*** & 24.5 (<0.01)*** \\
  3 & -3.5 (0.59) & -8.2 (0.13) & 4.7 (0.62) & 8.2 (0.29) & -16.3 (<0.01)*** & 24.5 (0.02)** & 9.0 (0.18) & -17.2 (<0.01)*** & 26.2 (<0.01)*** & 8.5 (0.18) & -15.4 (<0.01)*** & 23.9 (0.01)*** & 8.7 (0.13) & -16.0 (<0.01)*** & 24.7 (<0.01)*** \\
  4 & -2.6 (0.68) & -7.4 (0.14) & 4.8 (0.61) & 9.7 (0.23) & -15.0 (0.01)*** & 24.7 (0.02)** & 10.6 (0.14) & -16.2 (<0.01)*** & 26.7 (<0.01)*** & 10.5 (0.13) & -14.5 (<0.01)*** & 24.9 (0.01)*** & 10.5 (0.10) & -15.4 (<0.01)*** & 25.9 (<0.01)*** \\
  5 & -3.1 (0.63) & -6.1 (0.14) & 3.0 (0.74) & 10.5 (0.18) & -16.0 (<0.01)*** & 26.6 (0.01)*** & 11.5 (0.10) & -17.2 (<0.01)*** & 28.7 (<0.01)*** & 11.1 (0.09)* & -16.0 (<0.01)*** & 27.1 (<0.01)*** & 10.9 (0.07)* & -16.9 (<0.01)*** & 27.8 (<0.01)*** \\
  6 & -3.9 (0.55) & -7.2 (0.11) & 3.4 (0.71) & 8.2 (0.29) & -17.2 (<0.01)*** & 25.4 (0.02)** & 9.1 (0.18) & -17.6 (<0.01)*** & 26.7 (<0.01)*** & 9.0 (0.16) & -16.4 (<0.01)*** & 25.4 (<0.01)*** & 8.9 (0.15) & -17.2 (<0.01)*** & 26.1 (<0.01)*** \\
  7 & -2.5 (0.71) & -9.1 (0.09)* & 6.6 (0.50) & 9.8 (0.22) & -18.3 (0.01)*** & 28.1 (0.01)** & 9.9 (0.16) & -17.9 (<0.01)*** & 27.9 (0.01)*** & 9.6 (0.15) & -16.0 (<0.01)*** & 25.6 (0.01)*** & 9.2 (0.15) & -16.2 (<0.01)*** & 25.4 (0.01)*** \\
  8 & -2.6 (0.70) & -11.1 (0.06)* & 8.5 (0.40) & 11.0 (0.18) & -19.2 (0.01)*** & 30.2 (0.01)*** & 11.0 (0.12) & -19.0 (<0.01)*** & 30.0 (<0.01)*** & 10.4 (0.12) & -16.8 (<0.01)*** & 27.2 (<0.01)*** & 10.1 (0.11) & -17.4 (<0.01)*** & 27.5 (<0.01)*** \\
  9 & -3.5 (0.60) & -9.4 (0.07)* & 5.9 (0.55) & 10.2 (0.20) & -17.2 (0.01)*** & 27.3 (0.01)*** & 10.8 (0.13) & -17.8 (<0.01)*** & 28.6 (<0.01)*** & 10.6 (0.12) & -16.2 (<0.01)*** & 26.8 (<0.01)*** & 10.4 (0.11) & -16.5 (<0.01)*** & 27.0 (<0.01)*** \\
 10 & -3.7 (0.56) & -9.2 (0.06)* & 5.5 (0.54) & 10.6 (0.16) & -16.6 (0.01)** & 27.3 (0.01)*** & 11.4 (0.10) & -17.3 (<0.01)*** & 28.7 (<0.01)*** & 11.0 (0.09)* & -16.0 (0.01)*** & 27.0 (<0.01)*** & 11.0 (0.08)* & -16.4 (0.01)*** & 27.4 (<0.01)*** \\
 11 & -5.8 (0.36) & -7.9 (0.10)* & 2.2 (0.79) & 11.1 (0.13) & -15.9 (0.01)** & 27.0 (<0.01)*** & 12.9 (0.06)* & -16.6 (0.01)*** & 29.5 (<0.01)*** & 13.2 (0.05)* & -15.5 (0.01)** & 28.7 (<0.01)*** & 13.6 (0.04)** & -15.6 (0.01)** & 29.2 (<0.01)*** \\
 12 & -4.3 (0.49) & -7.3 (0.14) & 3.0 (0.72) & 13.4 (0.07)* & -14.0 (0.03)** & 27.3 (0.01)*** & 14.9 (0.03)** & -14.1 (0.02)** & 29.0 (<0.01)*** & 14.5 (0.04)** & -13.1 (0.04)** & 27.5 (<0.01)*** & 15.0 (0.03)** & -13.2 (0.04)** & 28.2 (<0.01)*** \\
 13 & -5.2 (0.42) & -7.6 (0.13) & 2.4 (0.77) & 12.2 (0.10)* & -15.4 (0.02)** & 27.6 (<0.01)*** & 13.7 (0.05)* & -15.7 (0.01)** & 29.4 (<0.01)*** & 13.1 (0.06)* & -14.6 (0.02)** & 27.7 (<0.01)*** & 13.9 (0.04)** & -14.8 (0.01)** & 28.7 (<0.01)*** \\
 14 & -3.7 (0.55) & -8.2 (0.11) & 4.5 (0.59) & 12.6 (0.09)* & -16.7 (0.01)** & 29.3 (<0.01)*** & 13.4 (0.05)* & -16.7 (0.01)*** & 30.1 (<0.01)*** & 12.8 (0.06)* & -15.1 (0.02)** & 27.9 (<0.01)*** & 13.4 (0.04)** & -15.4 (0.01)** & 28.8 (<0.01)*** \\
 15 & -4.5 (0.50) & -9.8 (0.07)* & 5.3 (0.53) & 11.6 (0.13) & -19.8 (<0.01)*** & 31.4 (<0.01)*** & 12.8 (0.07)* & -19.5 (<0.01)*** & 32.3 (<0.01)*** & 12.3 (0.07)* & -17.7 (0.01)*** & 30.0 (<0.01)*** & 12.8 (0.06)* & -17.6 (0.01)*** & 30.4 (<0.01)*** \\
\hline
Constant & \multicolumn{3}{c}{0.3 (0.23)} & \multicolumn{3}{c}{0.2 (0.58)} & \multicolumn{3}{c}{0.2 (0.44)} & \multicolumn{3}{c}{0.2 (0.44)} & \multicolumn{3}{c}{0.3 (0.41)} \\
\hline\hline
\end{tabular}

        }
        \caption{US treasury bond yield event study regression estimates (basis points)}
        \begin{tablenotes}
        \small
        \item Notes: * p$<$0.10, ** p$<$0.05, *** p$<$0.01. Estimates of cumulative returns (in basis points) at each day relative to 15 days before event. Constant captures baseline daily return. OLS regression using equation~\eqref{eq:cum_return_paired}. P-values in parentheses, computed using HAC standard errors (maxlags=30). Same estimates are plotted in Figure~\ref{fig:regression_difference}.
        \end{tablenotes}
        \label{tab:regression_estimates_bonds}
    \end{table}

    \begin{table}
        \resizebox{9in}{!}{            \begin{tabular}{lccccccccccccccc}
\hline\hline
Day & \multicolumn{3}{c}{3-5Y Corp} & \multicolumn{3}{c}{5-7Y Corp} & \multicolumn{3}{c}{7-10Y Corp} & \multicolumn{3}{c}{10-15Y Corp} & \multicolumn{3}{c}{15+ Y Corp} \\
     & Open & Closed & Diff. & Open & Closed & Diff. & Open & Closed & Diff. & Open & Closed & Diff. & Open & Closed & Diff. \\
\hline
-15 & 1.1 (0.46) & 1.5 (0.24) & -0.4 (0.82) & 1.2 (0.40) & 1.8 (0.19) & -0.5 (0.76) & 1.2 (0.34) & 1.5 (0.27) & -0.3 (0.86) & 1.1 (0.39) & 1.4 (0.26) & -0.4 (0.80) & 1.0 (0.37) & 1.1 (0.36) & -0.1 (0.97) \\
-14 & 1.9 (0.42) & -0.4 (0.86) & 2.2 (0.46) & 2.0 (0.37) & -0.1 (0.95) & 2.2 (0.46) & 1.9 (0.35) & -0.5 (0.81) & 2.4 (0.36) & 1.8 (0.38) & -0.3 (0.87) & 2.1 (0.42) & 2.0 (0.31) & -0.3 (0.84) & 2.3 (0.33) \\
-13 & 4.8 (0.07)* & -1.0 (0.66) & 5.8 (0.11) & 5.0 (0.06)* & -1.3 (0.58) & 6.3 (0.07)* & 4.7 (0.05)** & -2.1 (0.33) & 6.8 (0.03)** & 4.5 (0.06)* & -1.7 (0.38) & 6.3 (0.04)** & 4.2 (0.07)* & -1.7 (0.33) & 6.0 (0.03)** \\
-12 & 4.7 (0.09)* & -1.3 (0.67) & 6.0 (0.13) & 5.0 (0.08)* & -1.4 (0.63) & 6.5 (0.10)* & 4.8 (0.08)* & -2.8 (0.31) & 7.6 (0.04)** & 4.6 (0.08)* & -2.5 (0.36) & 7.1 (0.04)** & 4.4 (0.10) & -2.7 (0.25) & 7.1 (0.03)** \\
-11 & 5.1 (0.15) & -1.7 (0.59) & 6.8 (0.12) & 5.2 (0.15) & -1.7 (0.59) & 6.9 (0.12) & 4.6 (0.19) & -2.8 (0.33) & 7.4 (0.08)* & 4.4 (0.20) & -2.5 (0.37) & 6.9 (0.09)* & 3.8 (0.25) & -2.2 (0.39) & 6.1 (0.12) \\
-10 & 5.5 (0.15) & -1.8 (0.61) & 7.3 (0.13) & 5.8 (0.14) & -2.3 (0.51) & 8.1 (0.09)* & 5.5 (0.15) & -3.7 (0.27) & 9.2 (0.05)** & 5.3 (0.15) & -3.4 (0.29) & 8.7 (0.05)* & 4.8 (0.18) & -3.6 (0.23) & 8.4 (0.06)* \\
 -9 & 5.9 (0.17) & -1.1 (0.76) & 7.0 (0.18) & 6.0 (0.17) & -1.6 (0.65) & 7.6 (0.14) & 5.4 (0.20) & -3.2 (0.35) & 8.5 (0.09)* & 5.2 (0.21) & -2.6 (0.42) & 7.8 (0.11) & 4.6 (0.26) & -3.4 (0.26) & 8.0 (0.10)* \\
 -8 & 6.6 (0.17) & 0.9 (0.82) & 5.7 (0.31) & 6.9 (0.15) & 0.1 (0.98) & 6.8 (0.22) & 6.1 (0.18) & -1.7 (0.66) & 7.8 (0.15) & 5.9 (0.18) & -1.2 (0.73) & 7.1 (0.17) & 5.0 (0.23) & -2.2 (0.52) & 7.2 (0.16) \\
 -7 & 6.6 (0.22) & -1.1 (0.79) & 7.7 (0.20) & 7.2 (0.19) & -2.1 (0.61) & 9.3 (0.12) & 6.5 (0.22) & -4.0 (0.29) & 10.5 (0.07)* & 6.4 (0.21) & -3.7 (0.33) & 10.1 (0.07)* & 5.2 (0.30) & -5.2 (0.15) & 10.3 (0.06)* \\
 -6 & 8.6 (0.16) & -1.7 (0.68) & 10.3 (0.14) & 9.6 (0.11) & -2.7 (0.51) & 12.4 (0.07)* & 8.9 (0.12) & -4.7 (0.24) & 13.6 (0.04)** & 8.7 (0.12) & -4.5 (0.25) & 13.2 (0.04)** & 7.6 (0.16) & -5.4 (0.16) & 13.0 (0.04)** \\
 -5 & 8.4 (0.15) & -2.5 (0.63) & 10.9 (0.15) & 9.6 (0.11) & -3.9 (0.45) & 13.5 (0.07)* & 9.1 (0.12) & -6.2 (0.20) & 15.4 (0.04)** & 8.8 (0.12) & -5.9 (0.21) & 14.6 (0.04)** & 7.7 (0.16) & -7.2 (0.11) & 14.9 (0.03)** \\
 -4 & 7.5 (0.20) & -2.8 (0.59) & 10.3 (0.18) & 8.9 (0.13) & -3.7 (0.46) & 12.7 (0.10)* & 8.5 (0.14) & -6.1 (0.21) & 14.6 (0.05)** & 8.2 (0.14) & -5.8 (0.22) & 14.0 (0.05)** & 7.6 (0.15) & -7.0 (0.13) & 14.5 (0.04)** \\
 -3 & 7.0 (0.25) & -4.2 (0.39) & 11.1 (0.16) & 8.5 (0.17) & -4.7 (0.34) & 13.2 (0.09)* & 8.1 (0.17) & -6.5 (0.17) & 14.6 (0.06)* & 7.7 (0.18) & -6.4 (0.16) & 14.1 (0.06)* & 6.4 (0.24) & -7.2 (0.10) & 13.7 (0.06)* \\
 -2 & 6.7 (0.30) & -6.5 (0.15) & 13.2 (0.11) & 8.5 (0.20) & -7.2 (0.12) & 15.8 (0.06)* & 8.5 (0.19) & -8.5 (0.06)* & 17.0 (0.04)** & 7.9 (0.20) & -8.3 (0.06)* & 16.2 (0.04)** & 7.1 (0.23) & -8.8 (0.04)** & 15.9 (0.04)** \\
 -1 & 5.3 (0.43) & -8.7 (0.05)* & 14.0 (0.10) & 7.1 (0.30) & -9.2 (0.05)* & 16.3 (0.06)* & 7.1 (0.28) & -9.6 (0.05)** & 16.7 (0.05)* & 6.6 (0.30) & -9.8 (0.04)** & 16.3 (0.05)** & 5.9 (0.32) & -9.4 (0.05)* & 15.4 (0.06)* \\
  0 & 7.2 (0.30) & -9.1 (0.07)* & 16.3 (0.06)* & 8.9 (0.21) & -10.4 (0.04)** & 19.3 (0.03)** & 8.7 (0.19) & -11.2 (0.03)** & 20.0 (0.03)** & 8.3 (0.20) & -11.1 (0.03)** & 19.3 (0.02)** & 7.7 (0.20) & -10.8 (0.03)** & 18.5 (0.03)** \\
  1 & 9.4 (0.21) & -10.7 (0.03)** & 20.1 (0.03)** & 11.4 (0.13) & -12.7 (0.01)** & 24.1 (0.01)*** & 11.0 (0.12) & -13.1 (0.01)** & 24.2 (0.01)*** & 10.8 (0.11) & -13.2 (0.01)*** & 24.0 (0.01)*** & 10.4 (0.11) & -12.5 (0.01)** & 22.9 (0.01)*** \\
  2 & 8.7 (0.24) & -10.1 (0.07)* & 18.8 (0.04)** & 11.2 (0.13) & -12.0 (0.03)** & 23.2 (0.01)** & 11.4 (0.11) & -13.1 (0.01)** & 24.5 (0.01)*** & 11.2 (0.10) & -13.2 (0.01)*** & 24.3 (0.01)*** & 11.4 (0.09)* & -12.8 (0.01)** & 24.2 (0.01)*** \\
  3 & 9.6 (0.21) & -10.6 (0.04)** & 20.2 (0.03)** & 12.0 (0.12) & -12.5 (0.02)** & 24.4 (0.01)** & 11.9 (0.11) & -13.5 (0.01)** & 25.4 (0.01)*** & 11.6 (0.10)* & -13.5 (0.01)*** & 25.2 (0.01)*** & 11.5 (0.09)* & -12.9 (0.02)** & 24.4 (0.01)*** \\
  4 & 11.0 (0.17) & -9.8 (0.06)* & 20.8 (0.04)** & 13.6 (0.10)* & -11.5 (0.03)** & 25.0 (0.01)** & 13.4 (0.09)* & -12.9 (0.02)** & 26.3 (0.01)*** & 13.4 (0.07)* & -12.7 (0.01)** & 26.1 (0.01)*** & 13.3 (0.07)* & -12.2 (0.02)** & 25.4 (0.01)*** \\
  5 & 12.0 (0.13) & -11.6 (0.04)** & 23.6 (0.02)** & 14.4 (0.08)* & -13.3 (0.02)** & 27.7 (0.01)*** & 14.2 (0.07)* & -15.1 (0.01)*** & 29.3 (<0.01)*** & 14.2 (0.05)* & -14.5 (0.01)*** & 28.6 (<0.01)*** & 13.9 (0.05)** & -14.3 (0.01)** & 28.2 (<0.01)*** \\
  6 & 9.7 (0.21) & -14.0 (0.02)** & 23.7 (0.02)** & 12.0 (0.13) & -15.4 (0.01)** & 27.4 (0.01)*** & 11.9 (0.12) & -16.7 (0.01)*** & 28.7 (0.01)*** & 11.7 (0.11) & -16.0 (0.01)*** & 27.8 (0.01)*** & 12.0 (0.09)* & -15.4 (0.01)** & 27.5 (0.01)*** \\
  7 & 11.3 (0.16) & -14.1 (0.03)** & 25.3 (0.02)** & 13.2 (0.11) & -15.4 (0.02)** & 28.6 (0.01)** & 12.7 (0.11) & -16.1 (0.01)** & 28.8 (0.01)** & 12.7 (0.09)* & -15.4 (0.01)** & 28.2 (0.01)*** & 12.5 (0.09)* & -14.2 (0.03)** & 26.7 (0.01)** \\
  8 & 12.3 (0.13) & -14.2 (0.03)** & 26.5 (0.01)** & 14.5 (0.08)* & -15.5 (0.02)** & 30.0 (0.01)*** & 14.1 (0.08)* & -16.2 (0.02)** & 30.3 (0.01)*** & 14.0 (0.06)* & -15.6 (0.02)** & 29.6 (0.01)*** & 13.9 (0.06)* & -14.2 (0.03)** & 28.1 (0.01)*** \\
  9 & 11.7 (0.14) & -12.1 (0.07)* & 23.8 (0.02)** & 14.1 (0.08)* & -13.9 (0.04)** & 28.0 (0.01)*** & 14.2 (0.07)* & -14.9 (0.04)** & 29.1 (0.01)*** & 14.2 (0.06)* & -14.3 (0.04)** & 28.5 (0.01)*** & 14.5 (0.05)* & -13.3 (0.06)* & 27.8 (0.01)*** \\
 10 & 12.1 (0.12) & -11.2 (0.10) & 23.3 (0.02)** & 14.4 (0.07)* & -12.7 (0.08)* & 27.1 (0.01)** & 14.4 (0.07)* & -14.0 (0.07)* & 28.3 (0.01)*** & 14.4 (0.06)* & -13.4 (0.06)* & 27.8 (0.01)*** & 14.7 (0.05)** & -12.5 (0.10)* & 27.2 (0.01)** \\
 11 & 12.7 (0.10)* & -10.4 (0.14) & 23.1 (0.02)** & 15.5 (0.05)* & -12.0 (0.11) & 27.4 (0.01)** & 16.1 (0.04)** & -13.4 (0.09)* & 29.5 (0.01)*** & 16.3 (0.03)** & -12.8 (0.08)* & 29.2 (0.01)*** & 17.3 (0.02)** & -12.2 (0.12) & 29.4 (0.01)*** \\
 12 & 14.6 (0.06)* & -9.2 (0.20) & 23.8 (0.02)** & 17.1 (0.03)** & -10.4 (0.17) & 27.5 (0.01)** & 17.4 (0.03)** & -11.8 (0.14) & 29.2 (0.01)** & 17.7 (0.02)** & -11.1 (0.14) & 28.9 (0.01)*** & 18.3 (0.02)** & -10.3 (0.19) & 28.5 (0.01)** \\
 13 & 14.0 (0.07)* & -10.7 (0.13) & 24.6 (0.02)** & 15.9 (0.05)** & -12.0 (0.11) & 27.9 (0.01)** & 16.0 (0.05)** & -13.2 (0.09)* & 29.2 (0.01)** & 15.9 (0.04)** & -12.2 (0.10)* & 28.1 (0.01)** & 16.9 (0.03)** & -11.8 (0.12) & 28.6 (0.01)** \\
 14 & 14.4 (0.07)* & -12.3 (0.09)* & 26.7 (0.02)** & 16.1 (0.05)** & -13.7 (0.08)* & 29.8 (0.01)** & 15.9 (0.05)** & -14.4 (0.08)* & 30.2 (0.01)** & 16.0 (0.04)** & -13.3 (0.08)* & 29.3 (0.01)** & 16.3 (0.03)** & -12.2 (0.12) & 28.6 (0.02)** \\
 15 & 13.6 (0.09)* & -15.2 (0.04)** & 28.8 (0.01)** & 15.8 (0.05)* & -16.8 (0.04)** & 32.5 (0.01)*** & 15.6 (0.05)* & -17.1 (0.04)** & 32.7 (0.01)*** & 15.6 (0.04)** & -16.1 (0.04)** & 31.7 (0.01)*** & 16.2 (0.04)** & -14.6 (0.06)* & 30.8 (0.01)*** \\
\hline
Constant & \multicolumn{3}{c}{-0.1 (0.74)} & \multicolumn{3}{c}{-0.1 (0.74)} & \multicolumn{3}{c}{-0.1 (0.84)} & \multicolumn{3}{c}{-0.1 (0.81)} & \multicolumn{3}{c}{-0.1 (0.80)} \\
\hline\hline
\end{tabular}

        }
        \caption{Corporate bond yield event study regression estimates (basis points)}
        
        \begin{tablenotes}
        \small
        \item Notes: * p$<$0.10, ** p$<$0.05, *** p$<$0.01. Estimates of cumulative returns (in basis points) at each day relative to 15 days before event. Constant captures baseline daily return. OLS regression using equation~\eqref{eq:cum_return_paired}. P-values in parentheses, computed using HAC standard errors (maxlags=30). Same estimates are plotted in Figure~\ref{fig:regression_difference_corporate}.
        \end{tablenotes}
        \label{tab:regression_estimates_corporate}
    \end{table}
    
    \begin{table}
        \resizebox{9in}{!}{            \begin{tabular}{lcccccccccccc}
\hline\hline
Day & \multicolumn{3}{c}{5Y TIPS} & \multicolumn{3}{c}{10Y TIPS} & \multicolumn{3}{c}{20Y TIPS} & \multicolumn{3}{c}{30Y TIPS} \\
     & Open & Closed & Diff. & Open & Closed & Diff. & Open & Closed & Diff. & Open & Closed & Diff. \\
\hline
-15 & 1.1 (0.43) & 0.9 (0.30) & 0.2 (0.86) & 0.9 (0.41) & 0.9 (0.28) & 0.0 (0.98) & 1.0 (0.29) & 0.6 (0.50) & 0.4 (0.65) & 0.8 (0.35) & 0.6 (0.49) & 0.3 (0.78) \\
-14 & 1.7 (0.37) & 0.1 (0.95) & 1.6 (0.48) & 1.9 (0.28) & -0.0 (1.00) & 1.9 (0.35) & 1.8 (0.24) & -0.1 (0.95) & 1.9 (0.30) & 1.4 (0.35) & 0.0 (0.97) & 1.3 (0.45) \\
-13 & 4.2 (0.03)** & -0.6 (0.74) & 4.8 (0.06)* & 3.7 (0.05)** & -1.5 (0.35) & 5.2 (0.03)** & 3.3 (0.05)** & -1.8 (0.22) & 5.1 (0.02)** & 2.8 (0.09)* & -1.5 (0.26) & 4.4 (0.03)** \\
-12 & 3.2 (0.13) & -1.0 (0.68) & 4.2 (0.18) & 2.8 (0.19) & -2.5 (0.21) & 5.3 (0.06)* & 2.3 (0.27) & -2.9 (0.10) & 5.2 (0.04)** & 2.0 (0.33) & -2.7 (0.11) & 4.7 (0.06)* \\
-11 & 3.8 (0.15) & -1.2 (0.60) & 5.1 (0.12) & 2.8 (0.28) & -2.2 (0.27) & 5.0 (0.10) & 2.3 (0.37) & -2.4 (0.18) & 4.7 (0.11) & 1.7 (0.49) & -2.3 (0.18) & 4.0 (0.15) \\
-10 & 4.1 (0.15) & -2.0 (0.40) & 6.1 (0.08)* & 2.8 (0.33) & -3.3 (0.12) & 6.1 (0.07)* & 2.4 (0.38) & -3.7 (0.06)* & 6.2 (0.05)** & 2.2 (0.40) & -3.6 (0.06)* & 5.8 (0.05)* \\
 -9 & 4.4 (0.16) & -1.0 (0.67) & 5.4 (0.14) & 3.2 (0.31) & -2.7 (0.20) & 5.9 (0.09)* & 2.9 (0.35) & -3.5 (0.08)* & 6.4 (0.06)* & 2.9 (0.34) & -3.5 (0.07)* & 6.4 (0.06)* \\
 -8 & 4.6 (0.21) & 1.3 (0.64) & 3.4 (0.43) & 3.2 (0.36) & -0.8 (0.74) & 4.1 (0.30) & 2.6 (0.43) & -2.0 (0.41) & 4.6 (0.23) & 2.6 (0.40) & -2.3 (0.33) & 4.9 (0.19) \\
 -7 & 5.4 (0.21) & -0.5 (0.86) & 5.9 (0.21) & 3.8 (0.36) & -2.8 (0.29) & 6.6 (0.13) & 3.1 (0.43) & -4.2 (0.09)* & 7.3 (0.08)* & 3.0 (0.43) & -4.4 (0.07)* & 7.4 (0.07)* \\
 -6 & 6.2 (0.19) & -1.4 (0.62) & 7.6 (0.14) & 5.0 (0.26) & -3.5 (0.20) & 8.5 (0.07)* & 4.5 (0.27) & -4.8 (0.07)* & 9.3 (0.04)** & 4.3 (0.27) & -4.7 (0.07)* & 9.0 (0.04)** \\
 -5 & 5.2 (0.29) & -0.7 (0.85) & 5.9 (0.30) & 4.6 (0.31) & -3.7 (0.25) & 8.3 (0.11) & 4.3 (0.30) & -5.3 (0.09)* & 9.6 (0.05)* & 4.1 (0.31) & -5.5 (0.07)* & 9.6 (0.05)** \\
 -4 & 3.8 (0.44) & 0.6 (0.87) & 3.1 (0.60) & 3.7 (0.40) & -2.5 (0.48) & 6.2 (0.24) & 3.7 (0.34) & -3.9 (0.24) & 7.7 (0.12) & 3.7 (0.32) & -4.2 (0.20) & 7.9 (0.10)* \\
 -3 & 4.0 (0.42) & -1.1 (0.77) & 5.1 (0.41) & 3.5 (0.44) & -3.6 (0.30) & 7.1 (0.20) & 3.3 (0.40) & -4.7 (0.15) & 8.0 (0.11) & 3.3 (0.37) & -4.7 (0.13) & 8.0 (0.09)* \\
 -2 & 4.3 (0.41) & -3.4 (0.31) & 7.7 (0.21) & 3.1 (0.52) & -5.6 (0.07)* & 8.7 (0.13) & 3.1 (0.48) & -6.4 (0.04)** & 9.5 (0.07)* & 2.9 (0.47) & -6.3 (0.04)** & 9.2 (0.07)* \\
 -1 & 2.9 (0.60) & -5.7 (0.09)* & 8.6 (0.21) & 2.0 (0.69) & -7.2 (0.04)** & 9.2 (0.14) & 2.1 (0.63) & -7.4 (0.03)** & 9.5 (0.10)* & 2.1 (0.60) & -6.8 (0.05)** & 9.0 (0.10)* \\
  0 & 5.4 (0.36) & -7.8 (0.03)** & 13.1 (0.06)* & 4.0 (0.42) & -9.0 (0.01)*** & 13.0 (0.04)** & 3.7 (0.39) & -8.8 (0.01)** & 12.5 (0.03)** & 3.7 (0.35) & -8.0 (0.02)** & 11.6 (0.04)** \\
  1 & 7.3 (0.24) & -10.6 (0.01)*** & 17.9 (0.02)** & 5.9 (0.26) & -11.8 (<0.01)*** & 17.7 (0.01)*** & 5.6 (0.21) & -11.4 (<0.01)*** & 16.9 (<0.01)*** & 5.5 (0.17) & -10.2 (<0.01)*** & 15.8 (0.01)*** \\
  2 & 5.9 (0.35) & -9.8 (0.02)** & 15.7 (0.04)** & 5.1 (0.33) & -11.6 (<0.01)*** & 16.7 (0.01)** & 5.4 (0.23) & -11.5 (<0.01)*** & 16.9 (0.01)*** & 5.6 (0.18) & -10.6 (0.01)*** & 16.2 (0.01)*** \\
  3 & 5.0 (0.42) & -9.0 (0.03)** & 14.0 (0.07)* & 4.7 (0.38) & -10.7 (0.01)*** & 15.3 (0.03)** & 5.0 (0.27) & -10.4 (0.01)*** & 15.4 (0.02)** & 5.0 (0.23) & -9.4 (0.02)** & 14.4 (0.02)** \\
  4 & 6.4 (0.34) & -7.7 (0.06)* & 14.1 (0.09)* & 5.9 (0.32) & -9.6 (0.01)** & 15.5 (0.03)** & 6.4 (0.22) & -9.6 (0.01)*** & 16.1 (0.02)** & 6.4 (0.19) & -8.8 (0.02)** & 15.2 (0.02)** \\
  5 & 6.8 (0.31) & -8.8 (0.05)** & 15.6 (0.06)* & 6.8 (0.23) & -10.6 (0.01)*** & 17.4 (0.02)** & 6.9 (0.17) & -10.8 (0.01)*** & 17.6 (0.01)*** & 6.7 (0.14) & -10.0 (0.01)*** & 16.6 (0.01)*** \\
  6 & 5.1 (0.44) & -11.4 (0.02)** & 16.5 (0.05)* & 5.0 (0.37) & -11.8 (0.01)*** & 16.8 (0.02)** & 5.3 (0.28) & -11.5 (0.01)*** & 16.8 (0.01)** & 5.2 (0.26) & -10.8 (0.01)** & 16.0 (0.02)** \\
  7 & 7.1 (0.30) & -12.2 (0.02)** & 19.3 (0.04)** & 6.3 (0.29) & -11.9 (0.01)*** & 18.1 (0.02)** & 6.0 (0.25) & -10.8 (0.01)** & 16.8 (0.02)** & 5.6 (0.24) & -9.7 (0.02)** & 15.4 (0.03)** \\
  8 & 8.5 (0.22) & -12.1 (0.02)** & 20.6 (0.02)** & 7.6 (0.20) & -12.5 (0.01)*** & 20.2 (0.01)** & 7.1 (0.17) & -11.6 (0.01)*** & 18.7 (0.01)** & 6.6 (0.17) & -10.3 (0.02)** & 17.0 (0.02)** \\
  9 & 8.0 (0.25) & -10.1 (0.04)** & 18.2 (0.03)** & 7.1 (0.24) & -11.3 (0.02)** & 18.5 (0.02)** & 6.9 (0.20) & -10.6 (0.02)** & 17.5 (0.01)** & 6.7 (0.19) & -9.4 (0.03)** & 16.1 (0.02)** \\
 10 & 7.9 (0.25) & -9.2 (0.08)* & 17.0 (0.04)** & 6.9 (0.24) & -10.7 (0.04)** & 17.6 (0.02)** & 6.7 (0.20) & -10.1 (0.04)** & 16.8 (0.02)** & 6.6 (0.19) & -9.0 (0.05)* & 15.5 (0.02)** \\
 11 & 8.2 (0.21) & -9.3 (0.08)* & 17.5 (0.03)** & 8.4 (0.14) & -10.8 (0.04)** & 19.2 (0.01)** & 8.8 (0.09)* & -10.2 (0.04)** & 19.1 (0.01)*** & 8.6 (0.08)* & -9.0 (0.06)* & 17.7 (0.01)** \\
 12 & 8.6 (0.18) & -7.1 (0.21) & 15.7 (0.06)* & 8.6 (0.12) & -8.5 (0.12) & 17.1 (0.03)** & 8.9 (0.09)* & -8.1 (0.11) & 17.0 (0.02)** & 8.7 (0.08)* & -6.8 (0.16) & 15.5 (0.03)** \\
 13 & 6.6 (0.30) & -8.4 (0.15) & 15.0 (0.07)* & 6.9 (0.21) & -9.7 (0.08)* & 16.6 (0.03)** & 7.3 (0.16) & -9.1 (0.07)* & 16.4 (0.02)** & 7.2 (0.15) & -7.8 (0.10)* & 15.0 (0.03)** \\
 14 & 6.7 (0.30) & -10.0 (0.09)* & 16.8 (0.06)* & 6.7 (0.23) & -10.8 (0.05)* & 17.5 (0.03)** & 6.8 (0.19) & -9.6 (0.06)* & 16.4 (0.03)** & 6.6 (0.17) & -8.1 (0.09)* & 14.8 (0.04)** \\
 15 & 6.4 (0.33) & -11.7 (0.05)* & 18.2 (0.04)** & 6.5 (0.25) & -12.4 (0.03)** & 18.9 (0.02)** & 6.7 (0.20) & -11.2 (0.03)** & 17.9 (0.02)** & 6.8 (0.17) & -9.6 (0.05)** & 16.4 (0.02)** \\
\hline
Constant & \multicolumn{3}{c}{0.1 (0.73)} & \multicolumn{3}{c}{0.2 (0.43)} & \multicolumn{3}{c}{0.2 (0.34)} & \multicolumn{3}{c}{0.2 (0.37)} \\
\hline\hline
\end{tabular}

        }
        \caption{TIPS yield event study regression estimates (basis points)}
        \begin{tablenotes}
        \small
        \item Notes: * p$<$0.10, ** p$<$0.05, *** p$<$0.01. Estimates of cumulative returns (in basis points) at each day relative to 15 days before event. Constant captures baseline daily return. OLS regression using equation~\eqref{eq:cum_return_paired}. P-values in parentheses, computed using HAC standard errors (maxlags=30). Same estimates are plotted in Figure~\ref{fig:regression_difference_tips}.
        \end{tablenotes}
        \label{tab:regression_estimates_tips}
    \end{table}

\end{landscape}

\end{document}